\documentclass[12pt,english]{article}
\usepackage{mathpazo}
\usepackage[latin9]{inputenc}
\usepackage[letterpaper]{geometry}
\usepackage{babel}
\usepackage{verbatim}
\usepackage{amsmath}
\usepackage{amssymb}
\usepackage{graphicx}
\usepackage[unicode=true,
 bookmarks=true,bookmarksnumbered=false,bookmarksopen=false,
 breaklinks=false,pdfborder={0 0 1},backref=false,colorlinks=false]
 {hyperref}

\makeatletter

\providecommand{\tabularnewline}{\\}

\numberwithin{equation}{section}

\@ifundefined{date}{}{\date{}}

\usepackage[svgnames]{xcolor}
\usepackage{tikz}
\usetikzlibrary{decorations.markings}
\usetikzlibrary{shapes.geometric}
\usetikzlibrary{calc}
\usetikzlibrary{arrows}

\pgfdeclarelayer{edgelayer}
\pgfdeclarelayer{nodelayer}
\pgfsetlayers{edgelayer,nodelayer,main}

\tikzstyle{none}=[inner sep=0pt]

\tikzstyle{Vertex}=[circle,fill=black,draw=black]
\tikzstyle{Node}=[rectangle,fill=blue,draw=blue]
\tikzstyle{Point}=[circle,fill=red,draw=red,scale=0.33]

\tikzstyle{Edge}=[-,draw=black]
\tikzstyle{Link}=[-,dashed,draw=blue]
\tikzstyle{Segment}=[-,dotted,very thick,draw=red]
\tikzstyle{LinkArrow}=[-,draw=blue,
  postaction={decorate},decoration={markings,
  mark=at position .66 with {\arrow{latex}}}]
\tikzstyle{SegmentArrow}=[-,draw=red,
  postaction={decorate},decoration={markings,
  mark=at position .66 with {\arrow{latex}}}]
\def\centerarc [#1] (#2) (#3:#4:#5); { \draw[#1] ($(#2)+({#5*cos(#3)},{#5*sin(#3)})$) arc (#3:#4:#5); }

\renewcommand\[{\begin{equation}}
\renewcommand\]{\end{equation}}

\usepackage[margin=1cm]{caption}

\DeclareMathOperator{\e}{e}

\DeclareMathOperator{\ii}{i}

\DeclareMathOperator{\sign}{sign}

\makeatother

\begin{document}

\global\long\def\A{\mathbf{A}}%
\global\long\def\B{\mathbf{B}}%
\global\long\def\C{\mathbf{C}}%
\global\long\def\D{\mathbf{D}}%
\global\long\def\E{\mathbf{E}}%
\global\long\def\F{\mathbf{F}}%
\global\long\def\G{\mathbf{G}}%
\global\long\def\H{\mathbf{H}}%
\global\long\def\I{\mathbf{I}}%
\global\long\def\J{\mathbf{J}}%
\global\long\def\K{\mathbf{K}}%
\global\long\def\LL{\mathbf{L}}%
\global\long\def\M{\mathbf{M}}%
\global\long\def\N{\mathbf{N}}%
\global\long\def\OO{\mathbf{O}}%
\global\long\def\P{\mathbf{P}}%
\global\long\def\Q{\mathbf{Q}}%
\global\long\def\RR{\mathbf{R}}%
\global\long\def\SS{\mathbf{S}}%
\global\long\def\T{\mathbf{T}}%
\global\long\def\U{\mathbf{U}}%
\global\long\def\V{\mathbf{V}}%
\global\long\def\W{\mathbf{W}}%
\global\long\def\X{\mathbf{X}}%
\global\long\def\Y{\mathbf{Y}}%
\global\long\def\Z{\mathbf{Z}}%

\global\long\def\a{\mathbf{a}}%
\global\long\def\b{\mathbf{b}}%
\global\long\def\c{\mathbf{c}}%
\global\long\def\dd{\mathbf{d}}%
\global\long\def\ee{\mathbf{e}}%
\global\long\def\f{\mathbf{f}}%
\global\long\def\g{\mathbf{g}}%
\global\long\def\h{\mathbf{h}}%
\global\long\def\iii{\mathbf{i}}%
\global\long\def\j{\mathbf{j}}%
\global\long\def\k{\mathbf{k}}%
\global\long\def\l{\boldsymbol{l}}%
\global\long\def\el{\boldsymbol{\ell}}%
\global\long\def\m{\mathbf{m}}%
\global\long\def\n{\mathbf{n}}%
\global\long\def\o{\mathbf{o}}%
\global\long\def\p{\mathbf{p}}%
\global\long\def\q{\mathbf{q}}%
\global\long\def\r{\mathbf{r}}%
\global\long\def\s{\mathbf{s}}%
\global\long\def\t{\mathbf{t}}%
\global\long\def\u{\mathbf{u}}%
\global\long\def\v{\mathbf{v}}%
\global\long\def\w{\mathbf{w}}%
\global\long\def\x{\mathbf{x}}%
\global\long\def\y{\mathbf{y}}%
\global\long\def\z{\mathbf{z}}%

\global\long\def\Ga{\boldsymbol{\Gamma}}%
\global\long\def\De{\boldsymbol{\Delta}}%
\global\long\def\Th{\boldsymbol{\Theta}}%
\global\long\def\La{\boldsymbol{\Lambda}}%
\global\long\def\Xii{\boldsymbol{\Xi}}%
\global\long\def\Pii{\boldsymbol{\Pi}}%
\global\long\def\Si{\boldsymbol{\Sigma}}%
\global\long\def\Ph{\boldsymbol{\Phi}}%
\global\long\def\Ps{\boldsymbol{\Psi}}%
\global\long\def\Om{\boldsymbol{\Omega}}%

\global\long\def\al{\boldsymbol{\alpha}}%
\global\long\def\be{\boldsymbol{\beta}}%
\global\long\def\ga{\boldsymbol{\gamma}}%
\global\long\def\de{\boldsymbol{\delta}}%
\global\long\def\ep{\boldsymbol{\epsilon}}%
\global\long\def\vep{\boldsymbol{\varepsilon}}%
\global\long\def\ze{\boldsymbol{\zeta}}%
\global\long\def\et{\boldsymbol{\eta}}%
\global\long\def\th{\boldsymbol{\theta}}%
\global\long\def\io{\boldsymbol{\iota}}%
\global\long\def\ka{\boldsymbol{\kappa}}%
\global\long\def\la{\boldsymbol{\lambda}}%
\global\long\def\muu{\boldsymbol{\mu}}%
\global\long\def\nuu{\boldsymbol{\nu}}%
\global\long\def\xii{\boldsymbol{\xi}}%
\global\long\def\pii{\boldsymbol{\pi}}%
\global\long\def\rhh{\boldsymbol{\rho}}%
\global\long\def\si{\boldsymbol{\sigma}}%
\global\long\def\ta{\boldsymbol{\tau}}%
\global\long\def\ups{\boldsymbol{\upsilon}}%
\global\long\def\ph{\boldsymbol{\phi}}%
\global\long\def\vph{\boldsymbol{\varphi}}%
\global\long\def\ch{\boldsymbol{\chi}}%
\global\long\def\ps{\boldsymbol{\psi}}%
\global\long\def\om{\boldsymbol{\omega}}%

\global\long\def\AAb{\boldsymbol{\mathcal{A}}}%
\global\long\def\BBb{\boldsymbol{\mathcal{B}}}%
\global\long\def\CCb{\boldsymbol{\mathcal{C}}}%
\global\long\def\DDb{\boldsymbol{\mathcal{D}}}%
\global\long\def\EEb{\boldsymbol{\mathcal{E}}}%
\global\long\def\FFb{\boldsymbol{\mathcal{F}}}%
\global\long\def\GGb{\boldsymbol{\mathcal{G}}}%
\global\long\def\HHb{\boldsymbol{\mathcal{H}}}%
\global\long\def\IIb{\boldsymbol{\mathcal{I}}}%
\global\long\def\JJb{\boldsymbol{\mathcal{J}}}%
\global\long\def\KKb{\boldsymbol{\mathcal{K}}}%
\global\long\def\LLb{\boldsymbol{\mathcal{L}}}%
\global\long\def\MMb{\boldsymbol{\mathcal{M}}}%
\global\long\def\NNb{\boldsymbol{\mathcal{N}}}%
\global\long\def\OOb{\boldsymbol{\mathcal{O}}}%
\global\long\def\PPb{\boldsymbol{\mathcal{P}}}%
\global\long\def\QQb{\boldsymbol{\mathcal{Q}}}%
\global\long\def\RRb{\boldsymbol{\mathcal{R}}}%
\global\long\def\SSb{\boldsymbol{\mathcal{S}}}%
\global\long\def\TTb{\boldsymbol{\mathcal{T}}}%
\global\long\def\UUb{\boldsymbol{\mathcal{U}}}%
\global\long\def\VVb{\boldsymbol{\mathcal{V}}}%
\global\long\def\WWb{\boldsymbol{\mathcal{W}}}%
\global\long\def\XXb{\boldsymbol{\mathcal{X}}}%
\global\long\def\YYb{\boldsymbol{\mathcal{Y}}}%
\global\long\def\ZZb{\boldsymbol{\mathcal{Z}}}%

\global\long\def\Ab{\bar{A}}%
\global\long\def\Bb{\bar{B}}%
\global\long\def\Cb{\bar{C}}%
\global\long\def\Db{\bar{D}}%
\global\long\def\Eb{\bar{E}}%
\global\long\def\Fb{\bar{F}}%
\global\long\def\Gb{\bar{G}}%
\global\long\def\Hb{\bar{H}}%
\global\long\def\Ib{\bar{I}}%
\global\long\def\Jb{\bar{J}}%
\global\long\def\Kb{\bar{K}}%
\global\long\def\Lb{\bar{L}}%
\global\long\def\Mb{\bar{M}}%
\global\long\def\Nb{\bar{N}}%
\global\long\def\Ob{\bar{O}}%
\global\long\def\Pb{\bar{P}}%
\global\long\def\Qb{\bar{Q}}%
\global\long\def\Rb{\bar{R}}%
\global\long\def\Sb{\bar{S}}%
\global\long\def\Tb{\bar{T}}%
\global\long\def\Ub{\bar{U}}%
\global\long\def\Vb{\bar{V}}%
\global\long\def\Wb{\bar{W}}%
\global\long\def\Xb{\bar{X}}%
\global\long\def\Yb{\bar{Y}}%
\global\long\def\Zb{\bar{Z}}%

\global\long\def\ab{\bar{a}}%
\global\long\def\bb{\bar{b}}%
\global\long\def\cb{\bar{c}}%
\global\long\def\db{\bar{d}}%
\global\long\def\eb{\bar{e}}%
\global\long\def\fb{\bar{f}}%
\global\long\def\gb{\bar{g}}%
\global\long\def\hb{\bar{h}}%
\global\long\def\ib{\bar{i}}%
\global\long\def\jb{\bar{j}}%
\global\long\def\kb{\bar{k}}%
\global\long\def\lb{\bar{l}}%
\global\long\def\elb{\bar{\ell}}%
\global\long\def\mb{\bar{m}}%
\global\long\def\nb{\bar{n}}%
\global\long\def\ob{\bar{o}}%
\global\long\def\pb{\bar{p}}%
\global\long\def\qb{\bar{q}}%
\global\long\def\rb{\bar{r}}%
\global\long\def\ssb{\bar{s}}%
\global\long\def\tb{\bar{t}}%
\global\long\def\ub{\bar{u}}%
\global\long\def\vb{\bar{v}}%
\global\long\def\wb{\bar{w}}%
\global\long\def\xb{\bar{x}}%
\global\long\def\yb{\bar{y}}%
\global\long\def\zb{\bar{z}}%

\global\long\def\Gab{\bar{\Gamma}}%
\global\long\def\Deb{\bar{\Delta}}%
\global\long\def\Thb{\bar{\Theta}}%
\global\long\def\Lab{\bar{\Lambda}}%
\global\long\def\Xib{\bar{\Xi}}%
\global\long\def\Pib{\bar{\Pi}}%
\global\long\def\Sib{\bar{\Sigma}}%
\global\long\def\Phb{\bar{\Phi}}%
\global\long\def\Psb{\bar{\Psi}}%
\global\long\def\Thb{\bar{\Theta}}%

\global\long\def\alb{\bar{\alpha}}%
\global\long\def\beb{\bar{\beta}}%
\global\long\def\gab{\bar{\gamma}}%
\global\long\def\deb{\bar{\delta}}%
\global\long\def\epb{\bar{\epsilon}}%
\global\long\def\vepb{\bar{\varepsilon}}%
\global\long\def\zeb{\bar{\zeta}}%
\global\long\def\etb{\bar{\eta}}%
\global\long\def\thb{\bar{\theta}}%
\global\long\def\iob{\bar{\iota}}%
\global\long\def\kab{\bar{\kappa}}%
\global\long\def\lab{\bar{\lambda}}%
\global\long\def\mub{\bar{\mu}}%
\global\long\def\nub{\bar{\nu}}%
\global\long\def\xib{\bar{\xi}}%
\global\long\def\pib{\bar{\pi}}%
\global\long\def\rhb{\bar{\rho}}%
\global\long\def\sib{\bar{\sigma}}%
\global\long\def\tab{\bar{\tau}}%
\global\long\def\upb{\bar{\upsilon}}%
\global\long\def\phb{\bar{\phi}}%
\global\long\def\vphb{\bar{\varphi}}%
\global\long\def\chb{\bar{\chi}}%
\global\long\def\psb{\bar{\psi}}%
\global\long\def\omb{\bar{\omega}}%

\global\long\def\adt{\dot{a}}%
\global\long\def\add{\ddot{a}}%
\global\long\def\bd{\dot{b}}%
\global\long\def\bdd{\ddot{b}}%
\global\long\def\cd{\dot{c}}%
\global\long\def\cdd{\ddot{c}}%
\global\long\def\ddd{\dot{d}}%
\global\long\def\dddd{\ddot{d}}%
\global\long\def\ed{\dot{e}}%
\global\long\def\edd{\ddot{e}}%
\global\long\def\fd{\dot{f}}%
\global\long\def\fdd{\ddot{f}}%
\global\long\def\gd{\dot{g}}%
\global\long\def\gdd{\ddot{g}}%
\global\long\def\hd{\dot{h}}%
\global\long\def\hdd{\ddot{h}}%
\global\long\def\kd{\dot{k}}%
\global\long\def\kdd{\ddot{k}}%
\global\long\def\ld{\dot{l}}%
\global\long\def\ldd{\ddot{l}}%
\global\long\def\eld{\dot{\ell}}%
\global\long\def\eldd{\ddot{\ell}}%
\global\long\def\md{\dot{m}}%
\global\long\def\mdd{\ddot{m}}%
\global\long\def\nd{\dot{n}}%
\global\long\def\ndd{\ddot{n}}%
\global\long\def\od{\dot{o}}%
\global\long\def\odd{\ddot{o}}%
\global\long\def\pd{\dot{p}}%
\global\long\def\pdd{\ddot{p}}%
\global\long\def\qd{\dot{q}}%
\global\long\def\qdd{\ddot{q}}%
\global\long\def\rd{\dot{r}}%
\global\long\def\rdd{\ddot{r}}%
\global\long\def\sd{\dot{s}}%
\global\long\def\sdd{\ddot{s}}%
\global\long\def\td{\dot{t}}%
\global\long\def\tdd{\ddot{t}}%
\global\long\def\ud{\dot{u}}%
\global\long\def\udd{\ddot{u}}%
\global\long\def\vd{\dot{v}}%
\global\long\def\vdd{\ddot{v}}%
\global\long\def\wdt{\dot{w}}%
\global\long\def\wdd{\ddot{w}}%
\global\long\def\xd{\dot{x}}%
\global\long\def\xdd{\ddot{x}}%
\global\long\def\yd{\dot{y}}%
\global\long\def\ydd{\ddot{y}}%
\global\long\def\zd{\dot{z}}%
\global\long\def\zdd{\ddot{z}}%

\global\long\def\Adt{\dot{A}}%
\global\long\def\Add{\ddot{A}}%
\global\long\def\Bd{\dot{B}}%
\global\long\def\Bdd{\ddot{B}}%
\global\long\def\Cd{\dot{C}}%
\global\long\def\Cdd{\ddot{C}}%
\global\long\def\Dd{\dot{D}}%
\global\long\def\Ddd{\ddot{D}}%
\global\long\def\Ed{\dot{E}}%
\global\long\def\Edd{\ddot{E}}%
\global\long\def\Fd{\dot{F}}%
\global\long\def\Fdd{\ddot{F}}%
\global\long\def\Gd{\dot{G}}%
\global\long\def\Gdd{\ddot{G}}%
\global\long\def\Hd{\dot{H}}%
\global\long\def\Hdd{\ddot{H}}%
\global\long\def\Id{\dot{I}}%
\global\long\def\Idd{\ddot{I}}%
\global\long\def\Jd{\dot{J}}%
\global\long\def\Jdd{\ddot{J}}%
\global\long\def\Kd{\dot{K}}%
\global\long\def\Kdd{\ddot{K}}%
\global\long\def\Ld{\dot{L}}%
\global\long\def\Ldd{\ddot{L}}%
\global\long\def\Md{\dot{M}}%
\global\long\def\Mdd{\ddot{M}}%
\global\long\def\Nd{\dot{N}}%
\global\long\def\Ndd{\ddot{N}}%
\global\long\def\Od{\dot{O}}%
\global\long\def\Odd{\ddot{O}}%
\global\long\def\Pd{\dot{P}}%
\global\long\def\Pdd{\ddot{P}}%
\global\long\def\Qd{\dot{Q}}%
\global\long\def\Qdd{\ddot{Q}}%
\global\long\def\Rd{\dot{R}}%
\global\long\def\Rdd{\ddot{R}}%
\global\long\def\Sd{\dot{S}}%
\global\long\def\Sdd{\ddot{S}}%
\global\long\def\Td{\dot{T}}%
\global\long\def\Tdd{\ddot{T}}%
\global\long\def\Ud{\dot{U}}%
\global\long\def\Udd{\ddot{U}}%
\global\long\def\Vd{\dot{R}}%
\global\long\def\Vdd{\ddot{R}}%
\global\long\def\Wd{\dot{W}}%
\global\long\def\Wdd{\ddot{W}}%
\global\long\def\Xd{\dot{X}}%
\global\long\def\Xdd{\ddot{X}}%
\global\long\def\Yd{\dot{Y}}%
\global\long\def\Ydd{\ddot{Y}}%
\global\long\def\Zd{\dot{Z}}%
\global\long\def\Zdd{\ddot{Z}}%

\global\long\def\Gad{\dot{\Gamma}}%
\global\long\def\Gadd{\ddot{\Gamma}}%
\global\long\def\Ded{\dot{\Delta}}%
\global\long\def\Dedd{\ddot{\Delta}}%
\global\long\def\Thd{\dot{\Theta}}%
\global\long\def\Thdd{\ddot{\Theta}}%
\global\long\def\Lad{\dot{\Lambda}}%
\global\long\def\Ladd{\ddot{\Lambda}}%
\global\long\def\Xid{\dot{\Xi}}%
\global\long\def\Xidd{\ddot{\Xi}}%
\global\long\def\Pid{\dot{\Pi}}%
\global\long\def\Pidd{\ddot{\Pi}}%
\global\long\def\Sid{\dot{\Sigma}}%
\global\long\def\Sidd{\ddot{\Sigma}}%
\global\long\def\Phd{\dot{\Phi}}%
\global\long\def\Phdd{\ddot{\Phi}}%
\global\long\def\Psd{\dot{\Psi}}%
\global\long\def\Psdd{\ddot{\Psi}}%
\global\long\def\Thd{\dot{\Theta}}%
\global\long\def\Thdd{\ddot{\Theta}}%

\global\long\def\ald{\dot{\alpha}}%
\global\long\def\aldd{\ddot{\alpha}}%
\global\long\def\bed{\dot{\beta}}%
\global\long\def\bedd{\ddot{\beta}}%
\global\long\def\gad{\dot{\gamma}}%
\global\long\def\gadd{\ddot{\gamma}}%
\global\long\def\ded{\dot{\delta}}%
\global\long\def\dedd{\ddot{\delta}}%
\global\long\def\epd{\dot{\epsilon}}%
\global\long\def\epdd{\ddot{\epsilon}}%
\global\long\def\vepd{\dot{\varepsilon}}%
\global\long\def\vepdd{\ddot{\varepsilon}}%
\global\long\def\zed{\dot{\zeta}}%
\global\long\def\zedd{\ddot{\zeta}}%
\global\long\def\etd{\dot{\eta}}%
\global\long\def\etdd{\ddot{\eta}}%
\global\long\def\thd{\dot{\theta}}%
\global\long\def\thdd{\ddot{\theta}}%
\global\long\def\iod{\dot{\iota}}%
\global\long\def\iodd{\ddot{\iota}}%
\global\long\def\kad{\dot{\kappa}}%
\global\long\def\kadd{\ddot{\kappa}}%
\global\long\def\lad{\dot{\lambda}}%
\global\long\def\ladd{\ddot{\lambda}}%
\global\long\def\mud{\dot{\mu}}%
\global\long\def\mudd{\ddot{\mu}}%
\global\long\def\nud{\dot{\nu}}%
\global\long\def\nudd{\ddot{\nu}}%
\global\long\def\xid{\dot{\xi}}%
\global\long\def\xidd{\ddot{\xi}}%
\global\long\def\pid{\dot{\pi}}%
\global\long\def\pidd{\ddot{\pi}}%
\global\long\def\rhod{\dot{\rho}}%
\global\long\def\rhodd{\ddot{\rho}}%
\global\long\def\sid{\dot{\sigma}}%
\global\long\def\sidd{\ddot{\sigma}}%
\global\long\def\tad{\dot{\tau}}%
\global\long\def\tadd{\ddot{\tau}}%
\global\long\def\upd{\dot{\upsilon}}%
\global\long\def\updd{\ddot{\upsilon}}%
\global\long\def\phd{\dot{\phi}}%
\global\long\def\phdd{\ddot{\phi}}%
\global\long\def\vpd{\dot{\varphi}}%
\global\long\def\vpdd{\ddot{\varphi}}%
\global\long\def\chd{\dot{\chi}}%
\global\long\def\chdd{\ddot{\chi}}%
\global\long\def\psd{\dot{\psi}}%
\global\long\def\psdd{\ddot{\psi}}%
\global\long\def\omd{\dot{\omega}}%
\global\long\def\omdd{\ddot{\omega}}%

\global\long\def\BBA{\mathbb{A}}%
\global\long\def\BBB{\mathbb{B}}%
\global\long\def\BBC{\mathbb{C}}%
\global\long\def\BBD{\mathbb{D}}%
\global\long\def\BBE{\mathbb{E}}%
\global\long\def\BBF{\mathbb{F}}%
\global\long\def\BBG{\mathbb{G}}%
\global\long\def\BBH{\mathbb{H}}%
\global\long\def\BBI{\mathbb{I}}%
\global\long\def\BBJ{\mathbb{J}}%
\global\long\def\BBK{\mathbb{K}}%
\global\long\def\BBL{\mathbb{L}}%
\global\long\def\BBM{\mathbb{M}}%
\global\long\def\BBN{\mathbb{N}}%
\global\long\def\BBO{\mathbb{O}}%
\global\long\def\BBP{\mathbb{P}}%
\global\long\def\BBQ{\mathbb{Q}}%
\global\long\def\BBR{\mathbb{R}}%
\global\long\def\BBS{\mathbb{S}}%
\global\long\def\BBT{\mathbb{T}}%
\global\long\def\BBU{\mathbb{U}}%
\global\long\def\BBV{\mathbb{V}}%
\global\long\def\BBW{\mathbb{W}}%
\global\long\def\BBX{\mathbb{X}}%
\global\long\def\BBY{\mathbb{Y}}%
\global\long\def\BBZ{\mathbb{Z}}%

\global\long\def\AA{\mathcal{A}}%
\global\long\def\BB{\mathcal{B}}%
\global\long\def\CC{\mathcal{C}}%
\global\long\def\DD{\mathcal{D}}%
\global\long\def\EE{\mathcal{E}}%
\global\long\def\FF{\mathcal{F}}%
\global\long\def\GG{\mathcal{G}}%
\global\long\def\HH{\mathcal{H}}%
\global\long\def\II{\mathcal{I}}%
\global\long\def\JJ{\mathcal{J}}%
\global\long\def\KK{\mathcal{K}}%
\global\long\def\LLL{\mathcal{L}}%
\global\long\def\MM{\mathcal{M}}%
\global\long\def\NN{\mathcal{N}}%
\global\long\def\OOO{\mathcal{O}}%
\global\long\def\PP{\mathcal{P}}%
\global\long\def\QQ{\mathcal{Q}}%
\global\long\def\RRR{\mathcal{R}}%
\global\long\def\SSS{\mathcal{S}}%
\global\long\def\TT{\mathcal{T}}%
\global\long\def\UU{\mathcal{U}}%
\global\long\def\VV{\mathcal{V}}%
\global\long\def\WW{\mathcal{W}}%
\global\long\def\XX{\mathcal{X}}%
\global\long\def\YY{\mathcal{Y}}%
\global\long\def\ZZ{\mathcal{Z}}%

\global\long\def\At{\tilde{A}}%
\global\long\def\Bt{\tilde{B}}%
\global\long\def\Ct{\tilde{C}}%
\global\long\def\Dt{\tilde{D}}%
\global\long\def\Et{\tilde{E}}%
\global\long\def\Ft{\tilde{F}}%
\global\long\def\Gt{\tilde{G}}%
\global\long\def\Ht{\tilde{H}}%
\global\long\def\It{\tilde{I}}%
\global\long\def\Jt{\tilde{J}}%
\global\long\def\Kt{\tilde{K}}%
\global\long\def\Lt{\tilde{L}}%
\global\long\def\Mt{\tilde{M}}%
\global\long\def\Nt{\tilde{N}}%
\global\long\def\Ot{\tilde{O}}%
\global\long\def\Pt{\tilde{P}}%
\global\long\def\Qt{\tilde{Q}}%
\global\long\def\Rt{\tilde{R}}%
\global\long\def\St{\tilde{S}}%
\global\long\def\Tt{\tilde{T}}%
\global\long\def\Ut{\tilde{U}}%
\global\long\def\Vt{\tilde{V}}%
\global\long\def\Wt{\tilde{W}}%
\global\long\def\Xt{\tilde{X}}%
\global\long\def\Yt{\tilde{Y}}%
\global\long\def\Zt{\tilde{Z}}%

\global\long\def\at{\tilde{a}}%
\global\long\def\bt{\tilde{b}}%
\global\long\def\ct{\tilde{c}}%
\global\long\def\dt{\tilde{d}}%
\global\long\def\eet{\tilde{e}}%
\global\long\def\ft{\tilde{f}}%
\global\long\def\gt{\tilde{g}}%
\global\long\def\hht{\tilde{h}}%
\global\long\def\it{\tilde{i}}%
\global\long\def\jt{\tilde{j}}%
\global\long\def\kt{\tilde{k}}%
\global\long\def\lt{\tilde{l}}%
\global\long\def\elt{\tilde{\ell}}%
\global\long\def\mt{\tilde{m}}%
\global\long\def\nt{\tilde{n}}%
\global\long\def\ot{\tilde{o}}%
\global\long\def\pt{\tilde{p}}%
\global\long\def\qt{\tilde{q}}%
\global\long\def\rt{\tilde{r}}%
\global\long\def\st{\tilde{s}}%
\global\long\def\tt{\tilde{t}}%
\global\long\def\ut{\tilde{u}}%
\global\long\def\vt{\tilde{v}}%
\global\long\def\wt{\tilde{w}}%
\global\long\def\xt{\tilde{x}}%
\global\long\def\yt{\tilde{y}}%
\global\long\def\zt{\tilde{z}}%

\global\long\def\mfA{\mathfrak{A}}%
\global\long\def\mfB{\mathfrak{B}}%
\global\long\def\mfC{\mathfrak{C}}%
\global\long\def\mfD{\mathfrak{D}}%
\global\long\def\mfE{\mathfrak{E}}%
\global\long\def\mfF{\mathfrak{F}}%
\global\long\def\mfG{\mathfrak{G}}%
\global\long\def\mfH{\mathfrak{H}}%
\global\long\def\mfI{\mathfrak{I}}%
\global\long\def\mfJ{\mathfrak{J}}%
\global\long\def\mfK{\mathfrak{K}}%
\global\long\def\mfL{\mathfrak{L}}%
\global\long\def\mfM{\mathfrak{M}}%
\global\long\def\mfN{\mathfrak{N}}%
\global\long\def\mfO{\mathfrak{O}}%
\global\long\def\mfP{\mathfrak{P}}%
\global\long\def\mfQ{\mathfrak{Q}}%
\global\long\def\mfR{\mathfrak{R}}%
\global\long\def\mfS{\mathfrak{S}}%
\global\long\def\mfT{\mathfrak{T}}%
\global\long\def\mfU{\mathfrak{U}}%
\global\long\def\mfV{\mathfrak{V}}%
\global\long\def\mfW{\mathfrak{W}}%
\global\long\def\mfX{\mathfrak{X}}%
\global\long\def\mfY{\mathfrak{Y}}%
\global\long\def\mfZ{\mathfrak{Z}}%
\global\long\def\mfa{\mathfrak{a}}%
\global\long\def\mfb{\mathfrak{b}}%
\global\long\def\mfc{\mathfrak{c}}%
\global\long\def\mfd{\mathfrak{d}}%
\global\long\def\mfe{\mathfrak{e}}%
\global\long\def\mff{\mathfrak{f}}%
\global\long\def\mfg{\mathfrak{g}}%
\global\long\def\mfh{\mathfrak{h}}%
\global\long\def\mfi{\mathfrak{i}}%
\global\long\def\mfj{\mathfrak{j}}%
\global\long\def\mfk{\mathfrak{k}}%
\global\long\def\mfl{\mathfrak{l}}%
\global\long\def\mfm{\mathfrak{m}}%
\global\long\def\mfn{\mathfrak{n}}%
\global\long\def\mfo{\mathfrak{o}}%
\global\long\def\mfp{\mathfrak{p}}%
\global\long\def\mfq{\mathfrak{q}}%
\global\long\def\mfr{\mathfrak{r}}%
\global\long\def\mfs{\mathfrak{s}}%
\global\long\def\mft{\mathfrak{t}}%
\global\long\def\mfu{\mathfrak{u}}%
\global\long\def\mfv{\mathfrak{v}}%
\global\long\def\mfw{\mathfrak{w}}%
\global\long\def\mfx{\mathfrak{x}}%
\global\long\def\mfy{\mathfrak{y}}%
\global\long\def\mfz{\mathfrak{z}}%

\global\long\def\d{\mathrm{d}}%
\global\long\def\DDD{\mathrm{D}}%
\global\long\def\EEE{\mathrm{E}}%
\global\long\def\i{\ii}%
\global\long\def\MMM{\mathrm{M}}%
\global\long\def\OOOO{\mathrm{O}}%
\global\long\def\RRRR{\mathrm{R}}%
\global\long\def\TTT{\mathrm{T}}%
\global\long\def\UUU{\mathrm{U}}%

\global\long\def\GL{\mathrm{GL}}%
\global\long\def\ISU{\mathrm{ISU}}%
\global\long\def\ISUT{\mathrm{ISU}\left(2\right)}%
\global\long\def\SL{\mathrm{SL}}%
\global\long\def\SO{\mathrm{SO}}%
\global\long\def\SOH{\mathrm{SO}\left(3\right)}%
\global\long\def\SOT{\mathrm{SO}\left(2\right)}%
\global\long\def\Sp{\mathrm{Sp}}%
\global\long\def\SU{\mathrm{SU}}%
\global\long\def\SUT{\mathrm{SU}\left(2\right)}%
\global\long\def\UO{\mathrm{U}\left(1\right)}%
\global\long\def\gl{\mathfrak{gl}}%
\global\long\def\sl{\mathfrak{sl}}%
\global\long\def\sso{\mathfrak{so}}%
\global\long\def\soh{\mathfrak{so}\left(3\right)}%
\global\long\def\su{\mathfrak{su}}%
\global\long\def\sut{\mathfrak{su}\left(2\right)}%
\global\long\def\isut{\mathfrak{isu}\left(2\right)}%

\global\long\def\so{\Rightarrow}%
\global\long\def\os{\Leftarrow}%
\global\long\def\to{\rightarrow}%
\global\long\def\ot{\leftarrow}%
\global\long\def\soo{\Longrightarrow}%
\global\long\def\oos{\Longleftarrow}%
\global\long\def\too{\longrightarrow}%
\global\long\def\oot{\longleftarrow}%
\global\long\def\sos{\Leftrightarrow}%
\global\long\def\tot{\leftrightarrow}%
\global\long\def\soos{\Longleftrightarrow}%
\global\long\def\toot{\longleftrightarrow}%
\global\long\def\mt{\mapsto}%
\global\long\def\mtt{\longmapsto}%
\global\long\def\dn{\downarrow}%
\global\long\def\up{\uparrow}%
\global\long\def\updn{\updownarrow}%
\global\long\def\sea{\searrow}%
\global\long\def\nea{\nearrow}%
\global\long\def\nwa{\nwarrow}%
\global\long\def\swa{\swarrow}%
\global\long\def\hk{\hookrightarrow}%
\global\long\def\kh{\hookleftarrow}%
\global\long\def\soosp{\quad\Longrightarrow\quad}%
\global\long\def\oossp{\quad\Longleftarrow\quad}%
\global\long\def\soossp{\quad\Longleftrightarrow\quad}%

\global\long\def\multibrl#1{\left(#1\right.}%
\global\long\def\multibrr#1{\left.#1\right)}%
\global\long\def\multisql#1{\left[#1\right.}%
\global\long\def\multisqr#1{\left.#1\right]}%
\global\long\def\multicul#1{\left\{  #1\right.}%
\global\long\def\multicur#1{\left.#1\right\}  }%

\global\long\def\bl{\bigl|}%
\global\long\def\bll{\Bigl|}%
\global\long\def\blll{\biggl|}%
\global\long\def\bllll{\Biggl|}%

\global\long\def\ma#1#2{\left\langle #1\thinspace\middle|\thinspace#2\right\rangle }%
\global\long\def\mma#1#2#3{\left\langle #1\thinspace\middle|\thinspace#2\thinspace\middle|\thinspace#3\right\rangle }%
\global\long\def\mc#1#2{\left\{  #1\thinspace\middle|\thinspace#2\right\}  }%
\global\long\def\mmc#1#2#3{\left\{  #1\thinspace\middle|\thinspace#2\thinspace\middle|\thinspace#3\right\}  }%
\global\long\def\mr#1#2{\left(#1\thinspace\middle|\thinspace#2\right) }%
\global\long\def\mmr#1#2#3{\left(#1\thinspace\middle|\thinspace#2\thinspace\middle|\thinspace#3\right)}%

\global\long\def\pr{\parallel}%
\global\long\def\xx{\times}%
\global\long\def\dg{\lyxmathsym{\textdegree}}%
\global\long\def\sp{,\qquad}%
\global\long\def\sq{\square}%
\global\long\def\pt{\propto}%
\global\long\def\lrc{\lrcorner\thinspace}%
\global\long\def\pexp{\overrightarrow{\exp}}%
\global\long\def\dui#1#2#3{#1_{#2}{}^{#3}}%
\global\long\def\udi#1#2#3{#1^{#2}{}_{#3}}%
\global\long\def\pab{\bar{\partial}}%
\global\long\def\zr{\mathbf{0}}%
\global\long\def\on{\mathbf{1}}%
\global\long\def\na{\boldsymbol{\nabla}}%
\global\long\def\hf{\frac{1}{2}}%
\global\long\def\trd{\frac{1}{3}}%
\global\long\def\fr{\frac{1}{4}}%
\global\long\def\ei{\frac{1}{8}}%
\global\long\def\ap{\approx}%
\global\long\def\eqm{\overset{?}{=}}%
\global\long\def\fa{\forall}%
\global\long\def\ex{\exists}%
\global\long\def\xxd{\dot{\boldsymbol{x}}}%
\global\long\def\xxdd{\ddot{\boldsymbol{x}}}%
\global\long\def\ept{\tilde{\epsilon}}%

\global\long\def\AAh{\hat{\mathbf{A}}}%
\global\long\def\eeh{\hat{\mathbf{e}}}%
\global\long\def\FFh{\hat{\mathbf{F}}}%
\global\long\def\TTh{\hat{\mathbf{T}}}%
\global\long\def\Thh{\hat{\Theta}}%
\global\long\def\XXt{\tilde{\mathbf{X}}}%
\global\long\def\hr{\mathring{h}}%
\global\long\def\xxr{\mathring{\boldsymbol{x}}}%
\global\long\def\hh{\hat{h}}%
\global\long\def\xxh{\hat{\boldsymbol{x}}}%
\global\long\def\hrh{\hat{\mathring{h}}}%
\global\long\def\xxrh{\hat{\mathring{\boldsymbol{x}}}}%
\global\long\def\Sh{\hat{S}}%
\global\long\def\sxt{\frac{1}{6}}%
\global\long\def\se{\subseteq}%
\global\long\def\tHH{\widetilde{\mathcal{H}}}%
\global\long\def\bgl{\biggl|}%

\title{Spin Networks and Cosmic Strings in 3+1 Dimensions}
\author{\textbf{Barak Shoshany}\thanks{bshoshany@perimeterinstitute.ca}\\
\emph{ Perimeter Institute for Theoretical Physics,}\\
\emph{31 Caroline Street North, Waterloo, Ontario, Canada, N2L 2Y5}}
\maketitle
\begin{center}
\textbf{Abstract}
\par\end{center}

Spin networks, the quantum states of discrete geometry in loop quantum
gravity, are directed graphs whose links are labeled by irreducible
representations of SU(2), or spins. Cosmic strings are 1-dimensional
topological defects carrying distributional curvature in an otherwise
flat spacetime. In this paper we prove that the classical phase space
of spin networks coupled to cosmic strings may obtained as a straightforward
discretization of general relativity in 3+1 spacetime dimensions.
We decompose the continuous spatial geometry into 3-dimensional cells,
which are dual to a spin network graph in a unique and well-defined
way. Assuming that the geometry may only be probed by holonomies (or
Wilson loops) located on the spin network, we truncate the geometry
such that the cells become flat and the curvature is concentrated
at the edges of the cells, which we then interpret as a network of
cosmic strings. The discrete phase space thus describes a spin network
coupled to cosmic strings. This work proves that the relation between
gravity and spin networks exists not only at the quantum level, but
already at the classical level. Two appendices provide detailed derivations
of the Ashtekar formulation of gravity as a Yang-Mills theory and
the distributional geometry of cosmic strings in this formulation.

\tableofcontents{}

\section{Introduction}

\subsection{Loop Quantum Gravity and Spin Networks}

When perturbatively quantizing gravity, one obtains a low-energy effective
theory, which breaks down at high energies. There are several different
approaches to solving this problem and obtaining a theory of quantum
gravity. String theory, for example, attempts to do so by postulating
entirely new degrees of freedom, which can then be shown to reduce
to general relativity (or some modification thereof) at the low-energy
limit. \emph{Loop quantum gravity} \cite{Rovelli:2004tv} instead
tries to quantize gravity \emph{non-perturbatively}, by quantizing
\emph{holonomies }(or \emph{Wilson loops}) instead of the metric,
in an attempt to avoid the issues arising from perturbative quantization.

The starting point of the canonical version of loop quantum gravity
\cite{thiemann_2007}, which is the one we will discuss here, is the
reformulation of general relativity as a \emph{non-abelian Yang-Mills
gauge theory }on a 3-dimensional spatial slice of a 3+1-dimensional
spacetime, with the \emph{gauge group }$\SUT$ related to spatial
rotations, the \emph{Yang-Mills connection }$\A$ related to the usual
connection and extrinsic curvature, and the \emph{``electric field''}
$\E$ related to the metric (or more precisely, the frame field).
Spatial and time diffeomorphisms appear as additional gauge symmetries,
generated by the appropriate constraints. Once gravity is reformulated
in this way, one can utilize the existing arsenal of techniques from
Yang-Mills theory, and in particular \emph{lattice gauge theory},
to tackle the problem of quantum gravity \cite{Henneaux:1992ig}.

This theory is quantized by considering \emph{graphs}, that is, sets
of \emph{nodes }connected by \emph{links}. One defines \emph{holonomies},
or \emph{path-ordered exponentials} of the connection, which implement
\emph{parallel transport} along each link. The curvature on the spatial
slice can then be probed by looking at holonomies along loops on the
graph. Without going into the technical details, the general idea
is that if we know the curvature inside every possible loop, then
this is equivalent to knowing the curvature at every point.

The \emph{kinematical Hilbert space }of loop quantum gravity is obtained
from the set of all wave-functionals for all possible graphs, together
with an appropriate $\SUT$-invariant and diffeomorphism-invariant
inner product. The \emph{physical Hilbert space }is a subset of the
kinematical one, containing only the states invariant under all gauge
transformations -- or in other words, annihilated by all of the constraints.
Since gravity is a totally constrained system -- in the Hamiltonian
formulation, the action is just a sum of constraints -- a quantum
state annihilated by all of the constraints is analogous to a metric
which solves Einstein's equations in the classical Lagrangian formulation.

Specifically, to get from the kinematical to the physical Hilbert
space, three steps must be taken:
\begin{enumerate}
\item First, we apply the \emph{Gauss constraint} to the kinematical Hilbert
space. Since the Gauss constraint generates $\SUT$ gauge transformation,
we obtain a space of $\SUT$-invariant states, called \emph{spin network
states} \cite{Ashtekar:1991kc}, which are the graphs mentioned above,
but with their links colored by irreducible representations of $\SUT$,
that is, \emph{spins }$j\in\left\{ \hf,1,\frac{3}{2},2,\ldots\right\} $.
\item Then, we apply the \emph{spatial diffeomorphism constraint}. We obtain
a space of equivalence classes of spin networks under spatial diffeomorphisms,
a.k.a. \emph{knots}. These states are now abstract graphs, not localized
in space. This is analogous to how a classical geometry is an equivalence
class of metrics under diffeomorphisms.
\item Lastly, we apply the \emph{Hamiltonian constraint}. This step is still
not entirely well-understood, and is one of the main open problems
of the theory.
\end{enumerate}
One of loop quantum gravity's most celebrated results is the existence
of area and volume operators. They are derived by taking the usual
integrals of area and volume forms and promoting the ``electric field''
$\E$, which is conjugate to the connection $\A$, to a functional
derivative $\delta/\delta\A$. The spin network states turn out to
be eigenstates of these operators, and they have \emph{discrete spectra
}which depend on the spins of the links. This means that loop quantum
gravity contains a \emph{quantum geometry}, which is a feature one
would expect a quantum theory of spacetime to have. It also hints
that spacetime is discrete at the Planck scale.

However, it is not clear how to rigorously define the classical geometry
related to a particular spin network state. In this paper, we will
try to answer that question. In our previous two papers \cite{FirstPaper,Shoshany:2019ymo}
we showed how to obtain the phase space of spin networks coupled to
point particles by discretizing gravity in 2+1D; this paper generalizes
that result to 3+1D.

\subsection{Quantization, Discretization, Subdivision, and Truncation}

One of the key challenges in trying to define a theory of quantum
gravity at the quantum level is to find a regularization that does
not drastically break the fundamental symmetries of the theory. This
is a challenge in any gauge theory, but gravity is especially challenging,
for two reasons. First, one expects that the quantum theory possesses
a fundamental length scale; and second, the gauge group contains diffeomorphism
symmetry, which affects the nature of the space on which the regularization
is applied.

In gauge theories such as \emph{quantum chromodynamics} (QCD), the
only known way to satisfy these requirements, other than gauge-fixing
before regularization, is to put the theory on a \emph{lattice}, where
an effective finite-dimensional gauge symmetry survives at each scale.
One would like to devise such a scheme in the gravitational context
as well. In this paper, we develop a step-by-step procedure to achieve
this, exploiting, among other things, the fact that first-order gravity
in 3+1 dimensions with the Ashtekar variables closely resembles other
gauge theories, as discussed above. We find not only the spin network
(or \emph{holonomy-flux}) phase space, but also additional string-like
degrees of freedom coupled to the curvature and torsion.

As explained above, in \textit{\emph{canonical loop quantum gravity}},
one can show that the geometric operators possess a discrete spectrum.
This is, however, only possible after one chooses the quantum spin
network states to have support on a graph. Spin network states can
be understood as describing a quantum version of \textit{discretized}
spatial geometry \cite{Rovelli:2004tv}, and the Hilbert space associated
to a graph can be related, in the classical limit, to a set of discrete
piecewise-flat geometries \cite{Dittrich:2008ar,Freidel:2010aq}.

This means that the\textit{\emph{ loop quantum gravity}} quantization
scheme consists at the same time of a \emph{quantization} and a \emph{discretization};
moreover, the quantization of the geometric spectrum is entangled
with the discretization of the fundamental variables. It has been
argued that it is essential to disentangle these two different features
\cite{Freidel:2011ue}, especially when one wants to address dynamical
issues.

In \cite{Freidel:2011ue,Freidel:2013bfa}, it was suggested that one
should understand the discretization as a two-step process: a \emph{subdivision}
followed by a \emph{truncation}. In the first step one subdivides
the systems into fundamental cells, and in the second step one chooses
a truncation of degrees of freedom in each cell, which is consistent
with the symmetries of the theory. By focusing first on the classical
discretization, before any quantization takes place, several aspects
of the theory can be clarified, as we discussed in \cite{FirstPaper}.

The separation of discretization into two distinct steps in our formalism
work as follows. First we perform a \emph{subdivision}, or decomposition
into subsystems. More precisely, we define a \emph{cellular decomposition
}on our 3D spatial manifold, where the cells can be any (convex) polyhedra.
This structure has a dual structure, which as we will see, is the
spin network graph, with each cell dual to a node, and each side dual
to a link connected to that node.

Then, we perform a \emph{truncation}, or coarse-graining of the subsystems.
In this step, we assume that there is arbitrary curvature and torsion
inside each loop of the spin network. We then ``compress'' the information
about the geometry into singular codimension-2 excitations, which
is 3 spatial dimensions means 1-dimensional (string) excitation. Crucially,
since the only way to probe the geometry is by looking at the holonomies
on the loops of the spin network, the observables before and after
this truncation are the same.

Another way to interpret this step is to instead assume that spacetime
is flat everywhere, with matter sources being distributive, i.e.,
given by Dirac delta functions, which then generate singular curvature
and torsion by virtue of the Einstein equation. We interpret these
distributive matter sources as strings, and this is entirely equivalent
to truncating a continuous geometry, since holonomies cannot distinguish
between continuous and distributive geometries.

Once we performed subdivision and truncation, we can now define discrete
variables on each cell and integrate the continuous symplectic potential
in order to obtain a discrete potential, which represents the discrete
phase space. In this step, we will see that the mathematical structures
we are using conspire to cancel many terms in the potential, allowing
us to fully integrate it.

\subsection{Outline}

This paper is organized as follows. First, in Chapter \ref{subsec:Basic-Definitions-Notations},
we provide a comprehensive list of basic definitions, notations, and
conventions which will be used throughout the paper. It is recommended
that the reader not skip this chapter, as some of our notation is
slightly non-standard.

In Chapter \ref{sec:Gravity-as-a} and Appendix \ref{sec:Derivation-of-Ashtekar}
we provide a detailed and self-contained derivation of the Ashtekar
variables and the loop gravity Hamiltonian action, including the constraints.
Special care is taken to write everything in terms of index-free Lie-algebra-valued
differential forms, which -- in addition to being more elegant --
will greatly simplify our derivation.

Chapter \ref{sec:The-Discrete-Geometry-3P1} introduces the cellular
decomposition and explains how an arbitrary geometry is truncated
into a piecewise-flat geometry, or alternatively, a flat geometry
with matter degrees of freedom in the form of cosmic strings. Appendix
\ref{sec:Cosmic-Strings-in} discusses cosmic strings in detail, and
derives their representation in the Ashtekar formulation. Chapter
\ref{subsec:Classical-Spin-Networks} presents the classical spin
network phase space, and provides a detailed calculation of the Poisson
brackets.

Chapter \ref{sec:Discretizing-the-Symplectic} is the main part of
the paper, where we will use the structures and results of the previous
chapters to obtain the spin network phase space coupled to cosmic
strings from the continuous phase space of 3+1-dimensional gravity.
Finally, Chapter \ref{sec:Conclusions} summarizes the results of
this paper and presents several avenues for potential future research.

The content of this paper, together with that of our previous two
papers \cite{FirstPaper,Shoshany:2019ymo}, may also be found in the
author's PhD thesis \cite{PhDThesis}. The thesis relates and compares
the 2+1D and 3+1D calculations, adds a simpler way to perform the
2+1D calculation which did not appear in the previous papers, includes
an extensive discussion of the role of edge modes in our formalism,
and contains much more detailed derivations of some results.

\section{\label{subsec:Basic-Definitions-Notations}Basic Definitions, Notations,
and Conventions}

The following definitions, notations, and conventions will be used
throughout the paper.

\subsection{\label{subsec:Lie-Group-and}Lie Group and Algebra Elements}

Let $G$ be a \emph{Lie group}, let $\mfg$ be its associated \emph{Lie
algebra}, and let $\mfg^{*}$ be the dual to that Lie algebra. The
\emph{cotangent bundle} of $G$ is the Lie group $T^{*}G\cong G\ltimes\mfg^{*}$,
where $\ltimes$ is the \emph{semidirect product}, and it has the
associated Lie algebra $\mfg\oplus\mfg^{*}$. We assume that this
group is the \emph{Euclidean} \emph{group}, or a generalization thereof,
and its algebra takes the form
\[
\left[\P_{i},\P_{j}\right]=0\sp\left[\J_{i},\J_{j}\right]=\dui f{ij}k\J_{k}\sp\left[\J_{i},\P_{j}\right]=\dui f{ij}k\P_{k},
\]
where:
\begin{itemize}
\item $\dui f{ij}k$ are the \emph{structure constants}, which satisfy anti-symmetry
$\dui f{ij}k=-\dui f{ji}k$ and the Jacobi identity $\dui f{[ij}l\dui f{k]l}m=0$.
\item $\J_{i}\in\mfg$ are the \emph{rotation }generators,
\item $\P_{i}\in\mfg^{*}$ are the \emph{translation }generators,
\item The indices $i,j,k$ take the values $1,2,3$, since they are internal
spatial indices.
\end{itemize}
Note that sometimes we will use the notation $\ta_{i}$ to indicate
generators which could be in either $\mfg$ or $\mfg^{*}$.

Usually in the loop quantum gravity literature we take $G=\SUT$ such
that $\mfg^{*}=\BBR^{3}$ and
\[
\ISUT\cong\SUT\ltimes\BBR^{3}\cong T^{*}\SUT.
\]
However, here we will keep $G$ abstract for brevity and in order
for the discussion to be more general.

Throughout this paper, different fonts and typefaces will distinguish
elements of different groups and algebras, or differential forms valued
in those groups and algebras, as follows:
\begin{itemize}
\item $G\ltimes\mfg^{*}$-valued forms will be written in Calligraphic font:
$\AA,\BB,\CC,...$
\item $\mfg\oplus\mfg^{*}$-valued forms will be written in bold Calligraphic
font: $\AAb,\BBb,\CCb,...$
\item $G$-valued forms will be written in regular font: $a,b,c,...$
\item $\mfg$ or $\mfg^{*}$-valued forms will be written in bold font:
$\a,\b,\c,...$
\end{itemize}
The Calligraphic notation will only be used in this introduction;
elsewhere we will only talk about $G$, $\mfg$, or $\mfg^{*}$ elements.

\subsection{\label{subsec:Spacetime-and-Spatial}Indices and Differential Forms}

Throughout this paper, we will use the following conventions for indices\footnote{The usage of lowercase Latin letters for both spatial and internal
spatial indices is somewhat confusing, but seems to be standard in
the literature, so we will use it here as well.}:
\begin{itemize}
\item $\mu,\nu,\ldots\in\left\{ 0,1,2,3\right\} $ represent 3+1D spacetime
components.
\item $A,B,\ldots,I,J,\ldots\in\left(0,1,2,3\right)$ represent 3+1D internal
components.
\item $a,b,\ldots\in\left\{ 1,2,3\right\} $ represent 3D spatial components:
$\overbrace{0,\underbrace{1,2,3}_{a}}^{\mu}$.
\item $i,j,\ldots\in\left\{ 1,2,3\right\} $ represent 3D internal / Lie
algebra components: $\overbrace{0,\underbrace{1,2,3}_{i}}^{I}$.
\end{itemize}
We consider a 3+1D manifold $M$ with topology $\Sigma\xx\BBR$ where
$\Sigma$ is a 3-dimensional spatial manifold and $\BBR$ represents
time. Our metric signature convention is $\left(-,+,+,+\right)$.
In index-free notation, we denote a \emph{Lie-algebra-valued differential
form of degree $p$} (or \emph{$p$-form}) on $\Sigma$, with one
algebra index $i$ and $p$ spatial indices $a_{1},\ldots,a_{p}$,
as
\begin{equation}
\A\equiv\frac{1}{p!}A_{a_{1}\cdots a_{p}}^{i}\ta_{i}\thinspace\d x^{a_{1}}\wedge\cdots\wedge\d x^{a_{p}}\in\Omega^{p}\left(\Sigma,\mfg\right),\label{eq:index-free-notation}
\end{equation}
where $A_{a_{1}\cdots a_{p}}^{i}$ are the components and $\ta_{i}$
are the generators of the algebra $\mfg$ in which the form is valued.

Sometimes we will only care about the algebra index, and write $\A\equiv A^{i}\ta_{i}$
with the spatial indices implied, such that $A^{i}\equiv\frac{1}{p!}A_{a_{1}\cdots a_{p}}^{i}\d x^{a_{1}}\wedge\cdots\wedge\d x^{a_{p}}$
are real-valued $p$-forms. Other times we will only care about the
spacetime indices, and write $\A\equiv\frac{1}{p!}\A_{a_{1}\cdots a_{p}}\d x^{a_{1}}\wedge\cdots\wedge\d x^{a_{p}}$
with the algebra index implied, such that $\A_{a_{1}\cdots a_{p}}\equiv A_{a_{1}\cdots a_{p}}^{i}\ta_{i}$
are algebra-valued 0-forms.

\subsection{The Graded Commutator}

Given any two Lie-algebra-valued forms $\A$ and $\B$ of degrees\emph{
}$\deg\A$ and $\deg\B$ respectively, we define the \emph{graded
commutator}:
\[
\left[\A,\B\right]\equiv\A\wedge\B-\left(-1\right)^{\deg\A\deg\B}\B\wedge\A,
\]
which satisfies
\[
\left[\A,\B\right]=-\left(-1\right)^{\deg\A\deg\B}\left[\B,\A\right].
\]
If at least one of the forms has even degree, this reduces to the
usual anti-symmetric commutator; if we then interpret $\A$ and $\B$
as vectors in $\BBR^{3}$, then this is none other than the vector
cross product $\A\xx\B$. Note that $\left[\A,\B\right]$ is a Lie-algebra-valued
$\left(\deg\A+\deg\B\right)$-form.

The graded commutator satisfies the \emph{graded Leibniz rule}:
\[
\d\left[\A,\B\right]=\left[\d\A,\B\right]+\left(-1\right)^{\deg\A}\left[\A,\d\B\right].
\]
In terms of indices, with $\deg\A=p$ and $\deg\B=q$, we have
\[
\left[\A,\B\right]=\frac{1}{\left(p+q\right)!}\left[\A,\B\right]_{a_{1}\cdots a_{p}b_{1}\cdots b_{q}}^{k}\ta_{k}\d x^{a_{1}}\wedge\cdots\wedge\d x^{a_{p}}\wedge\d x^{b_{1}}\wedge\cdots\wedge\d x^{b_{q}},
\]
where
\[
\left[\A,\B\right]_{a_{1}\cdots a_{p}b_{1}\cdots b_{q}}^{k}\equiv\frac{\left(p+q\right)!}{p!q!}\udi{\epsilon}k{ij}A_{[a_{1}\cdots a_{p}}^{i}B_{b_{1}\cdots b_{q}]}^{j}.
\]
In terms of spatial indices alone, we have
\[
\left[\A,\B\right]_{a_{1}\cdots a_{p}b_{1}\cdots b_{q}}\equiv\frac{\left(p+q\right)!}{p!q!}\udi{\epsilon}k{ij}A_{[a_{1}\cdots a_{p}}^{i}B_{b_{1}\cdots b_{q}]}^{j}\ta_{k},
\]
and in terms of Lie algebra indices alone, we simply have
\[
\left[\A,\B\right]^{k}=\udi{\epsilon}k{ij}A^{i}B^{j}.
\]

\subsection{\label{subsec:The-Graded-Dot}The Graded Dot Product and the Triple
Product}

We define a \emph{dot (inner) product}, also known as the \emph{Killing
form}, on the generators of the Lie group as follows:
\begin{equation}
\J_{i}\cdot\P_{j}=\delta_{ij}\sp\J_{i}\cdot\J_{j}=\P_{i}\cdot\P_{j}=0,\label{eq:dot-product}
\end{equation}
where $\delta_{ij}$ is the Kronecker delta. Given two Lie-algebra-valued
forms $\A$ and $\B$ of degrees\emph{ }$\deg\A$ and $\deg\B$ respectively,
such that $\A\equiv A^{i}\J_{i}$ is a pure rotation and $\B\equiv B^{i}\P_{i}$
is a pure translation, we define the \emph{graded dot product}:
\[
\A\cdot\B\equiv\delta_{ij}A^{i}\wedge B^{j},
\]
where $\wedge$ is the usual \emph{wedge product}\footnote{Given any two differential forms $A$ and $B$, the wedge product
$A\wedge B$ is the $\left(\deg A+\deg B\right)$-form satisfying
$A\wedge B=\left(-1\right)^{\deg A\deg B}B\wedge A$ and $\d\left(A\wedge B\right)=\d A\wedge B+\left(-1\right)^{\deg A}A\wedge\deg B$.}\emph{ }of differential forms. The dot product satisfies
\[
\A\cdot\B=\left(-1\right)^{\deg\A\deg\B}\B\cdot\A.
\]
Again, if at least one of the forms has even degree, this reduces
to the usual symmetric dot product. Note that $\A\cdot\B$ is a real-valued
$\left(\deg\A+\deg\B\right)$-form.

The graded dot product satisfies the graded Leibniz rule:
\[
\d\left(\A\cdot\B\right)=\d\A\cdot\B+\left(-1\right)^{\deg\A}\A\cdot\d\B.
\]
In terms of indices, with $\deg\A=p$ and $\deg\B=q$, we have
\[
\A\cdot\B=\frac{1}{\left(p+q\right)!}\left(\A\cdot\B\right)_{a_{1}\cdots a_{p}b_{1}\cdots b_{q}}\d x^{a_{1}}\wedge\cdots\wedge\d x^{a_{p}}\wedge\d x^{b_{1}}\wedge\cdots\wedge\d x^{b_{q}},
\]
where
\[
\left(\A\cdot\B\right)_{a_{1}\cdots a_{p}b_{1}\cdots b_{q}}=\frac{\left(p+q\right)!}{p!q!}\delta_{ij}A_{[a_{1}\cdots a_{p}}^{i}B_{b_{1}\cdots b_{q}]}^{j}.
\]
Since the graded dot product is a trace, and thus cyclic, it satisfies
\begin{equation}
\left(g^{-1}\A g\right)\cdot\left(g^{-1}\B g\right)=\A\cdot\B,\label{eq:cyclic}
\end{equation}
where $g$ is any group element. We will use this identity many times
throughout the paper to simplify expressions.

Finally, by combining the dot product and the commutator, we obtain
the \emph{triple product}:
\begin{equation}
\left[\A,\B\right]\cdot\C=\A\cdot\left[\B,\C\right]=\epsilon_{ijk}A^{i}\wedge B^{j}\wedge C^{k}.\label{eq:triple-product}
\end{equation}
Note that this is a real-valued $\left(\deg\A+\deg\B+\deg\C\right)$-form.
The triple product inherits the symmetry and anti-symmetry properties
of the dot product and the commutator.

\subsection{\label{subsec:Variational-Anti-Derivations-on}Variational Anti-Derivations
on Field Space}

In addition to the familiar \emph{exterior derivative} (or \emph{differential})
$\d$ and \emph{interior product} $\iota$ on spacetime, we introduce
a \emph{variational exterior derivative} (or \emph{variational differential})
$\delta$ and a \emph{variational interior product} $I$ on field
space. These operators act analogously to $\d$ and $\iota$, and
in particular they are nilpotent, e.g. $\delta^{2}=0$, and satisfy
the graded Leibniz rule as defined above.

Degrees of differential forms are counted with respect to spacetime
and field space separately; for example, if $f$ is a 0-form then
$\d\delta f$ is a 1-form on spacetime, due to $\d$, and independently
also a 1-form on field space, due to $\delta$. The dot product defined
above also includes an implicit wedge product with respect to field-space
forms, such that e.g. $\delta\A\cdot\delta\B=-\delta\B\cdot\delta\A$
if $\f$ and $\g$ are 0-forms on field space. In this paper, the
only place where one should watch out for the wedge product and graded
Leibniz rule on field space is when we will discuss the symplectic
form, which is a field-space 2-form; everywhere else, we will only
deal with field-space 0-forms and 1-forms.

We also define a convenient shorthand notation for the \emph{Maurer-Cartan
1-form }on field space:
\begin{equation}
\De g\equiv\delta gg^{-1},\label{eq:Maurer-Cartan}
\end{equation}
where $g$ is a $G$-valued 0-form, which satisfies
\begin{equation}
\De\left(gh\right)=\De g+g\De hg^{-1}=g\left(\De h-\De(g^{-1})\right)g^{-1},\label{eq:Delta-id-1}
\end{equation}
\begin{equation}
\De g^{-1}=-g^{-1}\De gg\sp\delta\left(\De g\right)=\hf\left[\De g,\De g\right].\label{eq:Delta-id-2}
\end{equation}
Note that $\De g$ is a $\mfg$-valued form; in fact, $\De$ can be
interpreted as a map from the Lie group $G$ to its Lie algebra $\mfg$.

\subsection{\label{subsec:Holonomies-and}$G\ltimes\protect\mfg^{*}$-valued
Holonomies and the Adjacent Subscript Rule}

A $G\ltimes\mfg^{*}$-valued holonomy from a point $a$ to a point
$b$ will be denoted as
\[
\HH_{ab}\equiv\pexp\int_{a}^{b}\AAb,
\]
where $\AAb$ is the $\mfg\oplus\mfg^{*}$-valued connection 1-form
and $\pexp$ is a \emph{path-ordered exponential}. Composition of
two holonomies works as follows:
\[
\HH_{ab}\HH_{bc}=\left(\pexp\int_{a}^{b}\AAb\right)\left(\pexp\int_{b}^{c}\AAb\right)=\pexp\int_{a}^{c}\AAb=\HH_{ac}.
\]
Therefore, in our notation, \textbf{adjacent holonomy subscripts must
always be identical}; a term such as $\HH_{ab}\HH_{cd}$ is illegal,
since one can only compose two holonomies if the second starts where
the first ends. Inversion of holonomies works as follows:
\[
\HH_{ab}^{-1}=\left(\pexp\int_{a}^{b}\AAb\right)^{-1}=\pexp\int_{b}^{a}\AAb=\HH_{ba}.
\]
For the Maurer-Cartan 1-form on field space, we \textbf{move the end
point of the holonomy to a superscript}:
\[
\De\HH_{a}^{b}\equiv\delta\HH_{ab}\HH_{ba}.
\]
On the right-hand side, the subscripts $b$ are adjacent, so the two
holonomies $\delta\HH_{ab}$ and $\HH_{ba}$ may be composed. However,
one can only compose $\De\HH_{a}^{b}$ with a holonomy that starts
at $a$, and $b$ is raised to a superscript to reflect that. For
example, $\De\HH_{a}^{b}\HH_{bc}$ is illegal, since this is actually
$\delta\HH_{ab}\HH_{ba}\HH_{bc}$ and the holonomies $\HH_{ba}$ and
$\HH_{bc}$ cannot be composed. However, $\De\HH_{a}^{b}\HH_{ac}$
is perfectly legal, and results in $\delta\HH_{ab}\HH_{ba}\HH_{ac}=\delta\HH_{ab}\HH_{bc}$.

Note that from (\ref{eq:Delta-id-1}) and (\ref{eq:Delta-id-2}) we
have
\[
\De\HH_{b}^{a}=-\HH_{ba}\De\HH_{a}^{b}\HH_{ab},
\]
\[
\De\HH_{a}^{c}=\De\left(\HH_{ab}\HH_{bc}\right)=\De\HH_{a}^{b}+\HH_{ab}\De\HH_{b}^{c}\HH_{ba}=\HH_{ab}\left(\De\HH_{b}^{c}-\De\HH_{b}^{a}\right)\HH_{ba},
\]
both of which are compatible with the adjacent subscripts rule.

\subsection{\label{subsec:The-Cartan-Decomposition}The Cartan Decomposition}

We can split a $G\ltimes\mfg^{*}$-valued (Euclidean) holonomy $\HH_{ab}$
into a \emph{rotational holonomy} $h_{ab}$, valued in $G$, and a
\emph{translational holonomy} $\x_{a}^{b}$, valued in $\mfg^{*}$.
We do this using the \emph{Cartan decomposition}
\[
\HH_{ab}\equiv\e^{\x_{a}^{b}}h_{ab}\sp\HH_{ab}\in\Omega^{0}\left(\Sigma,G\ltimes\mfg^{*}\right)\sp h_{ab}\in\Omega^{0}\left(\Sigma,G\right)\sp\x_{a}^{b}\in\Omega^{0}\left(\Sigma,\mfg^{*}\right).
\]
In the following, we will employ the useful identity
\begin{equation}
h\e^{\x}h^{-1}=\e^{h\x h^{-1}}\sp h\in\Omega^{0}\left(\Sigma,G\right)\sp\x\in\Omega^{0}\left(\Sigma,\mfg^{*}\right),\label{eq:holonomy-exp-id}
\end{equation}
which for matrix Lie algebras (such as the ones we use here) may be
proven by writing the exponential as a power series.

Taking the inverse of $\HH_{ab}$ and using (\ref{eq:holonomy-exp-id}),
we get
\[
\HH_{ab}^{-1}=\left(\e^{\x_{a}^{b}}h_{ab}\right)^{-1}=h_{ab}^{-1}\e^{-\x_{a}^{b}}=h_{ab}^{-1}\e^{-\x_{a}^{b}}\left(h_{ab}h_{ab}^{-1}\right)=\e^{-h_{ab}^{-1}\x_{a}^{b}h_{ab}}h_{ab}^{-1}.
\]
But on the other hand
\[
\HH_{ab}^{-1}=\HH_{ba}=\e^{\x_{b}^{a}}h_{ba}.
\]
Therefore, we conclude that
\begin{equation}
h_{ba}=h_{ab}^{-1}\sp\x_{b}^{a}=-h_{ab}^{-1}\x_{a}^{b}h_{ab}.\label{eq:holonomy-inv}
\end{equation}
Similarly, composing two $G\ltimes\mfg^{*}$-valued holonomies and
using (\ref{eq:holonomy-exp-id}) and (\ref{eq:holonomy-inv}), we
get
\begin{align*}
\HH_{ab}\HH_{bc} & =\left(\e^{\x_{a}^{b}}h_{ab}\right)\left(\e^{\x_{b}^{c}}h_{bc}\right)\\
 & =\e^{\x_{a}^{b}}h_{ab}\e^{\x_{b}^{c}}\left(h_{ba}h_{ab}\right)h_{bc}\\
 & =\e^{\x_{a}^{b}}\e^{h_{ab}\x_{b}^{c}h_{ba}}h_{ab}h_{bc}\\
 & =\e^{\x_{a}^{b}+h_{ab}\x_{b}^{c}h_{ba}}h_{ab}h_{bc},
\end{align*}
where we used the fact that $\mfg^{*}$ is abelian, and therefore
the exponentials may be combined linearly. On the other hand
\[
\HH_{ab}\HH_{bc}=\HH_{ac}=\e^{\x_{a}^{c}}h_{ac},
\]
so we conclude that
\begin{equation}
h_{ac}=h_{ab}h_{bc}\sp\x_{a}^{c}=\x_{a}^{b}\oplus\x_{b}^{c}\equiv\x_{a}^{b}+h_{ab}\x_{b}^{c}h_{ba}=h_{ab}\left(\x_{b}^{c}-\x_{b}^{a}\right)h_{ba},\label{eq:holonomy-comp}
\end{equation}
where in the second identity we denoted the composition of the two
translational holonomies with a $\oplus$, and used (\ref{eq:holonomy-inv})
to get the right-hand side. It is now clear why the end point of the
translational holonomy is a superscript -- again, this is for compatibility
with the adjacent subscript rule.

\section{\label{sec:Gravity-as-a}Gravity as a Gauge Theory in the Ashtekar
Formulation}

We will now present the action and phase space structure for classical
gravity in 3+1 spacetime dimensions, as formulated using the Ashtekar
variables. As explained above, this formulation allows us to describe
gravity as a Yang-Mills gauge theory. The interested reader may refer
to Appendix \ref{sec:Derivation-of-Ashtekar} and to \cite{PhDThesis}
for a complete and detailed derivation of these results. Here, we
just summarize them.

\subsection{The Ashtekar Action with Indices}

In Appendix \ref{sec:Derivation-of-Ashtekar}, we find that the \emph{Ashtekar
action }of classical loop gravity is:
\[
S=\frac{1}{\gamma}\int\d t\int_{\Sigma}\d^{3}x\left(\Et_{i}^{a}\partial_{t}A_{a}^{i}+\lambda^{i}G_{i}+N^{a}V_{a}+NC\right),
\]
where:
\begin{itemize}
\item $\gamma$ is the \emph{Barbero-Immirzi parameter},
\item $\Sigma$ is a 3-dimensional spatial slice,
\item $a,b,c,\ldots\in\left\{ 1,2,3\right\} $ are spatial indices on $\Sigma$,
\item $i,j,k,\ldots\in\left\{ 1,2,3\right\} $ are indices in the Lie algebra
$\mfg$,
\item $\Et_{i}^{a}\equiv\det\left(e\right)e_{i}^{a}$ is the \emph{densitized
triad}, a rank $\left(1,0\right)$ tensor of density weight $-1$,
where $e_{i}^{a}$ is the inverse \emph{frame field} (or \emph{triad}),
related to the inverse spatial metric $g^{ab}$ via $g^{ab}=e_{i}^{a}e_{j}^{b}\delta^{ij}$,
\item $A_{a}^{i}$ is the \emph{Ashtekar-Barbero connection},
\item $\partial_{t}$ is the derivative with respect to the time coordinate
$t$, such that each spatial slice is at a constant value of $t$,
\item $\lambda^{i}$, $N^{a}$ and $N$ are Lagrange multipliers,
\item $G_{i}\equiv\partial_{a}\Et_{i}^{a}+\epsilon_{ij}^{k}A_{a}^{j}\Et_{k}^{a}$
is the \emph{Gauss constraint},
\item $V_{a}\equiv\Et_{i}^{b}F_{ab}^{i}$ is the \emph{vector (or momentum
or diffeomorphism) constraint}, where $F_{ab}^{i}$ is the \emph{curvature
}of the Ashtekar-Barbero connection:
\[
F_{ab}^{i}\equiv\partial_{a}A_{b}^{i}-\partial_{b}A_{a}^{i}+\epsilon_{jk}^{i}A_{a}^{j}A_{b}^{k}.
\]
\item $C$ is the\emph{ scalar (or Hamiltonian) constraint}, defined as
\[
C\equiv\frac{\epsilon_{i}^{mn}\Et_{m}^{a}\Et_{n}^{b}}{2\sqrt{\det\left(\Et\right)}}\left(F_{ab}^{i}-\left(1+\gamma^{2}\right)\epsilon_{jk}^{i}K_{a}^{j}K_{b}^{k}\right),
\]
where $K_{a}^{i}$ is the \emph{extrinsic curvature}.
\end{itemize}
From the first term in the action, we see that the connection and
densitized triad are conjugate variables, and they form the Poisson
algebra
\[
\left\{ A_{a}^{i}\left(x\right),A_{b}^{j}\left(y\right)\right\} =\left\{ \Et_{i}^{a}\left(x\right),\Et_{j}^{b}\left(y\right)\right\} =0,
\]
\[
\left\{ A_{a}^{i}\left(x\right),\Et_{j}^{b}\left(y\right)\right\} =\gamma\delta_{j}^{i}\delta_{a}^{b}\delta\left(x,y\right).
\]

\subsection{The Ashtekar Action in Index-Free Notation}

In index-free notation, which we will use throughout this paper, the
action takes the form
\[
S=\frac{1}{\gamma}\int\d t\int_{\Sigma}\left(\E\cdot\partial_{t}\A+\la\cdot\left[\ee,\P\right]+\N\cdot\left[\ee,\F\right]+N\left(\ee\cdot\F-\hf\left(\frac{1}{\gamma}+\gamma\right)\K\cdot\P\right)\right),
\]
where now:
\begin{itemize}
\item $\E\equiv\hf\left[\ee,\ee\right]$ is the electric field 2-form, defined
in terms of the densitized triad and the frame field as 
\[
\E\equiv\hf E_{ab}^{i}\ta_{i}\d x^{a}\wedge\d x^{b}\soosp E_{ab}^{i}=\ept_{abc}\delta^{ij}\Et_{j}^{c}=\epsilon_{jk}^{i}e_{a}^{j}e_{b}^{k}.
\]
\item $\A\equiv A_{a}^{i}\ta_{i}\d x^{a}$ is the $\mfg$-valued Ashtekar-Barbero
connection 1-form.
\item $\ee\equiv e_{a}^{i}\ta_{i}\d x^{a}$ is the $\mfg$-valued frame
field 1-form.
\item $\P\equiv\d_{\A}\ee$ is a $\mfg$-valued 2-form.
\item The Gauss constraint is $\la\cdot\left[\ee,\P\right]=\la\cdot\d_{\A}\E$
where the Lagrange multiplier $\la$ is a $\mfg$-valued 0-form.
\item The vector (or momentum or diffeomorphism) constraint is $\N\cdot\left[\ee,\F\right]$
where the Lagrange multiplier $\N$ is a $\mfg$-valued 0-form and
$\F$ is the $\mfg$-valued curvature 2-form
\[
\F\equiv\d_{\A}\A\equiv\d\A+\hf\left[\A,\A\right].
\]
\item The scalar (or Hamiltonian) constraint is $N\left(\ee\cdot\F-\hf\left(\frac{1}{\gamma}+\gamma\right)\K\cdot\P\right)$
where the Lagrange multiplier $N$ is a 0-form, \textbf{not }valued
in $\mfg$, and $\K\equiv K_{a}^{i}\ta_{i}\d x^{a}$ is the $\mfg$-valued
extrinsic curvature 1-form.
\end{itemize}
Finally, the \emph{symplectic potential}, in index-free notation,
is
\begin{equation}
\Theta=\int_{\Sigma}\E\cdot\delta\A,\label{eq:symplectic-potential}
\end{equation}
and it corresponds to the \emph{symplectic form}
\[
\Omega\equiv\delta\Theta=\int_{\Sigma}\delta\E\cdot\delta\A.
\]
The reader is referred to Appendix \ref{sec:Derivation-of-Ashtekar}
for derivations of the index-free expressions.

\subsection{\label{subsec:The-Constraints-as}The Constraints as Generators of
Symmetries}

Let $\CC$ be the space of smooth connections on $\Sigma$. The kinematical
(unconstrained) phase space of 3+1-dimensional gravity is given by
the cotangent bundle $\PP\equiv T^{*}\CC$. To get the physical (that
is, gauge-invariant) phase space, we must perform symplectic reductions
with respect to the constraints. These constraints are best understood
in their smeared form as generators of gauge transformations. The
smeared \emph{Gauss constraint} can be written as
\[
\GG\left(\al\right)\equiv\hf\int_{\Sigma}\la\cdot\d_{\A}\E,
\]
where $\la$ is a $\mfg$-valued 0-form. This constraint generates
the infinitesimal $G$ gauge transformations:
\[
\left\{ \A,\GG\left(\la\right)\right\} \propto\d_{\A}\la\sp\left\{ \E,\GG\left(\la\right)\right\} \propto\left[\E,\la\right].
\]
The smeared \emph{vector constraint} is given by
\[
\VV\left(\xi\right)\equiv\int_{\Sigma}\N\cdot\left[\ee,\F\right],
\]
where $\xi^{a}$ is a spatial vector and the Lagrange multiplier $N^{i}\equiv\xi^{a}e_{a}^{i}$
is a $\mfg$-valued 0-form. From the Gauss and vector constraints
we may construct the \emph{diffeomorphism constraint}:
\[
\DD\left(\xi\right)\equiv\VV\left(\xi\right)-\GG\left(\xi\lrc\A\right),
\]
where $\xi\lrc\A\equiv\xi^{a}A_{a}^{i}\ta_{i}$ is an interior product
(see Footnote \ref{fn:The-interior-product}). This constraint generates
the infinitesimal spatial diffeomorphism transformations
\[
\left\{ \A,\DD\left(\xi\right)\right\} \propto\LLL_{\xi}\A\sp\left\{ \E,\DD\left(\xi\right)\right\} \propto\LLL_{\xi}\E,
\]
where $\LLL_{\xi}$ is the \emph{Lie derivative}.

\section{\label{sec:The-Discrete-Geometry-3P1}The Discrete Geometry}

In this chapter, we will present the discrete geometry under consideration:
a cellular decomposition, with curvature and torsion located only
on the edges of the cells, which we interpret as a network of cosmic
strings.

\subsection{The Cellular Decomposition and Its Dual}

We embed a \emph{cellular decomposition} $\Delta$ and a \emph{dual
cellular decomposition} $\Delta^{*}$ in our 3-dimensional spatial
manifold $\Sigma$. These structures consist of the following elements,
where each element of $\Delta$ is \textbf{uniquely dual }to an element
of $\Delta^{*}$. Each \emph{cell }$c\in\Delta$ is uniquely dual
to a \emph{node }$c^{*}\in\Delta^{*}$. The boundary of the cell $c$
is composed of \emph{sides }$s_{1},s_{2},\ldots\in\Delta$, which
are uniquely dual to \emph{links }$s_{1}^{*},s_{2}^{*},\ldots\in\Delta^{*}$;
these links are exactly all the links which are connected to the node
$c^{*}$. The boundary of each side is composed of \emph{edges} $e_{1},e_{2},\ldots\in\Delta$,
which are uniquely dual to \emph{faces} $e_{1}^{*},e_{2}^{*},\ldots\in\Delta^{*}$.
Finally, the boundary of each edge\footnote{In the 2+1D analysis of \cite{FirstPaper,Shoshany:2019ymo}, we regularized
the singularities, which were then at the vertices of $\Delta$, with
disks. In the 3+1-dimensional case considered here, it would make
sense to similarly regularize the singularities, which will now be
on the edges of $\Delta$, using cylinders. This construction is left
for future work; in this paper, we will not worry about regularizing
the singularities, and instead just use holonomies to probe the curvature
and torsion in an indirect manner, as will be shown below.} is composed of \emph{vertices }$v,v'\in\Delta$, which are uniquely
dual to \emph{volumes} $v^{*},v^{\prime*}\in\Delta^{*}$. This is
summarized in the following table:
\begin{center}
\begin{tabular}{|c|c|c|}
\hline 
$\Delta$ &  & $\Delta^{*}$\tabularnewline
\hline 
\hline 
0-cells (\emph{vertices}) $v$ & dual to & 3-cells (\emph{volumes}) $v^{*}$\tabularnewline
\hline 
1-cells (\emph{edges}) $e$ & dual to & 2-cells (\emph{faces}) $e^{*}$\tabularnewline
\hline 
2-cells \emph{(sides)} $s$ & dual to & 1-cells (\emph{links}) $s^{*}$\tabularnewline
\hline 
3-cells \emph{(cells)} $c$ & dual to & 0-cells (\emph{nodes}) $c^{*}$\tabularnewline
\hline 
\end{tabular}
\par\end{center}

\begin{figure}[p]
\begin{centering}
\includegraphics[width=1\textwidth]{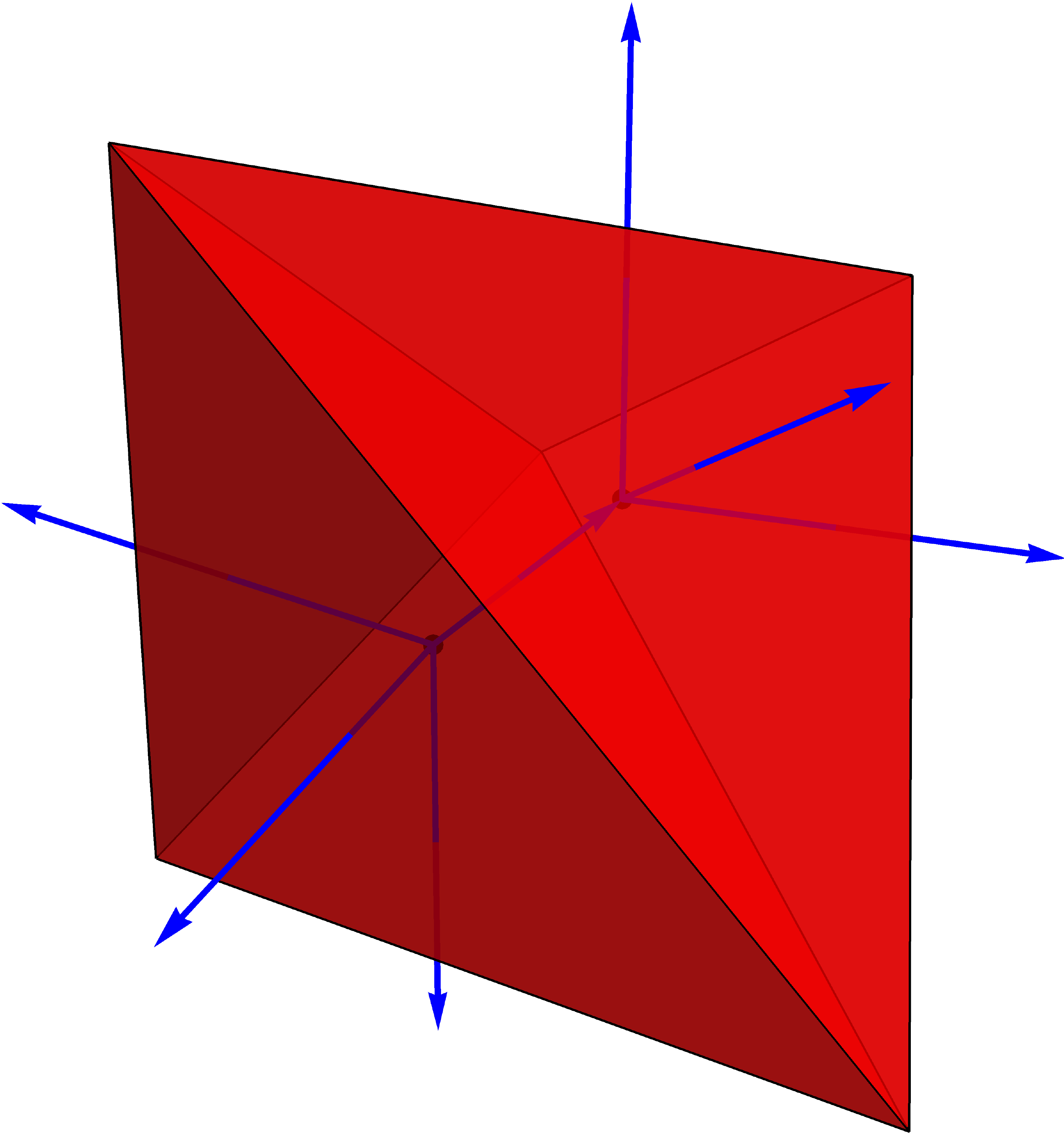}
\par\end{centering}
\caption{\label{fig:3P1TwoTetras}A simple discretization with two cells (in
red) and a dual spin network (in blue). Here, the discretization is
the simplest possible one: a \emph{simplicial complex}, where the
cells are 3-simplices (tetrahedrons) and their faces are 2-simplices
(triangles). The edges are 1-simplices (line segments), and the vertices
are, of course, 0-simplices (points). However, our formalism also
allows the cells to be arbitrary convex polyhedra and the faces to
be arbitrary convex polygons. From the figure, it should be clear
that each of the two cells is dual to a node (located inside it),
and each of the faces is dual to a link, with the face shared by the
two cells dual to the link connecting the two nodes. This is further
illustrated by Figure \ref{fig:3P1MiddleLink}.}
\end{figure}

We will write:
\begin{itemize}
\item $c=\left(s_{1},\ldots,s_{n}\right)$ to indicate that the boundary
of the cell $c$ is composed of the $n$ sides $s_{1},\ldots,s_{n}$.
\item $s=\left(e_{1},\ldots,e_{n}\right)$ to indicate that the boundary
of the side $s$ is composed of the $n$ edges $e_{1},\ldots,e_{n}$.
\item $s=\left(cc'\right)$ to indicate that the side $s$ is shared by
the two cells $c,c'$.
\item $s^{*}=\left(cc'\right)^{*}$ to indicate that the link $s^{*}$ (dual
to the side $s$) connects the two nodes $c^{*}$ and $c^{\prime*}$
(dual to the cells $c,c'$).
\item $e=\left(c_{1},\ldots,c_{n}\right)$ to indicate the the edge $e$
is shared by the $n$ cells $c_{1},\ldots,c_{n}$.
\item $e=\left(s_{1},\ldots,s_{n}\right)$ to indicate the the edge $e$
is shared by the $n$ sides $s_{1},\ldots,s_{n}$.
\item $e=\left(vv'\right)$ to indicate that the edge $e$ connects the
two vertices $v,v'$.
\end{itemize}
The \emph{1-skeleton graph} $\Gamma\subset\Delta$ is the set of all
vertices and edges of $\Delta$. It is dual to the \emph{spin network
graph} $\Gamma^{*}\subset\Delta^{*}$, the set of all nodes and links
of $\Delta^{*}$. Both graphs are oriented. This construction is illustrated
in Figure \ref{fig:3P1TwoTetras}.

\begin{figure}[p]
\begin{centering}
\includegraphics[width=1\textwidth]{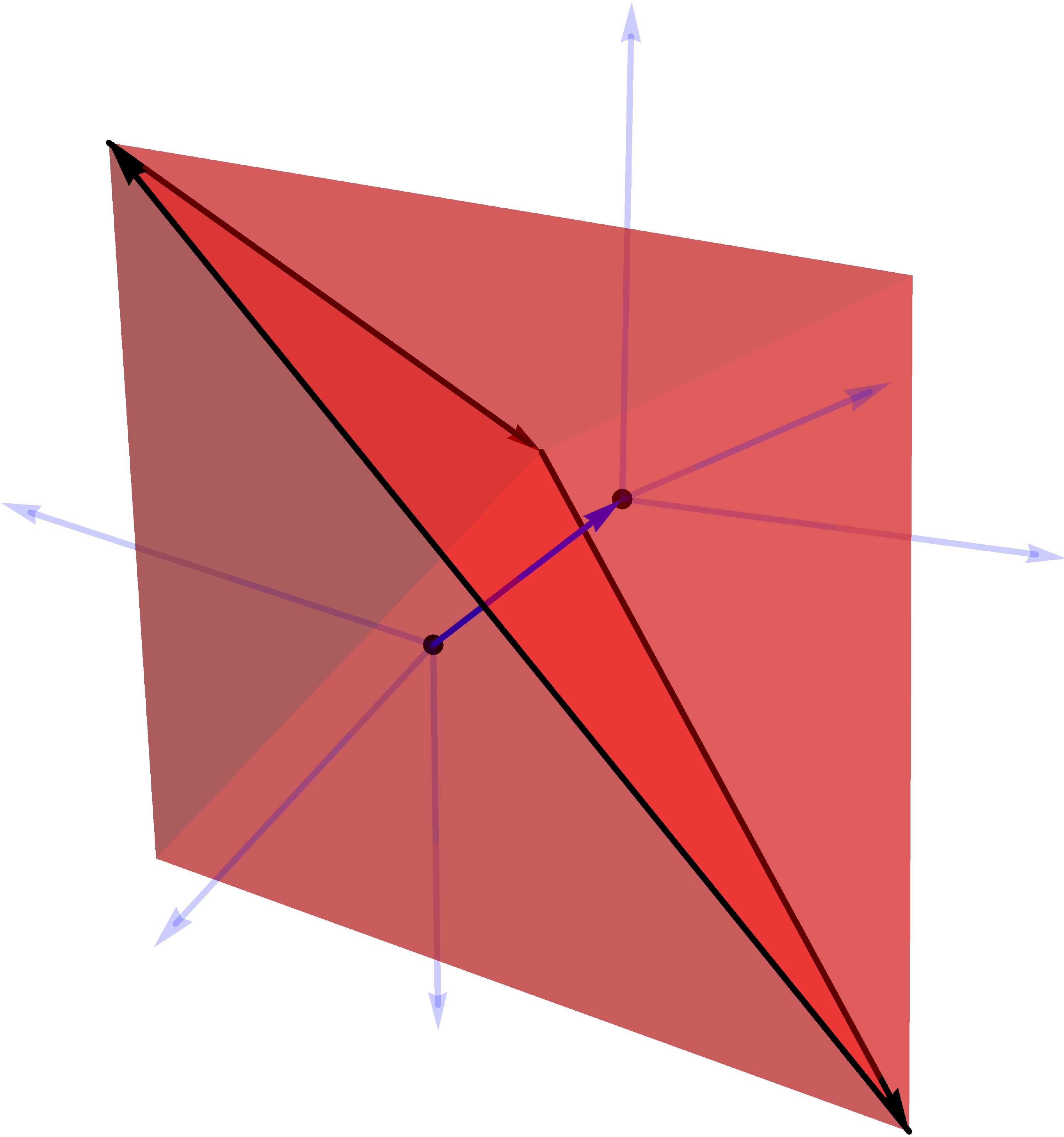}
\par\end{centering}
\caption{\label{fig:3P1MiddleLink}For clarity, we have highlighted the middle
link, the two nodes it connects, the (triangular) face shared by the
two cells, and the edges on the boundary on that face (which are,
as illustrated, oriented) -- and dimmed everything else.}
\end{figure}

\subsection{Truncating the Geometry to the Edges}

The connection $\A$ and frame field $\ee$ inside each cell $c$
-- that is, on the \emph{interior }of the cell, but \textbf{not }on
the edges and vertices -- are taken to be
\[
\A\bl_{c}=h_{c}^{-1}\d h_{c}\sp\ee\bl_{c}=h_{c}^{-1}\d\x_{c}h_{c}\soosp\E\bl_{c}=\hf\left[\ee,\ee\right]\bl_{c}=h_{c}^{-1}\left[\d\x_{c},\d\x_{c}\right]h_{c},
\]
in analogy with the 2+1D case discussed in \cite{FirstPaper,Shoshany:2019ymo}.
Since they correspond to
\[
\F\equiv\d_{\A}\A=0\sp\P\equiv\d_{\A}\ee=0,
\]
they trivially solve all of the constraints:
\[
\left[\ee,\P\right]=\left[\ee,\F\right]=\ee\cdot\F-\hf\left(\frac{1}{\gamma}+\gamma\right)\K\cdot\P=0.
\]
We also impose that $\F$ and $\P$ are non-zero only on the edges:
\begin{equation}
\F=\sum_{e}\p_{e}\delta^{\left(2\right)}\left(e\right)\sp\P=\sum_{e}\j_{e}\delta^{\left(2\right)}\left(e\right),\label{eq:F,T}
\end{equation}
where $\delta^{\left(2\right)}\left(e\right)$ is a delta 2-form such
that for any 1-form $f$
\[
\int_{\Sigma}f\wedge\delta^{\left(2\right)}\left(e\right)=\int_{e}f,
\]
as discussed in Appendix (\ref{sec:Cosmic-Strings-in}), and $\p_{e}$,
$\j_{e}$ are constant algebra elements encoding the curvature and
torsion on each edge\footnote{We absorbed the factor of $2\pi$ into $\p_{e}$ and $\j_{e}$ for
brevity.}. These distributional curvature and torsion describe a network of
\emph{cosmic strings}: 1-dimensional topological defects carrying
curvature and torsion in an otherwise flat spacetime. The expressions
(\ref{eq:F,T}) are derived from first principles in Appendix \ref{sec:Cosmic-Strings-in}.

This construction describes a \emph{piecewise-flat-and-torsionless
geometry}; the cells are flat and torsionless, and the curvature and
torsion are located only on the edges of the cells. We may interpret
the 1-skeleton $\Gamma$, the set of all edges in the cellular decomposition
$\Delta$, as a network of cosmic strings.

The reason for considering this particular geometry comes from the
assumption that the geometry can only be probed by taking loops of
holonomies along the spin network. Imagine a 3-dimensional slice $\Sigma$
with arbitrary geometry. We first embed a spin network $\Gamma^{*}$,
which can be any graph, in $\Sigma$. Then we draw a dual graph, $\Gamma$,
such that each edge of $\Gamma$ passes through exactly one loop of
$\Gamma^{*}$. We take a holonomy along each of the loops of $\Gamma^{*}$,
and encode the result on the edges of $\Gamma$. The resulting discrete
geometry is exactly the one we described above, and it is completely
equivalent to the continuous geometry with which we started, since
the holonomies along the spin network cannot tell the difference between
the continuous geometry and the discrete one.

In short, given a choice of a particular spin network, an arbitrary
continuous geometry may be reduced to an equivalent discrete geometry,
given by a network of cosmic strings, one for each loop of the spin
network.

\section{\label{subsec:Classical-Spin-Networks}Classical Spin Networks}

Our goal is to show that, by discretizing the continuous phase space
of gravity, we obtain the spin network phase space of loop quantum
gravity, coupled to cosmic strings. Therefore, before we perform our
discretization, let us study the spin network phase space.

\subsection{The Spin Network Phase Space}

In the previous chapter, we defined the spin network $\Gamma^{*}$
as a collection of links $e^{*}$ connecting nodes $c^{*}$. The kinematical
spin network phase space is isomorphic to a direct product of $T^{*}G$
cotangent bundles for each link $e^{*}\in\Gamma^{*}$:
\[
P_{\Gamma^{*}}\equiv\underset{e^{*}\in\Gamma^{*}}{\prod}T^{*}G.
\]
Since $T^{*}G\cong G\ltimes\mfg^{*}$, the phase space variables are
a group element $h_{e^{*}}\in G$ and a Lie algebra element $\X_{e^{*}}\in\mfg^{*}$
for each link $e^{*}\in\Gamma^{*}$. Under orientation reversal of
the link $e^{*}$ we have, according to (\ref{eq:holonomy-inv}),
\[
h_{e^{*-1}}=h_{e^{*}}^{-1}\sp\X_{e^{*-1}}=-h_{e^{*}}^{-1}\X_{e^{*}}h_{e^{*}}.
\]
These variables satisfy the Poisson algebra derived in the next section:
\[
\{h_{e^{*}},h_{e^{\prime*}}\}=0\sp\{X_{e^{*}}^{i},X_{e^{\prime*}}^{j}\}=\delta_{e^{*}e^{\prime*}}\udi{\epsilon}{ij}kX_{e^{*}}^{k}\sp\{X_{e^{*}}^{i},h_{e^{\prime*}}\}=\delta_{e^{*}e^{\prime*}}h_{e^{*}}\J^{i},
\]
where $e^{*}$ and $e^{\prime*}$ are two links and $\J^{i}$ are
the generators of $\mfg$.

The symplectic potential is
\begin{equation}
\Theta=\sum_{e^{*}\in\Gamma^{*}}\De h_{e^{*}}\cdot\X_{e^{*}},\label{eq:spin-network-symplectic-potential}
\end{equation}
where we used the graded dot product defined in Section \ref{subsec:The-Graded-Dot}
and the Maurer-Cartan form defined in (\ref{eq:Maurer-Cartan}). This
phase space enjoys the action of the gauge group $G^{N}$, where $N$
is the number of nodes in $\Gamma^{*}$. This action is generated
by the \emph{discrete Gauss constraint} at each node,
\[
\G_{c}\equiv\sum_{e^{*}\ni c^{*}}\X_{e^{*}},
\]
where $e^{*}\ni c^{*}$ means ``all links $e^{*}$ connected to the
node $c^{*}$''. This means that the sum of the fluxes vanishes when
summed over all the links connected to the node $c^{*}$. Given a
link $e^{*}=\left(cc'\right)^{*}$, the action of the Gauss constraint
is given in terms of two group elements $g_{c},g_{c'}\in G$, one
at each node, as
\[
h_{e^{*}}\mt g_{c}h_{e^{*}}g_{c'}^{-1}\sp\X_{e^{*}}\mt g_{c}\X_{e^{*}}g_{c}^{-1}.
\]

\subsection{Calculation of the Poisson Brackets}

Let us calculate the Poisson brackets of the spin network phase space
$T^{*}G$, for one link. For the Maurer-Cartan form, we use the notation
\[
\th\equiv-\De h\equiv-\delta hh^{-1}.
\]
This serves two purposes: first, we can talk about the components
$\theta^{i}$ of $\th$ without the notation getting too cluttered,
and second, from (\ref{eq:Delta-id-2}), the Maurer-Cartan form thus
defined satisfies the \emph{Maurer-Cartan structure equation}
\begin{equation}
\F\left(\th\right)\equiv\delta_{\th}\th\equiv\delta\th+\hf\left[\th,\th\right]=0,\label{eq:curv-th}
\end{equation}
where $\F\left(\th\right)$ is the curvature of $\th$. Note that
$\th$ is a $\mfg$-valued 1-form on field space, and we also have
a $\mfg^{*}$-valued 0-form $\X$, the flux. We take a set of vector
fields $\q_{i}$ and $\p_{i}$ for $i\in\left\{ 1,2,3\right\} $ which
are chosen to satisfy
\[
\q_{j}\lrc\theta^{i}=\p_{j}\lrc\delta X^{i}=\delta_{J}^{i}\sp\p_{j}\lrc\theta^{i}=\q_{j}\lrc\delta X^{i}=0,
\]
where $\lrc$ is the usual interior product on differential forms\footnote{\label{fn:The-interior-product}The \emph{interior product }$V\lrc A$
of a vector $V$ with a $p$-form $A$, sometimes written $\iota_{V}A$
and sometimes called the \emph{contraction }of $V$ with $A$, is
the $\left(p-1\right)$-form with components
\[
\left(V\lrc A\right)_{a_{2}\cdots a_{p}}\equiv V^{a_{1}}A_{a_{1}\cdots a_{p}}.
\]
}. The symplectic potential on one link is taken to be
\[
\Theta=-\th\cdot\X,
\]
and its symplectic form is
\begin{align*}
\Omega\equiv\delta\Theta & =-\delta\left(\th\cdot\X\right)=-\left(\delta\th\cdot\X-\th\cdot\delta\X\right)\\
 & =-\left(-\hf\left[\th,\th\right]\cdot\X-\th\cdot\delta\X\right)\\
 & =\delta\X\cdot\th+\hf\X\cdot\left[\th,\th\right],
\end{align*}
where in the first line we used the graded Leibniz rule (on field-space
forms) and in the second line we used (\ref{eq:curv-th}). In components,
we have
\[
\Omega=\delta X_{k}\wedge\theta^{k}+\hf\epsilon_{ijk}\theta^{i}\wedge\theta^{j}X^{k}.
\]
Now, recall the definition of the \emph{Hamiltonian vector field}
of $f$: it is the vector field $\H_{f}$ satisfying
\[
\H_{f}\lrc\Omega=-\delta f.
\]
Let us contract the vector field $\q_{i}$ with $\Omega$ using $\q_{j}\lrc\theta^{i}=\delta_{J}^{i}$
and $\q_{j}\lrc\delta X^{i}=0$:
\begin{align*}
\q_{l}\lrc\Omega & =\q_{l}\lrc\left(\delta X_{k}\wedge\theta^{k}+\hf\epsilon_{ijk}\theta^{i}\wedge\theta^{j}X^{k}\right)\\
 & =-\delta X_{k}\delta_{l}^{k}+\hf\epsilon_{ijk}\delta_{l}^{i}\theta^{j}X^{k}-\hf\epsilon_{ijk}\delta_{l}^{j}\theta^{i}X^{k}\\
 & =-\delta X_{l}+\epsilon_{ljk}\theta^{j}X^{k}.
\end{align*}
Similarly, let us contract $\p_{i}$ with $\Omega$ using $\p_{j}\lrc\delta X^{i}=\delta_{J}^{i}$
and $\p_{j}\lrc\theta^{i}=0$:
\[
\p_{l}\lrc\Omega=\p_{l}\lrc\left(\delta X_{k}\wedge\theta^{k}+\hf\epsilon_{ijk}\theta^{i}\wedge\theta^{j}X^{k}\right)=\delta_{kl}\theta^{k}=\theta_{l}.
\]
Note that
\[
-\delta X_{i}=\q_{i}\lrc\Omega-\epsilon_{ijk}\theta^{j}X^{k}=\q_{i}\lrc\Omega-\epsilon_{ijk}\left(\p^{j}\lrc\Omega\right)X^{k}.
\]
Thus, we can construct the Hamiltonian vector field for $X^{i}$:
\[
\H_{X_{i}}\equiv\q_{i}-\epsilon_{ijk}\p^{j}X^{k}\sp\H_{X_{i}}\lrc\Omega=-\delta X_{i}.
\]
As for $h$, we consider explicitly the matrix components in the fundamental
representation, $h_{\ B}^{A}$. The Hamiltonian vector field for the
component $h_{\ B}^{A}$ satisfies, by definition,
\[
\H_{h_{\ B}^{A}}\lrc\Omega=-\delta h_{\ B}^{A}.
\]
If we multiply by $\left(h^{-1}\right)_{\ C}^{B}$, we get
\[
\left(\H_{h_{\ B}^{A}}\left(h^{-1}\right)_{\ C}^{B}\right)\lrc\Omega=-\delta h_{\ B}^{A}\left(h^{-1}\right)_{\ C}^{B}=\left(-\delta hh^{-1}\right)_{\ C}^{A}=\theta^{i}\left(\J_{i}\right)_{\ C}^{A}=\left(\p^{i}\left(\J_{i}\right)_{\ C}^{A}\right)\lrc\Omega.
\]
Thus we conclude that the Hamiltonian vector field for $h_{\ B}^{A}$
is
\[
\H_{h_{\ B}^{A}}=\left(h\J_{i}\right)_{\ B}^{A}\p^{i}.
\]
Now that we have found $\H_{X_{i}}$ and $\H_{h_{\ B}^{A}}$, we can
finally calculate the Poisson brackets. First, we have
\begin{align*}
\left\{ h_{\ B}^{A},h_{\ D}^{C}\right\}  & =-\Omega\left(\H_{h_{\ B}^{A}},\H_{h_{\ D}^{C}}\right)\\
 & =-\left(\delta X_{k}\wedge\theta^{k}+\hf\epsilon_{ijk}\theta^{i}\wedge\theta^{j}X^{k}\right)\left(\left(h\J_{l}\right)_{\ B}^{A}\p^{l},\left(h\J_{m}\right)_{\ D}^{C}\p^{m}\right)\\
 & =0,
\end{align*}
since $\p_{j}\lrc\theta^{i}=0$. Thus
\[
\left\{ h,h\right\} =0.
\]
Next, we have
\begin{align*}
\left\{ X^{i},X^{j}\right\}  & =-\Omega\left(\H_{X^{i}},\H_{X^{j}}\right)\\
 & =-\left(\delta X_{k}\wedge\theta^{k}+\hf\epsilon_{pqk}\theta^{p}\wedge\theta^{q}X^{k}\right)\left(\q^{i}-\epsilon_{lm}^{i}\p^{l}X^{m},\q^{j}-\epsilon_{no}^{j}\p^{n}X^{o}\right)\\
 & =\epsilon_{lm}^{i}\delta_{k}^{l}X^{m}\delta^{jk}-\delta^{ik}\epsilon_{no}^{j}\delta_{k}^{n}X^{o}-\hf\epsilon_{pqk}X^{k}\left(\delta^{ip}\delta^{jq}-\delta^{iq}\delta^{jp}\right)\\
 & =\epsilon_{k}^{ij}X^{k}-\epsilon_{k}^{ji}X^{k}-\hf\epsilon_{k}^{ij}X^{k}+\hf\epsilon_{k}^{ji}X^{k}\\
 & =\epsilon_{k}^{ij}X^{k}.
\end{align*}
Finally, we have
\begin{align*}
\left\{ X^{i},h_{\ B}^{A}\right\}  & =-\Omega\left(\H_{X^{i}},\H_{h_{\ B}^{A}}\right)\\
 & =-\left(\delta X_{k}\wedge\theta^{k}+\hf\epsilon_{ijk}\theta^{i}\wedge\theta^{j}X^{k}\right)\left(\q^{i}-\epsilon_{mn}^{i}\p^{m}X^{n},\left(h\J_{l}\right)_{\ B}^{A}\p^{l}\right)\\
 & =\delta^{ki}\left(h\J_{l}\right)_{\ B}^{A}\delta_{k}^{l}\\
 & =\left(h\J^{i}\right)_{\ B}^{A},
\end{align*}
so
\[
\left\{ X^{i},h\right\} =h\J^{i}.
\]
We conclude that the Poisson brackets are
\[
\{h,h\}=0\sp\{X^{i},X^{j}\}=\epsilon_{k}^{ij}X^{k}\sp\{X^{i},h\}=h\J^{i}.
\]
All of this was calculated on one link $e^{*}$. To get the Poisson
brackets for two phase space variables which are not necessarily on
the same link, we simply add a Kronecker delta function:
\[
\left\{ h_{e^{*}},h_{e^{\prime*}}\right\} =0\sp\left\{ X_{e^{*}}^{i},X_{e^{\prime*}}^{j}\right\} =\delta_{e^{*}e^{\prime*}}\udi{\epsilon}{ij}kX_{e^{*}}^{k}\sp\left\{ X_{e^{*}}^{i},h_{e^{\prime*}}\right\} =\delta_{e^{*}e^{\prime*}}h_{e^{*}}\J^{i}.
\]
This concludes our discussion of the spin network phase space.

\section{\label{sec:Discretizing-the-Symplectic}Discretizing the Symplectic
Potential}

Now we are finally ready to discretize the continuous phase space.
To do this, we will integrate the symplectic potential, which is a
3-dimensional integral, one dimension at a time.

\subsection{First Step: From Continuous to Discrete Variables}

We start with the symplectic potential obtained in (\ref{eq:symplectic-potential}),
\[
\Theta=-\int_{\Sigma}\E\cdot\delta\A.
\]
Using the identity
\[
\delta\A\bl_{c}=h_{c}^{-1}\left(\d\De h_{c}\right)h_{c},
\]
the potential becomes
\[
\Theta=-\sum_{c}\int_{c}\left[\d\x_{c},\d\x_{c}\right]\cdot\d\De h_{c}.
\]
To use Stokes' theorem in the first integral, we note that
\[
\left[\d\x_{c},\d\x_{c}\right]\cdot\d\De h_{c}=\d\left(\left[\d\x_{c},\d\x_{c}\right]\cdot\De h_{c}\right)=\d\left(\left[\x_{c},\d\x_{c}\right]\cdot\d\De h_{c}\right),
\]
hence we can write
\[
\left[\d\x_{c},\d\x_{c}\right]\cdot\d\De h_{c}=\left(1-\lambda\right)\d\left(\left[\d\x_{c},\d\x_{c}\right]\cdot\De h_{c}\right)+\lambda\d\left(\left[\x_{c},\d\x_{c}\right]\cdot\d\De h_{c}\right),
\]
so
\[
\Theta=-\sum_{c}\int_{\partial c}\left(\left(1-\lambda\right)\left[\d\x_{c},\d\x_{c}\right]\cdot\De h_{c}+\lambda\left[\x_{c},\d\x_{c}\right]\cdot\d\De h_{c}\right).
\]
This describes a family of polarizations corresponding to different
values of $\lambda\in\left[0,1\right]$, just as we found in \cite{Shoshany:2019ymo}
for the 2+1D case. There, we interpreted the choice $\lambda=0$ as
the usual loop gravity polarization and $\lambda=1$ as a dual polarization.
We motivated a relation between this dual polarization and a dual
formulation of gravity called \emph{teleparallel gravity}\footnote{In general relativity, gravity is encoded as curvature degrees of
freedom. In teleparallel gravity \cite{2005physics...3046U,Maluf:2013gaa,Ferraro:2016wht},
gravity is instead encoded as torsion degrees of freedom. In 2+1 spacetime
dimensions, where gravity is topological \cite{Carlip:1998uc}, the
theory has two constraints: the Gauss (or torsion) constraint or the
curvature (or flatness) constraint. In the quantum theory, the first
constraint that we impose is used to define the kinematics of the
theory, while the second constraint will encode the dynamics. Thus,
it seems natural to identify general relativity with the quantization
in which the Gauss constraint is imposed first, and teleparallel gravity
with that in which the curvature constraint is imposed first.

In \cite{clement}, an alternative choice was suggested where the
order of constraints is reversed. The curvature constraint is imposed
first by employing the \emph{group network }basis of translation-invariant
states, and the Gauss constraint is the one which encodes the dynamics.
This \emph{dual loop quantum gravity} quantization is the quantum
counterpart of teleparallel gravity, and could be used to study the
dual vacua proposed in \cite{Dittrich:2014wpa,Dittrich:2014wda}.

The $\lambda=1$ case in 2+1D was first studied in \cite{Dupuis:2017otn},
but only in the simple case where there are no curvature or torsion
excitations. In \cite{Shoshany:2019ymo} we expanded the analysis
to include such excitations, and analyzed both the usual case --
first studied by us in \cite{FirstPaper} -- and the dual case in
great detail, including the discrete constraints and the symmetries
they generate. Another phase space of interest, corresponding to $\lambda=1/2$,
is a mixed phase space, containing both loop gravity and its dual.
In 2+1D it is intuitively related to Chern-Simons theory \cite{Horowitz:1989ng},
as we motivated in \cite{Shoshany:2019ymo}. In this case the formalism
of \cite{FirstPaper,Shoshany:2019ymo} is related to existing results
\cite{Alekseev:1993rj,Alekseev:1994pa,Alekseev:1994au,Meusburger:2003hc,Meusburger:2005mg,Meusburger:2003ta,Meusburger:2008dc}.}. The analysis of the relation between the $\lambda=1$ polarization
and teleparallel gravity in 3+1D is left for future work.

\subsection{Second Step: From Cells to Sides}

Next we decompose the boundary $\partial c$ of each cell $c=\left(s_{1},\ldots,s_{n}\right)$
into sides $s_{1},\ldots,s_{n}$. Each side $s=\left(cc'\right)$
will have exactly two contributions, one from the cell $c$ and another,
with opposite sign, from the cell $c'$. We thus rewrite $\Theta$
as 
\[
\Theta=-\sum_{\left(cc'\right)}\int_{\left(cc'\right)}\left(I_{c'}-I_{c}\right),
\]
where
\[
I_{c}\equiv\left(1-\lambda\right)\left[\d\x_{c},\d\x_{c}\right]\cdot\De h_{c}+\lambda\left[\x_{c},\d\x_{c}\right]\cdot\d\De h_{c}.
\]
Now, the connection and frame field must be continuous across cells:
\[
\A\bl_{c}=h_{c}^{-1}\d h_{c}=h_{c'}^{-1}\d h_{c'}=\A\bl_{c'}\sp\textrm{on }\left(cc'\right),
\]
\[
\ee\bl_{c}=h_{c}^{-1}\d\x_{c}h_{c}=h_{c'}^{-1}\d\x_{c'}h_{c'}=\ee\bl_{c'}\sp\textrm{on }\left(cc'\right).
\]
For this to be satisfied, we must impose the following \emph{continuity
conditions}: 
\[
h_{c'}=h_{c'c}h_{c}\sp\x_{c'}=h_{c'c}(\x_{c}-\x_{c}^{c'})h_{cc'}\sp\textrm{on }\left(cc'\right).
\]
From these conditions we derive the following identities:
\[
\De h_{c'}=\De\left(h_{c'c}h_{c}\right)=h_{c'c}\left(\De h_{c}-\De h_{c}^{c'}\right)h_{cc'}\sp\d\De h_{c'}=h_{c'c}\d\De h_{c}h_{cc'},
\]
\[
\d\x_{c'}=\d\left(h_{c'c}(\x_{c}-\x_{c}^{c'})h_{cc'}\right)=h_{c'c}\d\x_{c}h_{cc'},
\]
where all of the conditions are valid only on the side $s=\left(cc'\right)$.
Using these conditions, we find that
\begin{align*}
I_{c'} & =\left(1-\lambda\right)\left[\d\x_{c'},\d\x_{c'}\right]\cdot\De h_{c'}+\lambda\left[\x_{c'},\d\x_{c'}\right]\cdot\d\De h_{c'}\\
 & =\left(1-\lambda\right)\left[\d\x_{c},\d\x_{c}\right]\cdot\left(\De h_{c}-\De h_{c}^{c'}\right)+\lambda\left[\x_{c}-\x_{c}^{c'},\d\x_{c}\right]\cdot\d\De h_{c}.
\end{align*}
Comparing with $I_{c}$, we see that many terms cancel, and we are
left with
\[
\Theta=\sum_{\left(cc'\right)}\int_{\left(cc'\right)}\left(\left(1-\lambda\right)\left[\d\x_{c},\d\x_{c}\right]\cdot\De h_{c}^{c'}+\lambda\left[\x_{c}^{c'},\d\x_{c}\right]\cdot\d\De h_{c}\right).
\]
Since $\De h_{c}^{c'}$ and $\x_{c}^{c'}$ are constant (unlike $h_{c}$
and $\x_{c}$), we may take them out of the integral and rewrite the
potential as
\[
\Theta=\sum_{\left(cc'\right)}\left(\left(1-\lambda\right)\De h_{c}^{c'}\cdot\int_{\left(cc'\right)}\left[\d\x_{c},\d\x_{c}\right]+\lambda\x_{c}^{c'}\cdot\int_{\left(cc'\right)}\left[\d\x_{c},\d\De h_{c}\right]\right).
\]
Now, in order to use Stokes' theorem again, we can write
\[
\left[\d\x_{c},\d\x_{c}\right]=\d\left[\x_{c},\d\x_{c}\right],
\]
and
\[
\left[\d\x_{c},\d\De h_{c}\right]=\d\left[\x_{c},\d\De h_{c}\right]=-\d\left[\d\x_{c},\De h_{c}\right],
\]
which we write, defining an additional polarization parameter $\mu\in\left[0,1\right]$,
as
\[
\left[\d\x_{c},\d\De h_{c}\right]=\left(1-\mu\right)\d\left[\x_{c},\d\De h_{c}\right]-\mu\d\left[\d\x_{c},\De h_{c}\right].
\]
The symplectic potential now becomes
\[
\Theta=\sum_{\left(cc'\right)}\left(\left(1-\lambda\right)\De h_{c}^{c'}\cdot\int_{\partial\left(cc'\right)}\left[\x_{c},\d\x_{c}\right]+\lambda\x_{c}^{c'}\cdot\int_{\partial\left(cc'\right)}\left(\left(1-\mu\right)\left[\x_{c},\d\De h_{c}\right]-\mu\left[\d\x_{c},\De h_{c}\right]\right)\right),
\]
and it describes a two-parameter family of potentials for each value
of $\lambda\in\left[0,1\right]$ and $\mu\in\left[0,1\right]$.

\subsection{\label{subsec:Third-Step:-From}Third Step: From Sides to Edges}

The boundary $\partial s$ of each side $s=\left(e_{1},\ldots,e_{n}\right)$
is composed of edges $e$. Conversely, each edge $e=\left(s_{1},\ldots,s_{N_{e}}\right)$
is part of the boundary of $N_{e}$ different sides, which we label
in sequential order $s_{i}\equiv\left(c_{i}c_{i+1}\right)$ for $i\in\left\{ 1,\ldots,N_{e}\right\} $,
with the convention that $c_{N_{e}+1}$ is the same as $c_{1}$ after
encircling the edge $e$ once. Note that this sequence of sides is
dual to a loop of links $s_{i}^{*}=\left(c_{i}c_{i+1}\right)^{*}$
around the edge $e$. Then we can rearrange the integrals as follows:
\[
\sum_{\left(cc'\right)}\int_{\partial\left(cc'\right)}=\sum_{e}\int_{e}\sum_{i=1}^{N_{e}}.
\]
The potential becomes
\[
\Theta=\sum_{e}\int_{e}\sum_{i=1}^{N_{e}}I_{c_{i}c_{i+1}},
\]
where
\[
I_{c_{i}c_{i+1}}\equiv\left(1-\lambda\right)\De h_{c_{i}}^{c_{i+1}}\cdot\left[\x_{c_{i}},\d\x_{c_{i}}\right]+\lambda\x_{c_{i}}^{c_{i+1}}\cdot\left(\left(1-\mu\right)\left[\x_{c_{i}},\d\De h_{c_{i}}\right]-\mu\left[\d\x_{c_{i}},\De h_{c_{i}}\right]\right).
\]
We would like to perform a final integration using Stokes' theorem.
For this we again need to somehow cancel some elements, as we did
before. However, since there are now $N_{e}$ different contributions,
we cannot use the continuity conditions between each pair of adjacent
cells, since in order to get cancellations, all terms must have the
same base point (subscript).

One option is to choose a particular cell and trace everything back
to that cell. However, this forces us to choose a specific cell for
each edge. A more symmetric solution involves splitting each holonomy
$h_{c_{i}c_{i+1}}$, which goes from from $c_{i}^{*}$ to $c_{i+1}^{*}$,
into two holonomies -- first going from $c_{i}^{*}$ to (some arbitrary
point $e_{0}$ on) $e$ and then back to $c_{i+1}^{*}$, using the
recipe given in Section \ref{subsec:Holonomies-and}: 
\[
h_{c_{i}c_{i+1}}=h_{c_{i}e}h_{ec_{i+1}}\sp\x_{c_{i}}^{c_{i+1}}=\x_{c_{i}}^{e}\oplus\x_{e}^{c_{i+1}}=\x_{c_{i}}^{e}+h_{c_{i}e}\x_{e}^{c_{i+1}}h_{ec_{i}}=h_{c_{i}e}\left(\x_{e}^{c_{i+1}}-\x_{e}^{c_{i}}\right)h_{ec_{i}}.
\]
From this we find that
\[
\De h_{c_{i}}^{c_{i+1}}=h_{c_{i}e}\left(\De h_{e}^{c_{i+1}}-\De h_{e}^{c_{i}}\right)h_{ec_{i}}.
\]
Therefore
\begin{align*}
I_{c_{i}c_{i+1}} & =\left(1-\lambda\right)h_{c_{i}e}\left(\De h_{e}^{c_{i+1}}-\De h_{e}^{c_{i}}\right)h_{ec_{i}}\cdot\left[\x_{c_{i}},\d\x_{c_{i}}\right]+\\
 & \qquad+\lambda h_{c_{i}e}\left(\x_{e}^{c_{i+1}}-\x_{e}^{c_{i}}\right)h_{ec_{i}}\cdot\left(\left(1-\mu\right)\left[\x_{c_{i}},\d\De h_{c_{i}}\right]-\mu\left[\d\x_{c_{i}},\De h_{c_{i}}\right]\right).
\end{align*}
Furthermore, we again have continuity conditions\footnote{If we had a cylinder around the edge $e$ to regularize the divergences,
like the disks we had in \cite{FirstPaper,Shoshany:2019ymo}, then
these conditions would have been valid on the boundary between the
cylinder and the cell. However, in the case we are considering here,
the cylinder has zero radius, so these conditions are instead valid
on the edge $e$ itself.}, this time between each cell $c_{i}$ and the edge $e$:
\[
\x_{c_{i}}=h_{c_{i}e}\left(\x_{e}-\x_{e}^{c_{i}}\right)h_{ec_{i}}\sp\d\x_{c_{i}}=h_{c_{i}e}\d\x_{e}h_{ec_{i}},
\]
\[
h_{c_{i}}=h_{c_{i}e}h_{e}\sp\De h_{c_{i}}=h_{c_{i}e}\left(\De h_{e}-\De h_{e}^{c_{i}}\right)h_{ec_{i}}\sp\d\De h_{c_{i}}=h_{c_{i}e}\d\De h_{e}h_{ec_{i}}.
\]

Plugging in, we get
\begin{align*}
I_{c_{i}c_{i+1}} & =\left(1-\lambda\right)\left(\De h_{e}^{c_{i+1}}-\De h_{e}^{c_{i}}\right)\cdot\left[\x_{e}-\x_{e}^{c_{i}},\d\x_{e}\right]+\\
 & \qquad+\lambda\left(\x_{e}^{c_{i+1}}-\x_{e}^{c_{i}}\right)\cdot\left(\left(1-\mu\right)\left[\x_{e}-\x_{e}^{c_{i}},\d\De h_{e}\right]-\mu\left[\d\x_{e},\De h_{e}-\De h_{e}^{c_{i}}\right]\right).
\end{align*}
Now we sum over all the terms, and take anything that does not depend
on $i$ out of the sum and anything that is constant out of the integral.
We get
\[
\Theta=\sum_{e}\left(\Theta_{e}+\sum_{i=1}^{N_{e}}\Theta_{e}^{c_{i}c_{i+1}}\right),
\]
where
\begin{align*}
\Theta_{e} & \equiv\left(1-\lambda\right)\int_{e}\left[\x_{e},\d\x_{e}\right]\cdot\sum_{i=1}^{N_{e}}\left(\De h_{e}^{c_{i+1}}-\De h_{e}^{c_{i}}\right)+\\
 & \qquad+\lambda\left(\left(1-\mu\right)\int_{e}\left[\x_{e},\d\De h_{e}\right]-\mu\int_{e}\left[\d\x_{e},\De h_{e}\right]\right)\cdot\sum_{i=1}^{N_{e}}\left(\x_{e}^{c_{i+1}}-\x_{e}^{c_{i}}\right),
\end{align*}
\begin{align*}
\Theta_{e}^{c_{i}c_{i+1}} & \equiv-\left(1-\lambda\right)\left[\x_{e}^{c_{i}},\int_{e}\d\x_{e}\right]\cdot\left(\De h_{e}^{c_{i+1}}-\De h_{e}^{c_{i}}\right)+\\
 & \qquad-\lambda\left(\x_{e}^{c_{i+1}}-\x_{e}^{c_{i}}\right)\cdot\left(\left(1-\mu\right)\left[\x_{e}^{c_{i}},\int_{e}\d\De h_{e}\right]-\mu\left[\int_{e}\d\x_{e},\De h_{e}^{c_{i}}\right]\right).
\end{align*}
Note that $\Theta_{e}$ exists uniquely for each edge, while $\Theta_{e}^{c_{i}c_{i+1}}$
exists uniquely for each combination of edge $e$ and side $\left(c_{i}c_{i+1}\right)$.

\subsection{The Edge Potential}

\begin{figure}[p]
\begin{centering}
\includegraphics[width=1\textwidth]{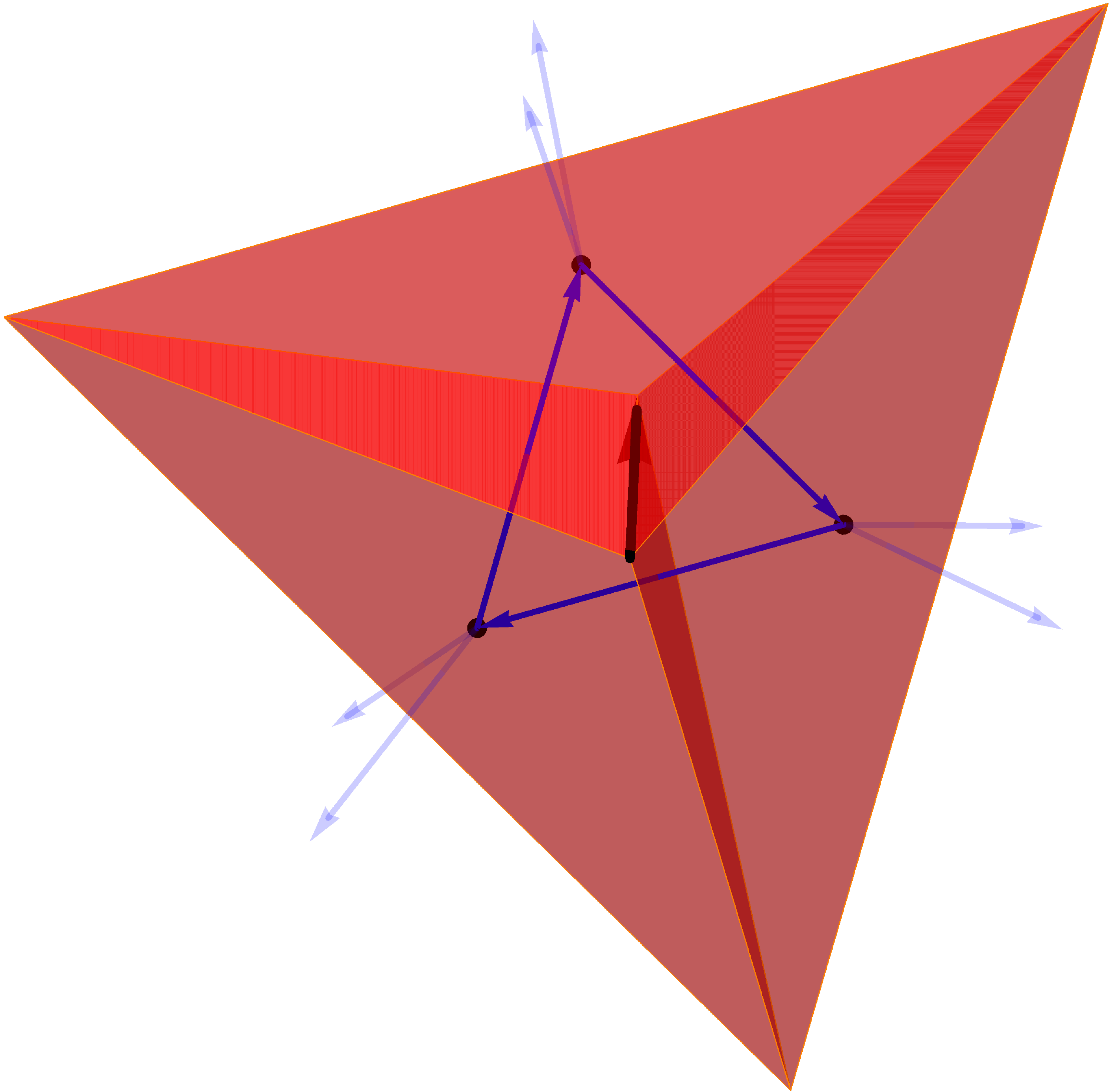}
\par\end{centering}
\caption{\label{fig:3P1Loop}Three cells (tetrahedrons, in red) dual to three
nodes (in black). The cells share three faces (highlighted) which
are dual to three oriented links (blue arrows) connecting the nodes.
The three links form a loop, which goes around the single edge shared
by all three cells (the thick black arrow in the middle). By taking
a holonomy around the loop $\left(c_{1}c_{2}c_{3}c_{1}\right)$, we
can detect the curvature and torsion encoded in the middle edge.}
\end{figure}

In $\Theta_{e}$, we notice that both sums are telescoping -- each
term cancels out one other term, and we are left with only the first
and last term:
\begin{align*}
\sum_{i=1}^{N_{e}}\left(\De h_{e}^{c_{i+1}}-\De h_{e}^{c_{i}}\right) & =\left(\De h_{e}^{c_{2}}-\De h_{e}^{c_{1}}\right)+\left(\De h_{e}^{c_{3}}-\De h_{e}^{c_{2}}\right)+\cdots+\left(\De h_{e}^{c_{N_{e}+1}}-\De h_{e}^{c_{N_{e}}}\right)\\
 & =\De h_{e}^{c_{N_{e}+1}}-\De h_{e}^{c_{1}},
\end{align*}
\begin{align*}
\sum_{i=1}^{N_{e}}\left(\x_{e}^{c_{i+1}}-\x_{e}^{c_{i}}\right) & =\left(\x_{e}^{c_{2}}-\x_{e}^{c_{1}}\right)+\left(\x_{e}^{c_{3}}-\x_{e}^{c_{2}}\right)+\cdots+\left(\x_{e}^{c_{N_{e}+1}}-\x_{e}^{c_{N_{e}}}\right)\\
 & =\x_{e}^{c_{N_{e}+1}}-\x_{e}^{c_{1}}.
\end{align*}
Now, $c_{N_{e}+1}$ is the same as $c_{1}$ after encircling $e$
once. So, if the geometry is completely flat and torsionless, we can
just say that $\Theta_{e}$ vanishes. However, if the edge carries
curvature and/or torsion, then after winding around the edge once,
the rotational and translational holonomies should detect them. This
is illustrated in Figure \ref{fig:3P1Loop}. We choose to label this
as follows:
\begin{equation}
\De h_{e}^{c_{N_{e}+1}}-\De h_{e}^{c_{1}}\equiv\delta\M_{e}\sp\x_{e}^{c_{N_{e}+1}}-\x_{e}^{c_{1}}\equiv\SS_{e}.\label{eq:holo-MS}
\end{equation}
The values of $\M_{e}$ and $\SS_{e}$ in (\ref{eq:holo-MS}) are
directly related\footnote{To find the exact relation, we should regularize the edges using cylinders,
just as we regularized the vertices using disks in the 2+1D case \cite{FirstPaper,Shoshany:2019ymo},
which then allowed us to find a relation between the holonomies and
the mass and spin of the particles. We leave this calculation for
future work.} to the values of $\p_{e}$ and $\j_{e}$ in (\ref{eq:F,T}), which
determine the momentum and angular momentum of the string that lies
on the edge $e$. We may interpret (\ref{eq:holo-MS}) in two ways.
Either we first find $\M_{e}$ and $\SS_{e}$ by calculating the difference
of holonomies, as defined in (\ref{eq:holo-MS}), and then define
$\p_{e}$ and $\j_{e}$ in (\ref{eq:F,T}) as functions of these quantities
-- or, conversely, we start with strings that have well-defined momentum
and angular momentum $\p_{e}$ and $\j_{e}$, and then define $\M_{e}$
and $\SS_{e}$ as appropriate functions of $\p_{e}$ and $\j_{e}$.

Unfortunately, aside from this simplification, it does not seem possible
to simplify $\Theta_{e}$ any further, since there is no obvious way
to write the integrands as exact 1-forms. The only thing left for
us to do, therefore, is to call the integrals by names\footnote{Our definition of $\X_{e}$ here alludes to the definition of ``angular
momentum'' in \cite{Freidel:2013bfa}, and is analogous to the ``vertex
flux'' $\X_{v}$ we defined in the 2+1D case in \cite{Shoshany:2019ymo}.
Similarly, the definition of $\De H_{e}$ (below) is analogous to
the ``vertex holonomy'' $H_{v}$ we defined in the 2+1D case.}\footnote{The definitions of $\De H_{e}$, $\De H_{e}^{1}$, and $\De H_{e}^{2}$,
which are 1-forms on field space, define the holonomies $H_{e}$,
$H_{e}^{1}$ and $H_{e}^{2}$ themselves only implicitly. Despite
the suggestive notation, in principle $\De H_{e}$ need not be of
the form $\delta H_{e}H_{e}^{-1}$ for some $G$-valued 0-form $H_{e}$.
It can instead be of the form $\delta\h_{e}$ for some $\mfg$-valued
0-form $\h_{e}$. Its precise form is left implicit, and we merely
assume that there exists some solution, either in the form $H_{e}$
or $\h_{e}$. The same applies to $H_{e}^{1}$ and $H_{e}^{2}$, and
also to $\M_{e}$ in (\ref{eq:holo-MS}).}:
\[
\X_{e}\equiv\int_{e}\left[\x_{e},\d\x_{e}\right]\sp\De H_{e}^{1}\equiv\int_{e}\left[\x_{e},\d\De h_{e}\right]\sp\De H_{e}^{2}\equiv\int_{e}\left[\d\x_{e},\De h_{e}\right],
\]
and write:
\[
\Theta_{e}=\left(1-\lambda\right)\X_{e}\cdot\delta\M_{e}+\lambda\SS_{e}\cdot\left(\left(1-\mu\right)\De H_{e}^{1}-\mu\De H_{e}^{2}\right).
\]
In fact, since both $H_{e}^{1}$ and $H_{e}^{2}$ are conjugate to
the same variable $\SS_{e}$, we might as well collect them into a
single variable:
\[
\De H_{e}\equiv\left(1-\mu\right)\De H_{e}^{1}-\mu\De H_{e}^{2},
\]
so that the choice of parameter $\mu\in\left[0,1\right]$ simply chooses
how much of $H_{e}^{1}$ compared to $H_{e}^{2}$ is used this variable.
We obtain:
\[
\Theta_{e}=\left(1-\lambda\right)\X_{e}\cdot\delta\M_{e}+\lambda\SS_{e}\cdot\De H_{e}.
\]
This term is remarkably similar to the vertex potential we found in
the 2+1D case \cite{Shoshany:2019ymo}, which represented the phase
space of a point particle with mass $\M_{e}$ and spin $\SS_{e}$.
This term encodes the dynamics of the curvature and torsion on each
edge $e$. In fact, if we perform a change of variables:
\[
\left(1-\lambda\right)\X_{e}\mt\X_{e}\sp\lambda\SS_{e}\mt-\left(\SS_{e}+\left[\M_{e},\X_{e}\right]\right),
\]
we obtain precisely the same term that we obtained in the 2+1D case:
\[
\Theta_{e}=\X_{e}\cdot\delta\M_{e}-\left(\SS_{e}+\left[\M_{e},\X_{e}\right]\right)\cdot\De H_{e}.
\]

\subsection{The Link Potential}

The term $\Theta_{e}^{c_{i}c_{i+1}}$, defined at the end of Section
\ref{subsec:Third-Step:-From}, is easily integrable. Since we don't
need the telescoping sum anymore, we can simplify this term by returning
to the original variables:
\[
\x_{e}^{c_{i+1}}-\x_{e}^{c_{i}}=h_{ec_{i}}\x_{c_{i}}^{c_{i+1}}h_{c_{i}e}\sp\De h_{e}^{c_{i+1}}-\De h_{e}^{c_{i}}=h_{ec_{i}}\De h_{c_{i}}^{c_{i+1}}h_{c_{i}e},
\]
so it becomes
\begin{align*}
\Theta_{e}^{c_{i}c_{i+1}} & =-\left(1-\lambda\right)\left[\x_{e}^{c_{i}},\int_{e}\d\x_{e}\right]\cdot h_{ec_{i}}\De h_{c_{i}}^{c_{i+1}}h_{c_{i}e}+\\
 & \qquad-\lambda h_{ec_{i}}\x_{c_{i}}^{c_{i+1}}h_{c_{i}e}\cdot\left(\left(1-\mu\right)\left[\x_{e}^{c_{i}},\int_{e}\d\De h_{e}\right]-\mu\left[\int_{e}\d\x_{e},\De h_{e}^{c_{i}}\right]\right).
\end{align*}
We also have the usual inversion relations (see Section \ref{subsec:The-Cartan-Decomposition})
\[
h_{c_{i}e}\x_{e}^{c_{i}}h_{ec_{i}}=-\x_{c_{i}}^{e}\sp,h_{c_{i}e}\De h_{e}^{c_{i}}h_{ec_{i}}=-\De h_{c_{i}}^{e},
\]
so we can further simplify to:
\begin{align*}
\Theta_{e}^{c_{i}c_{i+1}} & =\left(1-\lambda\right)\left[\x_{c_{i}}^{e},h_{c_{i}e}\left(\int_{e}\d\x_{e}\right)h_{ec_{i}}\right]\cdot\De h_{c_{i}}^{c_{i+1}}+\\
 & \qquad+\lambda\x_{c_{i}}^{c_{i+1}}\cdot\left(\left(1-\mu\right)\left[\x_{c_{i}}^{e},h_{c_{i}e}\left(\int_{e}\d\De h_{e}\right)h_{ec_{i}}\right]+\mu\left[h_{c_{i}e}\left(\int_{e}\d\x_{e}\right)h_{ec_{i}},\De h_{c_{i}}^{e}\right]\right).
\end{align*}
Next, we assume that the edge $e$ starts at the vertex $v$ and ends
at the vertex $v'$, i.e. $e=\left(vv'\right)$. Then we can evaluate
the integrals explicitly:
\[
\int_{e}\d\De h_{e}=\De h_{e}\left(v'\right)-\De h_{e}\left(v\right)\sp\int_{e}\d\x_{e}=\x_{e}\left(v'\right)-\x_{e}\left(v\right).
\]
Now, let $h_{vv'}$ and $\x_{v}^{v'}$ be the rotational and translational
holonomies along the edge $e$, that is, from $v$ to $v'$. Then
we can split them so that they also pass through a point $e_{0}$
on the edge $e$, as follows:
\[
h_{vv'}=h_{ve}h_{ev'}\soosp\De h_{v}^{v'}=h_{ve}\left(\De h_{e}^{v'}-\De h_{e}^{v}\right)h_{ev},
\]
\[
\x_{v}^{v'}=\x_{v}^{e}\oplus\x_{e}^{v'}=h_{ve}\left(\x_{e}^{v'}-\x_{e}^{v}\right)h_{ev}.
\]
Given that $h_{e}\left(v\right)=h_{ev}$ and $\x_{e}\left(v\right)=\x_{e}^{v}$,
the integrals may now be written as
\[
\int_{e}\d\De h_{e}=h_{ev}\De h_{v}^{v'}h_{ve}\sp\int_{e}\d\x_{e}=h_{ev}\x_{v}^{v'}h_{ve}.
\]
Moreover, since $h_{c_{i}e}h_{ev}=h_{c_{i}v}$, we have
\[
h_{c_{i}e}\left(\int_{e}\d\De h_{e}\right)h_{ec_{i}}=h_{c_{i}e}\left(h_{ev}\De h_{v}^{v'}h_{ve}\right)h_{ec_{i}}=h_{c_{i}v}\De h_{v}^{v'}h_{vc_{i}},
\]
\[
h_{c_{i}e}\left(\int_{e}\d\x_{e}\right)h_{ec_{i}}=h_{c_{i}e}\left(h_{ev}\x_{v}^{v'}h_{ve}\right)h_{ec_{i}}=h_{c_{i}v}\x_{v}^{v'}h_{vc_{i}}.
\]
With this, we may simplify $\Theta_{e}^{c_{i}c_{i+1}}$ to
\begin{align*}
\Theta_{e}^{c_{i}c_{i+1}} & =\left(1-\lambda\right)\left[\x_{c_{i}}^{e},h_{c_{i}v}\x_{v}^{v'}h_{vc_{i}}\right]\cdot\De h_{c_{i}}^{c_{i+1}}+\\
 & \qquad+\lambda\x_{c_{i}}^{c_{i+1}}\cdot\left(\left(1-\mu\right)\left[\x_{c_{i}}^{e},h_{c_{i}v}\De h_{v}^{v'}h_{vc_{i}}\right]+\mu\left[h_{c_{i}v}\x_{v}^{v'}h_{vc_{i}},\De h_{c_{i}}^{e}\right]\right).
\end{align*}

\subsection{Holonomies and Fluxes}

Finally, in order to relate this to the spin network phase space discussed
in Chapter \ref{subsec:Classical-Spin-Networks}, we need to identify
holonomies and fluxes. From the 2+1D case \cite{FirstPaper,Shoshany:2019ymo},
we know that the fluxes are in fact also holonomies -- but they are
translational, not rotational, holonomies. $h_{c_{i}c_{i+1}}$ is
by definition the rotational holonomy on the link $\left(c_{i}c_{i+1}\right)^{*}$,
and $\x_{c_{i}}^{c_{i+1}}$ is is by definition the translational
holonomy on the link $\left(c_{i}c_{i+1}\right)^{*}$, so it's natural
to simply define
\[
H_{c_{i}c_{i+1}}\equiv h_{c_{i}c_{i+1}}\sp\X_{c_{i}}^{c_{i+1}}\equiv\x_{c_{i}}^{c_{i+1}}.
\]
We should also define holonomies and fluxes on the sides dual to the
links. By inspection, the flux on the side $\left(c_{i}c_{i+1}\right)$
must be\footnote{Note that this expression depends only on the source cell $c_{i}$
and not on the target cell $c_{i+1}$, just as the analogous flux
in the 2+1D case only depended on the source cell. This is an artifact
of using the continuity conditions to write everything in terms of
the source cell in order to make the expression integrable; however,
the expression may be symmetrized, as we did in \cite{PhDThesis}. }
\[
\XXt_{c_{i}}^{c_{i+1}}\equiv\left[\x_{c_{i}}^{v},h_{c_{i}v}\x_{v}^{v'}h_{vc_{i}}\right].
\]
The first term in the commutator is $\x_{c_{i}}^{v}$, the translational
holonomy from the node $c_{i}^{*}$ to the vertex $v$, the starting
point of $e$. The second term contains $\x_{v}^{v'}$, the translational
holonomy along the edge $e$.

As for holonomies on the sides -- again, since we initially had two
ways to integrate, we also have two different ways to define holonomies.
However, as above, since both holonomies are conjugate to the same
flux, $\X_{c_{i}}^{c_{i+1}}$, there is really no reason to differentiate
them. Therefore we just define implicitly:
\[
\De\Ht_{c_{i}}^{c_{i+1}}\equiv\left(1-\mu\right)\left[\x_{c_{i}}^{v},h_{c_{i}v}\De h_{v}^{v'}h_{vc_{i}}\right]+\mu\left[h_{c_{i}v}\x_{v}^{v'}h_{vc_{i}},\De h_{c_{i}}^{v}\right],
\]
and the choice of parameter $\mu\in\left[0,1\right]$ simply determines
how much of this holonomy comes from each polarization. We finally
get:
\[
\Theta_{e}^{c_{i}c_{i+1}}=\left(1-\lambda\right)\XXt_{c_{i}}^{c_{i+1}}\cdot\De H_{c_{i}}^{c_{i+1}}+\lambda\X_{c_{i}}^{c_{i+1}}\cdot\De\Ht_{c_{i}}^{c_{i+1}}.
\]
This is exactly\footnote{Aside from the relative sign, which comes from the fact that in the
beginning we were writing a 3-form instead of a 2-form as an exact
form, and plays no role here since each term describes a separate
phase space.} the same term we obtained in the 2+1D case \cite{Shoshany:2019ymo}!
It represents a holonomy-flux phase space on each link. For $\lambda=0$
the holonomies are on links and the fluxes are on their dual sides,
while for the dual polarization $\lambda=1$ the fluxes are on the
links and the holonomies are on the sides, in analogy with the two
polarization we found in the 2+1D case.

\subsection{Summary}

We have obtained the following discrete symplectic potential:
\[
\Theta=\sum_{e}\left(\left(1-\lambda\right)\X_{e}\cdot\delta\M_{e}+\lambda\SS_{e}\cdot\De H_{e}+\sum_{i=1}^{N_{e}}\left(\left(1-\lambda\right)\XXt_{c_{i}}^{c_{i+1}}\cdot\De H_{c_{i}}^{c_{i+1}}+\lambda\X_{c_{i}}^{c_{i+1}}\cdot\De\Ht_{c_{i}}^{c_{i+1}}\right)\right),
\]
where for each edge $e$:
\begin{itemize}
\item $\left\{ c_{1},\ldots,c_{N_{e}}\right\} $ are the $N_{e}$ cells
around the edge,
\item $\X_{e}\equiv\int_{e}\left[\x_{e},\d\x_{e}\right]$ is the ``edge
flux'',
\item $\M_{e}$, defined implicitly by $\delta\M_{e}\equiv\De h_{e}^{c_{N_{e}+1}}-\De h_{e}^{c_{1}}$,
represents the curvature on the edge,
\item $\De H_{e}\equiv\int_{e}\left(\left(1-\mu\right)\left[\x_{e},\d\De h_{e}\right]-\mu\left[\d\x_{e},\De h_{e}\right]\right)$
is the ``edge holonomy'',
\item $\SS_{e}\equiv\x_{e}^{c_{N_{e}+1}}-\x_{e}^{c_{1}}$ represents the
torsion on the edge,
\item $\XXt_{c_{i}}^{c_{i+1}}\equiv\left[\x_{c_{i}}^{v},h_{c_{i}v}\x_{v}^{v'}h_{vc_{i}}\right]$
is the flux on the side $\left(c_{i}c_{i+1}\right)$ shared by the
cells $c_{i}$ and $c_{i+1}$,
\item $H_{c_{i}c_{i+1}}\equiv h_{c_{i}c_{i+1}}$ is the holonomy on the
link $\left(c_{i}c_{i+1}\right)^{*}$ dual to the side $\left(c_{i}c_{i+1}\right)$,
\item $\X_{c_{i}}^{c_{i+1}}\equiv\x_{c_{i}}^{c_{i+1}}$ is the flux on the
link $\left(c_{i}c_{i+1}\right)^{*}$,
\item $\Ht_{c_{i}}^{c_{i+1}}$, defined implicitly by $\De\Ht_{c_{i}}^{c_{i+1}}\equiv\left(1-\mu\right)\left[\x_{c_{i}}^{v},h_{c_{i}v}\De h_{v}^{v'}h_{vc_{i}}\right]+\mu\left[h_{c_{i}v}\x_{v}^{v'}h_{vc_{i}},\De h_{c_{i}}^{v}\right]$,
is the holonomy on the side $\left(c_{i}c_{i+1}\right)$.
\end{itemize}
We interpret this as the phase space of a spin network $\Gamma^{*}$
coupled to a network of cosmic strings $\Gamma$, with mass and spin
related to the curvature and torsion.

\section{\label{sec:Conclusions}Conclusions}

\subsection{Summary of Our Results}

In this paper, we performed a piecewise-flat-and-torsionless discretization
of 3+1D classical general relativity in the first-order formulation,
keeping track of curvature and torsion via holonomies. We showed that
the resulting phase space is precisely that of spin networks, the
quantum states of discrete spacetime in loop quantum gravity, coupled
to a network of cosmic strings, 1-dimensional topological defects
carrying curvature and torsion. Our results illustrate, for the first
time, a precise way in which spin network states can be assigned classical
spatial geometries and/or matter distributions.

Each node of the spin network is dual to a 3-dimensional cell, and
each link connecting two nodes is dual to the side shared by the two
corresponding cells. A loop of links (or a face) is dual to an edge
of the cellular decomposition. These edges are the locations where
strings reside, and by examining the value of the holonomies along
the loop dual to an edge, we learn about the curvature and torsion
induced by the string at the edge by virtue of the Einstein equation.

Equivalently, if we assume that the only way to detect curvature and
torsion is by looking at appropriate holonomies on the loops of the
spin networks, then we may interpret our result as taking some arbitrary
continuous geometry, not necessarily generated by strings, truncating
it, and encoding it on the edges. The holonomies cannot tell the difference
between a continuous geometry and a singular geometry; they can only
tell us about the total curvature and torsion inside the loop.

\subsection{Future Plans}

In previous papers \cite{FirstPaper,Shoshany:2019ymo} we presented
a very detailed analysis of the 2+1-dimensional toy model, which is,
of course, simpler than the realistic 3+1-dimensional case. This analysis
was performed with the philosophy that the 2+1D toy model can provide
deep insights about the 3+1D theory. Indeed, many structures from
the 2+1D case, such as the cellular decomposition and its relation
to the spin network, the rotational and translational holonomies and
their properties, and the singular matter sources, can be readily
generalized to the 3+1D with minimal modifications. Thus, other results
should be readily generalizable as well. As we have seen, we indeed
obtain the same symplectic potential in both cases, which is not surprising
-- since we used the same structures in both.

However, the 3+1-dimensional case presents many challenges which would
require much more work, far beyond the scope of this paper, to overcome.
Here we present some suggestions for possible research directions
in 3+1D. Note that there are also many things one could explore in
the 2+1D case, but we choose to focus on 3+1D since it is the physically
relevant case. Of course, in many cases it would be beneficial to
try introducing new structures (e.g. a cosmological constant) in the
2+1D case first, since the lessons learned from the toy theory may
then be employed in the realistic theory -- as we, indeed, did in
this paper.

\textbf{1. Proper treatment of the singularities}
\begin{quote}
In the 2+1D case, we carefully treated the 0-dimensional singularities,
the point particles, by regularizing them with disks. This introduced
many complications, but also ensured that our results were completely
rigorous. In the 3+1D case, we skipped this crucial part, and instead
jumped right to the end by assuming the results we had in 2+1D apply
to the 3+1D case as well.

It would be instructive to repeat this in 3+1D and carefully treat
the 1-dimensional singularities, the cosmic strings, by regularizing
them with cylinders. Of course, this calculation will be much more
involved than the one we did in 2+1D, as we now have to worry not
only about the boundary of the disk but about the various boundaries
of the cylinder. In particular, we must also regularize the vertices
by spheres such that the top and bottom of each cylinder start on
the surface of a sphere; this is further necessary in order to understand
what happens at the points where several strings meet.

In attempts to perform this calculation, we encountered many mathematical
and conceptual difficulties, which proved to be impossible to overcome
within the scope of this paper. Therefore, we leave it to future work.
\end{quote}
\textbf{2. Proper treatment of edge modes}
\begin{quote}
In the 2+1D case we discovered additional degrees of freedom called
edge modes, which result from the discretization itself and possess
their own unique symmetries. We analyzed them in detail, in particular
by studying their role in the symplectic potential in both the continuous
and discrete cases. However, in the 3+1D case we again skipped this
and instead assumed our results from 2+1D still hold. In future work,
we plan to perform a rigorous study of the edge modes in 3+1D, including
their role in the symplectic potential and the new symmetries they
generate.
\end{quote}
\textbf{3. Introducing a cosmological constant}
\begin{quote}
In this paper, we greatly simplified the calculation in 3+1 dimensions
by imposing that the geometry inside the cells is flat, mimicking
the 2+1-dimensional case. A more complicated case, but still probably
doable within our framework, is incorporating a cosmological constant,
which will then impose that the cells are homogeneously curved rather
than flat. In this case, it would be instructive to perform the calculation
in the 2+1D toy model first, and then generalize it to 3+1D.
\end{quote}
\textbf{4. Including point particles}
\begin{quote}
Cosmic strings in 3+1D have a very similar mathematical structure
to point particles in 2+1D \cite{PhDThesis}. For this reason, we
used string-like defects as our sources of curvature and torsion in
3+1D, which then allowed us to generalize our results from 2+1D in
a straightforward way. An important, but extremely complicated, modification
would be to allow point particles in 3+1D as well.

More precisely, in 2+1D, we added sources for the curvature and torsion
constraints, which are 2-forms. This is equivalent to adding matter
sources on the right-hand side of the Einstein equation. Since these
distributional sources are 2-form delta functions on a 2-dimensional
spatial slice, they pick out 0-dimensional points, which we interpreted
as a particle-like defects.

In 3+1D, we again added sources for the curvature and torsion, which
in this case are \textbf{not }constraints, but rather imposed by hand
to vanish. Since these distributional sources are 2-form delta functions
on a 3-dimensional spatial slice, they pick out 1-dimensional strings.
What we should actually do is add sources for the three constraints
-- Gauss, vector, and scalar -- which are 3-forms. The 3-form delta
functions will then pick out 0-dimensional points.

Unfortunately, doing this would introduce several difficulties, both
mathematical and conceptual. Perhaps the most serious problem would
be that in 3 dimensions, once cannot place a vertex inside a loop.
Indeed, in 2 dimensions, a loop encircling a vertex cannot be shrunk
to a point, as it would have to pass through the vertex. Similarly,
in 3 dimensions, a loop encircling an edge cannot be shrunk to a point
without passing through the edge. Therefore, in these cases it makes
sense to say that the vertex or edge is inside the loop.

However, in 3 dimensions there is no well-defined way in which a vertex
can be said to be inside a loop; any loop can always be shrunk to
a point without passing through any particular vertex. Hence, it is
unclear how holonomies on the loops of the spin network would be able
to detect the curvature induced by a point particle at a vertex. Solving
this problem might require generalizing the concept of spin networks
to allow for higher-dimensional versions of holonomies.
\end{quote}
\textbf{5. Taking Lorentz boosts into account}
\begin{quote}
In 2+1D, we split spacetime into 2-dimensional slices of equal time,
but we left the internal space 2+1-dimensional. The internal symmetry
group was then the full Lorentz group. However, in 3+1D, we not only
split spacetime into 3-dimensional slices of equal time, we did the
same to the internal space as well, and imposed the time gauge $e_{a}^{0}=0$.
The internal symmetry group thus reduced from the Lorentz group to
the rotation group.

Although this 3+1 split of the internal space is standard in 3+1D
canonical loop gravity, one may still wonder what happened to the
boosts, and whether we might be missing something important by assuming
that the variables on each cell are related to those on other cells
only by rotations, and not by a full Lorentz transformation. This
analysis might prove crucial for capturing the full theory of gravity
in 3+1D in our formalism, and in particular, for considering forms
of matter other than cosmic strings.
\end{quote}
\textbf{6. Motivating a relation to teleparallel gravity}
\begin{quote}
In both 2+1D and 3+1D, we found that the discrete phase space carries
two different polarizations. In 2+1D, we motivated an interpretation
where one polarization corresponds to usual general relativity and
the other to teleparallel gravity, an equivalent theory where gravity
is encoded in torsion instead of curvature degrees of freedom. In
the future we plan to motivate a similar relation between the two
polarizations in 3+1D.
\end{quote}
\textbf{7. Analyzing the discrete constraints}
\begin{quote}
In 2+1D, we provided a detailed analysis of the discrete Gauss and
curvature constraints, and the symmetries that they generate. We would
like to provide a similar analysis of the discrete Gauss, vector,
and scalar constraints in the 3+1D case. This will allow us to better
understand the discrete structure we have found, and in particular,
its relation to edge modes symmetries.
\end{quote}
\textbf{8. Quantizing the model}
\begin{quote}
In loop quantum gravity, spin networks arise during the quantization
process as the quantum states of discrete geometry, which are eigenstates
of the area and volume operators. One of the major motivations for
our formalism was to separate discretization from quantization in
this theory, and indeed, after discretization, the spin networks which
we found are not yet quantum states in a Hilbert space, but rather
classical entities which live in a classical phase space.

As the next step, the spin networks may be quantized, such that each
link is equipped with an irreducible group representation rather than
a group element. In the usual case of $\SUT$, this means equipping
the links with spin representations $j=0,1/2,1,3/2,\ldots$ such that
the eigenvalues of area obtain a contribution of $\gamma\ell_{P}^{2}\sqrt{j\left(j+1\right)}$
from each link, where $\ell_{P}$ is the Planck length and $\gamma$
is the Barbero-Immirzi parameter.

In this work, we found a precise way in which spin networks are dual
to piecewise-flat geometries. It would be interesting to investigate
the role of these dual geometries in the quantum theory, and use them
to explore the meaning of quantum properties such as superposition
and entanglement in the context of geometry.
\end{quote}

\subsection{Acknowledgments}

The author would like to thank Laurent Freidel and Florian Girelli
for their invaluable mentorship, and the anonymous referee for suggesting
interesting avenues for future investigation. This research was supported
in part by Perimeter Institute for Theoretical Physics. Research at
Perimeter Institute is supported by the Government of Canada through
the Department of Innovation, Science and Economic Development Canada
and by the Province of Ontario through the Ministry of Research, Innovation
and Science.

\appendix

\section{\label{sec:Derivation-of-Ashtekar}Derivation of the Ashtekar Variables}

In this appendix, we will derive the Ashtekar variables. We will start
by describing the first-order formulation of 3+1D gravity, introducing
the spin connection and frame field in Section \ref{subsec:The-Spin-Connection},
the Holst action in Section \ref{subsec:The-Holst-Action}, and the
Hamiltonian formulation in Section \ref{subsec:The-Hamiltonian-Formulation}.

In Section \ref{subsec:The-Ashtekar-Variables} we will define the
Ashtekar variables themselves, along with useful identities. We will
then proceed, in Section \ref{subsec:The-Action-in}, to rewrite the
Hamiltonian action of first-order gravity using these variables, and
define the Gauss, vector, and scalar constraints. Finally, we will
derive the symplectic potential in in Section \ref{subsec:The-Symplectic-Potential-Ashtekar}.

\subsection{\label{subsec:The-Spin-Connection}The Spin Connection and Frame
Field}

Let $M=\Sigma\xx\BBR$ be a 3+1-dimensional spacetime manifold, where
$\Sigma$ is a 3-dimensional spatial slice and $\BBR$ represents
time. Please see Section \ref{subsec:Spacetime-and-Spatial} for details
and conventions.

We define a spacetime $\mathfrak{so}\left(3,1\right)$ \emph{spin
connection} 1-form $\omega_{\mu}^{IJ}$ and a \emph{frame field }1-form
$e_{\mu}^{I}$. Here we will use \emph{partially index-free notation},
where only the internal-space indices of the forms are written explicitly:
\[
e^{I}\equiv e_{\mu}^{I}\thinspace\d x^{\mu}\sp\omega^{IJ}\equiv\omega_{\mu}^{IJ}\thinspace\d x^{\mu}.
\]
The frame field is related to the familiar metric by:
\[
g=\eta_{IJ}e^{I}\otimes e^{J}\soosp g_{\mu\nu}=\eta_{IJ}e_{\mu}^{I}e_{\nu}^{J},
\]
where $\eta_{IJ}$ is the Minkowski metric acting on the internal
space indices. Thus, the internal space is flat, and the curvature
is entirely encoded in the fields $e^{I}$; we will see below that
$\omega^{IJ}$ is completely determined by $e^{I}$. We also have
an \emph{inverse frame field}\footnote{Usually the vector $e_{I}^{\mu}$ is called the frame field and the
1-form $e_{\mu}^{I}$ is called the coframe field, but we will ignore
that subtlety here.} $e_{I}^{\mu}$, a vector, which satisfies:
\[
e_{I}^{\mu}e_{\nu}^{I}=\delta_{\nu}^{\mu}\sp e_{I}^{\mu}e_{\mu}^{J}=\delta_{I}^{J}\sp g_{\mu\nu}e_{I}^{\mu}e_{J}^{\nu}=\eta_{IJ}.
\]
We can view $e_{I}^{\mu}$ as a set of four 4-vectors, $e_{1}$, $e_{2}$,
$e_{3}$, and $e_{4}$, which form an \emph{orthonormal basis }(in
Lorentzian signature) with respect to the usual inner product:
\[
\langle x,y\rangle\equiv g_{\mu\nu}x^{\mu}y^{\mu}\soosp\langle e_{I},e_{J}\rangle=\eta_{IJ}.
\]
The familiar \emph{Levi-Civita connection} $\Gamma_{\mu\nu}^{\lambda}$
is related to the spin connection and frame field by
\[
\Gamma_{\mu\nu}^{\lambda}=\omega_{\mu J}^{I}e_{I}^{\lambda}e_{\nu}^{J}+e_{I}^{\lambda}\partial_{\mu}e_{\nu}^{I},
\]
such that there is a \emph{covariant derivative} $\nabla_{\mu}$,
which acts on both spacetime and internal indices, and is \emph{compatible}
with (i.e. annihilates) the frame field:
\begin{equation}
\nabla_{\mu}e_{\nu}^{I}\equiv\partial_{\mu}e_{\nu}^{I}-\Gamma_{\mu\nu}^{\lambda}e_{\lambda}^{I}+\omega_{\mu J}^{I}e_{\nu}^{J}=0.\label{eq:e-comp}
\end{equation}
Now, if we act with the covariant derivative on the internal-space
Minkowski metric $\eta_{IJ}$, we find:
\[
\nabla_{\mu}\eta^{IJ}=\partial_{\mu}\eta^{IJ}+\omega_{\mu K}^{I}\eta^{KJ}+\omega_{\mu K}^{J}\eta^{IK}.
\]
Of course, $\eta^{IJ}$ is constant in spacetime, so $\partial_{\mu}\eta^{IJ}=0$.
If we furthermore demand that the spin connection is metric-compatible
with respect to the internal-space metric, that is $\nabla_{\mu}\eta^{IJ}=0$,
then we get
\[
0=\omega_{\mu K}^{I}\eta^{KJ}+\omega_{\mu K}^{J}\eta^{IK}=\omega_{\mu}^{IJ}+\omega_{\mu}^{JI}=2\omega_{\mu}^{(IJ)}.
\]
We thus conclude that the spin connection must be anti-symmetric in
its internal indices:
\[
\omega_{\mu}^{(IJ)}=0\soosp\omega_{\mu}^{IJ}=\omega_{\mu}^{[IJ]}.
\]
Let us also define the \emph{covariant differential }$\d_{\omega}$
as follows:
\[
\d_{\omega}\phi\equiv\d\phi\sp\d_{\omega}X^{I}\equiv\d X^{I}+\udi{\omega}IJ\wedge X^{J},
\]
where $\phi$ is a scalar in the internal space and $X^{I}$ is a
vector in the internal space. With this we may define the \emph{torsion
2-form}:
\[
T^{I}\equiv\d_{\omega}e^{I}=\d e^{I}+\udi{\omega}IJ\wedge e^{J},
\]
and the \emph{curvature 2-form}:
\[
\udi FIJ\equiv\d_{\omega}\udi{\omega}IJ=\d\udi{\omega}IJ+\udi{\omega}IK\wedge\udi{\omega}KJ.
\]
Note that $\d_{\omega}$, unlike $\d$, is \textbf{not} nilpotent.
Instead, it satisfies the \emph{first Bianchi identity}
\begin{equation}
\d_{\omega}^{2}X^{I}=\udi FIK\wedge X^{K}.\label{eq:Bianchi}
\end{equation}

\subsection{\label{subsec:The-Holst-Action}The Holst Action}

\subsubsection{The Action and its Variation}

The action of 3+1D gravity (with zero cosmological constant) is given
by the \emph{Holst action}:\footnote{Usually there is also a factor of $1/\kappa$ in front of the action,
where $\kappa\equiv8\pi G$ and $G$ is \emph{Newton's constant}.
However, here we take $\kappa\equiv1$ for brevity.}
\begin{equation}
S\equiv\fr\int_{M}\left(\star+\frac{1}{\gamma}\right)e_{I}\wedge e_{J}\wedge F^{IJ},\label{eq:Holst-action}
\end{equation}
where $\star$ is the internal-space \emph{Hodge dual}\footnote{\label{fn:The-Hodge-dual}The \emph{Hodge dual} of a $p$-form $B$
on an $n$-dimensional manifold is the $\left(n-p\right)$-form $\star B$
defined such that, for any $p$-form $A$,
\[
A\wedge\star B=\langle A,B\rangle\epsilon,
\]
where $\epsilon$ is the volume $n$-form defined above, and $\langle A,B\rangle$
is the symmetric inner product of $p$-forms, defined as
\[
\langle A,B\rangle\equiv\frac{1}{p!}A^{a_{1}\cdots a_{p}}B_{a_{1}\cdots a_{p}}.
\]
$\star$ is called the \emph{Hodge star operator}. In terms of indices,
the Hodge dual is given by
\[
\left(\star B\right)_{b_{1}\cdots b_{n-p}}=\frac{1}{p!}B_{a_{1}\cdots a_{p}}\udi{\epsilon}{a_{1}\cdots a_{p}}{b_{1}\cdots b_{n-p}},
\]
and its action on basis $p$-forms is given by
\[
\star\left(\d x^{a_{1}}\wedge\cdots\wedge\d x^{a_{p}}\right)\equiv\frac{1}{\left(n-p\right)!}\udi{\epsilon}{a_{1}\cdots a_{p}}{b_{1}\cdots b_{n-p}}\d x^{b_{1}}\wedge\cdots\wedge\d x^{b_{n-p}}.
\]
Interestingly, we have that $\star1=\epsilon$. Also, if acting with
the Hodge star on a $p$-forms twice, we get
\[
\star^{2}=\sign\left(g\right)\left(-1\right)^{p\left(n-p\right)},
\]
where $\sign\left(g\right)$ is the signature of the metric: $+1$
for Euclidean or $-1$ for Lorentzian signature.} such that
\[
\star\left(e_{I}\wedge e_{J}\right)\equiv\hf\epsilon_{IJKL}e^{K}\wedge e^{L},
\]
$\gamma\in\BBR\backslash\left\{ 0\right\} $ is called the \emph{Barbero-Immirzi
parameter}, and
\[
\udi FIJ\equiv\d_{\omega}\udi{\omega}IJ=\d\udi{\omega}IJ+\udi{\omega}IK\wedge\udi{\omega}KJ.
\]
is the \emph{curvature 2-form }defined above. Let us derive the equation
of motion and symplectic potential from the Holst action. Taking the
variation, we get
\[
\delta S=\fr\int_{M}\left(2\left(\star+\frac{1}{\gamma}\right)\delta e_{I}\wedge e_{J}\wedge F^{IJ}+\left(\star+\frac{1}{\gamma}\right)e_{I}\wedge e_{J}\wedge\delta F^{IJ}\right).
\]
In the second term, we use the identity $\delta F^{IJ}=\d_{\omega}\left(\delta\omega^{IJ}\right)$
and integrate by parts to get
\[
\left(\star+\frac{1}{\gamma}\right)e_{I}\wedge e_{J}\wedge\delta F^{IJ}=\d_{\omega}\left(\left(\star+\frac{1}{\gamma}\right)e_{I}\wedge e_{J}\wedge\delta\omega^{IJ}\right)-2\left(\star+\frac{1}{\gamma}\right)\d_{\omega}e_{I}\wedge e_{J}\wedge\delta\omega^{IJ}.
\]
Thus the variation becomes
\[
\delta S=\hf\int_{M}\left(\left(\star+\frac{1}{\gamma}\right)\delta e_{I}\wedge e_{J}\wedge F^{IJ}-\left(\star+\frac{1}{\gamma}\right)\d_{\omega}e_{I}\wedge e_{J}\wedge\delta\omega^{IJ}\right)+\Theta,
\]
where the \emph{symplectic potential} $\Theta$ is the boundary term:
\begin{equation}
\Theta\equiv\fr\int_{\Sigma}\left(\star+\frac{1}{\gamma}\right)e_{I}\wedge e_{J}\wedge\delta\omega^{IJ}.\label{eq:symplectic-Holst}
\end{equation}

\subsubsection{The $\delta\omega$ Variation and the Definition of the Spin Connection}

From the variation with respect to $\delta\omega$ we see that the
\emph{torsion 2-form} must vanish\footnote{Note that in the usual metric formulation of general relativity, the
Levi-Civita connection $\Gamma_{\alpha\beta}^{\mu}$ is also taken
to be torsionless; however, there is also a dual formulation called
\emph{teleparallel gravity}, where we instead use a connection (the
Weitzenböck connection) which is flat but has torsion.}:
\begin{equation}
T^{I}\equiv\d_{\omega}e^{I}=\d e^{I}+\omega_{\ J}^{I}\wedge e^{J}=0.\label{eq:torsion-zero}
\end{equation}
In fact, we can take this equation of motion as a \textbf{definition
}of $\omega$. In other words, the only independent variable in our
theory is going to be the frame field $e^{I}$, and the spin connection
$\omega^{IJ}$ is going to be completely determined by $e^{I}$. Once
$\omega$ is defined in this way, it automatically satisfies this
equation of motion (or equivalently, there is no variation with respect
to $\delta\omega$ in the first place since $\omega$ is not an independent
variable). The formulation where $e$ and $\omega$ are independent
is called \emph{first-order}, and when $\omega$ depends on $e$ it
is called \emph{second-order}.

Let us look at the anti-symmetric part of the compatibility condition
(\ref{eq:e-comp}):
\[
\nabla_{[\mu}e_{\nu]I}=\partial_{[\mu}e_{\nu]I}+\omega_{[\mu|IL|}e_{\nu]}^{L}=0.
\]
Note that the term $\Gamma_{\mu\nu}^{\lambda}e_{\lambda}^{I}$ vanishes
automatically from this equation since $\Gamma_{\left[\mu\nu\right]}^{\lambda}=0$
from requiring that the Levi-Civita connection is torsion-free. Also,
the anti-symmetrizer in $\omega_{[\mu|IL|}e_{\nu]}^{L}$ acts on the
spacetime indices only (i.e. $\mu$ and $\nu$ are not inside the
anti-symmetrizer). Contracting with $e_{J}^{\mu}e_{K}^{\nu}$, we
get
\[
e_{J}^{\mu}e_{K}^{\nu}\left(\partial_{[\mu}e_{\nu]I}+\omega_{[\mu|IL|}e_{\nu]}^{L}\right)=0.
\]
We now permute the indices $I,J,K$ in this equation:
\[
e_{I}^{\mu}e_{J}^{\nu}\left(\partial_{[\mu}e_{\nu]K}+\omega_{[\mu|KL|}e_{\nu]}^{L}\right)=0,
\]
\[
e_{K}^{\mu}e_{I}^{\nu}\left(\partial_{[\mu}e_{\nu]J}+\omega_{[\mu|JL|}e_{\nu]}^{L}\right)=0.
\]
Taking the sum of the last two equations minus the first one, we get:
\[
e_{I}^{\mu}e_{J}^{\nu}\partial_{[\mu}e_{\nu]K}+e_{K}^{\mu}e_{I}^{\nu}\partial_{[\mu}e_{\nu]J}-e_{J}^{\mu}e_{K}^{\nu}\partial_{[\mu}e_{\nu]I}+\omega_{\mu\left(IJ\right)}e_{K}^{\mu}-\omega_{\mu\left(KI\right)}e_{J}^{\mu}-\omega_{\mu\left[JK\right]}e_{I}^{\mu}=0.
\]
Since $\omega_{\mu\left(IJ\right)}=0$, the two symmetric terms cancel,
and we get
\[
\omega_{\mu JK}e_{I}^{\mu}=e_{I}^{\mu}e_{J}^{\nu}\partial_{[\mu}e_{\nu]K}+e_{K}^{\mu}e_{I}^{\nu}\partial_{[\mu}e_{\nu]J}-e_{J}^{\mu}e_{K}^{\nu}\partial_{[\mu}e_{\nu]I}.
\]
Finally, we multiply by $e_{\lambda}^{I}$ to get 
\[
\omega_{\lambda JK}=e_{J}^{\nu}\partial_{[\lambda}e_{\nu]K}+e_{K}^{\mu}\partial_{[\mu}e_{\lambda]J}-e_{\lambda}^{I}e_{J}^{\mu}e_{K}^{\nu}\partial_{[\mu}e_{\nu]I}.
\]
Rearranging and relabeling the indices, we obtain the slightly more
elegant form:
\[
\omega_{\mu}^{IJ}=2e^{\lambda[I}\partial_{[\mu}e_{\lambda]}^{J]}-e_{\mu K}e^{\lambda I}e^{\sigma J}\partial_{[\lambda}e_{\sigma]}^{K},
\]
where the first term contains an anti-symmetrizer in both the spacetime
and internal space indices. Thus, $\omega$ is completely determined
by $e$, just as $\Gamma$ is completely determined by $g$ in the
usual metric formulation.

\subsubsection{The $\delta e$ Variation and the Einstein Equation}

From the variation with respect to $\delta e$ we get
\[
e_{J}\wedge\left(\star+\frac{1}{\gamma}\right)F^{IJ}=0.
\]
Note that, from the Bianchi identity (\ref{eq:Bianchi}), we have
$e_{J}\wedge F^{IJ}=\d_{\omega}^{2}e_{J}=0$ by the torsion condition
(\ref{eq:torsion-zero}). In other words, the $\gamma$-dependent
term vanishes on-shell, i.e., when the torsion vanishes. We are therefore
left with 
\[
e_{J}\wedge\star F^{IJ}=0,
\]
which is the \emph{Einstein equation} $R_{\mu\nu}-\hf g_{\mu\nu}R=0$
in first-order form. Note that this equation is independent of $\gamma$;
therefore, the $\gamma$-dependent term in the action does not affect
the physics, at least not at the level of the classical equation of
motion.

Let us prove that this is indeed the Einstein equation. We have
\[
0=e_{J}\wedge\star F^{IJ}=\hf\epsilon_{KLM}^{I}e_{\rho}^{K}F_{\mu\nu}^{LM}\d x^{\rho}\wedge\d x^{\mu}\wedge\d x^{\nu}.
\]
Taking the spacetime Hodge dual of this 3-form (see Footnote \ref{fn:The-Hodge-dual}),
we get
\[
0=\star\left(e_{J}\wedge\star F^{IJ}\right)=\frac{1}{3!\cdot2}\epsilon_{\alpha}^{\rho\mu\nu}\epsilon_{KLM}^{I}e_{\rho}^{K}F_{\mu\nu}^{LM}\d x^{\alpha}.
\]
Of course, we can throw away the numerical factor of $1/3!\cdot2$,
and look at the components of the 1-form:
\[
\epsilon_{\alpha}^{\rho\mu\nu}\epsilon_{KLM}^{I}e_{\rho}^{K}F_{\mu\nu}^{LM}=0.
\]
The relation between the \emph{Riemann tensor}\footnote{The Riemann tensor satisfies the symmetry $R_{\mu\nu\alpha\beta}=R_{\alpha\beta\mu\nu}$,
so we can write it as $R_{\mu\nu}^{\alpha\beta}$ with the convention
that, if the indices are lowered, each pair could be either the first
or second pair of indices, as long as they are adjacent. In other
words, $g_{\alpha\gamma}g_{\beta\delta}R_{\mu\nu}^{\gamma\delta}=R_{\mu\nu\alpha\beta}$
or equivalently $g_{\alpha\gamma}g_{\beta\delta}R_{\mu\nu}^{\gamma\delta}=R_{\alpha\beta\mu\nu}$.} on spacetime and the curvature 2-form is:
\[
F_{\mu\nu}^{LM}=e_{\gamma}^{L}e_{\delta}^{M}R_{\mu\nu}^{\gamma\delta}.
\]
Plugging in, we get
\[
\epsilon_{\alpha}^{\rho\mu\nu}\epsilon_{KLM}^{I}e_{\rho}^{K}e_{\gamma}^{L}e_{\delta}^{M}R_{\mu\nu}^{\gamma\delta}=0.
\]
Multiplying by $e_{I}^{\beta}$, and using the relation
\[
\epsilon_{KLM}^{I}e_{I}^{\beta}e_{\rho}^{K}e_{\gamma}^{L}e_{\delta}^{M}=\epsilon_{\rho\gamma\delta}^{\beta},
\]
we get, after raising $\alpha$ and lowering $\beta$,
\[
\epsilon^{\rho\mu\nu\alpha}\epsilon_{\rho\gamma\delta\beta}R_{\mu\nu}^{\gamma\delta}=0.
\]
Finally, we use the identity
\[
\epsilon^{\rho\mu\nu\alpha}\epsilon_{\rho\gamma\delta\beta}=-2\left(\delta_{\gamma}^{[\mu}\delta_{\delta}^{\nu]}\delta_{\beta}^{\alpha}+\delta_{\gamma}^{[\alpha}\delta_{\delta}^{\mu]}\delta_{\beta}^{\nu}+\delta_{\gamma}^{[\nu}\delta_{\delta}^{\alpha]}\delta_{\beta}^{\mu}\right),
\]
where the minus sign comes from the Lorentzian signature of the metric,
to get:
\[
R_{\beta}^{\alpha}-\hf\delta_{\beta}^{\alpha}R=0,
\]
where we defined the \emph{Ricci tensor} and \emph{Ricci scalar}:
\[
R_{\beta}^{\alpha}\equiv R_{\mu\beta}^{\mu\alpha}\sp R\equiv R_{\mu}^{\mu}.
\]
Lowering $\alpha$, we see that we have indeed obtained the Einstein
equation,
\[
R_{\alpha\beta}-\hf g_{\alpha\beta}R=0,
\]
as desired.

\subsection{\label{subsec:The-Hamiltonian-Formulation}The Hamiltonian Formulation}

\subsubsection{\label{subsec:The-3+1-Split}The 3+1 Split and the Time Gauge}

To go to the Hamiltonian formulation, we split our spacetime manifold
$M$ into space $\Sigma$ and time $\BBR$. We remind the reader that,
as detailed in Section \ref{subsec:Spacetime-and-Spatial}, the spacetime
and spatial indices on both real space and the internal space are
related as follows:
\[
\overbrace{0,\underbrace{1,2,3}_{a}}^{\mu}\sp\overbrace{0,\underbrace{1,2,3}_{i}}^{I}.
\]
Let us decompose the 1-form $e^{I}\equiv e_{\mu}^{I}\d x^{\mu}$:
\[
e^{0}\equiv e_{\mu}^{0}\thinspace\d x^{\mu}=e_{0}^{0}\thinspace\d x^{0}+e_{a}^{0}\thinspace\d x^{a}\sp e^{i}\equiv e_{\mu}^{i}\thinspace\d x^{\mu}=e_{0}^{i}\thinspace\d x^{0}+e_{a}^{i}\thinspace\d x^{a}.
\]
Here we merely changed notation from 3+1D spacetime indices $I,\mu$
to 3D spatial indices $i,a$. However, now we are going to impose
a partial gauge fixing, the \emph{time gauge}, given by
\begin{equation}
e_{a}^{0}=0.\label{eq:time-gauge}
\end{equation}
We also define
\begin{equation}
e_{0}^{0}\equiv N\sp e_{0}^{i}\equiv N^{i},\label{eq:lapse-shift}
\end{equation}
where $N$ is called the \emph{lapse} and $N^{i}$ is called the \emph{shift},
as in the ADM formalism. In other words, we have:
\[
e^{0}=N\thinspace\d x^{0}\sp e^{i}=N^{i}\thinspace\d x^{0}+e_{a}^{i}\thinspace\d x^{a},
\]
or in matrix form,
\[
e_{\mu}^{I}=\left(\begin{array}{cccc}
N &  & N^{i} & \thinspace\\
\\
0 &  & e_{a}^{i}\\
\\
\end{array}\right).
\]
As we will soon see, $N$ and $N^{i}$ are non-dynamical \emph{Lagrange
multipliers}, so we are left with $e_{a}^{i}$ as the only dynamical
degrees of freedom of the frame field -- although they will be further
reduced by the internal gauge symmetry.

\subsubsection{The Hamiltonian}

In order to derive the Hamiltonian, we are going to have to sacrifice
the elegant index-free differential form language (for now) and write
everything in terms of indices. This will allow us to perform the
3+1 split in those indices. Writing the differential forms explicitly
in coordinate basis, that is, $e^{I}\equiv e_{\mu}^{I}\d x^{\mu}$
and so on, we get:
\begin{align*}
e^{I}\wedge e^{J}\wedge F^{KL} & =\left(e_{\mu}^{I}\d x^{\mu}\right)\wedge\left(e_{\nu}^{J}\d x^{\nu}\right)\wedge\left(F_{\rho\sigma}^{KL}\d x^{\rho}\wedge\d x^{\sigma}\right)\\
 & =e_{\mu}^{I}e_{\nu}^{J}F_{\rho\sigma}^{KL}\d x^{\mu}\wedge\d x^{\nu}\wedge\d x^{\rho}\wedge\d x^{\sigma}.
\end{align*}
Note that $\d x^{\mu}\wedge\d x^{\nu}\wedge\d x^{\rho}\wedge\d x^{\sigma}$
is a wedge produce of 1-forms, and is therefore completely anti-symmetric
in the indices $\mu\nu\rho\sigma$, just like the Levi-Civita symbol\footnote{\label{fn:Density}The tilde on the Levi-Civita symbol signifies that
it is not a tensor but a \emph{tensor density}. The symbol is defined
as
\[
\ept_{\mu\nu\rho\sigma}\equiv\begin{cases}
+1 & \textrm{if }\left(\mu\nu\rho\sigma\right)\textrm{ is an even permutation of }\left(0123\right),\\
-1 & \textrm{if }\left(\mu\nu\rho\sigma\right)\textrm{ is an odd permutation of }\left(0123\right),\\
0 & \textrm{if any two indices are the same}.
\end{cases}
\]
By definition this quantity has the same values in every coordinate
system, and thus it cannot be a tensor. Let us define a \emph{tensor
density }$\Tt$ as a quantity related to a proper tensor $T$ by
\[
\Tt=\left|g\right|^{-w/2}T,
\]
where $g$ is the determinant of the metric and $w$ is called the
\emph{density weight}. It can be shown that
\[
\ept_{\mu\nu\rho\sigma}\equiv g^{-1/2}\epsilon_{\mu\nu\rho\sigma},
\]
and therefore the Levi-Civita symbol is a tensor density of weight
$+1$.} $\ept^{\mu\nu\rho\sigma}$. Thus we can write:
\[
\d x^{\mu}\wedge\d x^{\nu}\wedge\d x^{\rho}\wedge\d x^{\sigma}=-\ept^{\mu\nu\rho\sigma}\d x^{0}\wedge\d x^{1}\wedge\d x^{2}\wedge\d x^{3},
\]
where the minus sign comes from the fact that $\sign\left(g\right)=-1$,
and we defined $\ept^{\mu\nu\rho\sigma}\equiv\sign\left(g\right)\ept_{\mu\nu\rho\sigma}$.
To see that this relation is satisfied, simply plug in values for
$\mu,\nu,\rho,\sigma$ and compare both sides. For example, for $\left(\mu\nu\rho\sigma\right)=\left(0123\right)$
we have:
\[
-\ept^{0123}=\ept_{0123}=+1,
\]
and both sides are satisfied. We thus have
\[
e^{I}\wedge e^{J}\wedge F^{KL}=-\ept^{\mu\nu\rho\sigma}e_{\mu}^{I}e_{\nu}^{J}F_{\rho\sigma}^{KL}\d t\wedge\d^{3}x,
\]
where $\d t\equiv\d x^{0}$ and $\d^{3}x\equiv\d x^{1}\wedge\d x^{2}\wedge\d x^{3}$.
Plugging this into the Holst action (\ref{eq:Holst-action}), we get
after some careful manipulations\footnote{Here, the Levi-Civita symbol $\epsilon^{IJKL}$ is actually a tensor,
not a tensor density, since we are in a flat space -- so we omit
the tilde.}\footnote{We chose to write down the internal space Minkowski metric $\eta_{IJ}$
explicitly so that internal space indices $I,J,\ldots$ on differential
forms can always be upstairs and spacetime indices $\mu,\nu,\ldots$
can always be downstairs. This will also remind us that terms with
$I,J=0$ in the summation should get a minus sign, since $\eta_{00}=-1$.}
\begin{align*}
S & =-\hf\int\d t\int\d^{3}x\thinspace\ept^{abc}\multibrl{\hf\epsilon_{IJKL}e_{0}^{I}e_{a}^{J}F_{bc}^{KL}+\hf\epsilon_{IJKL}e_{a}^{I}e_{b}^{J}F_{0c}^{KL}+}\\
 & \qquad\qquad\multibrr{+\frac{1}{\gamma}\eta_{IK}\eta_{JL}e_{0}^{I}e_{a}^{J}F_{bc}^{KL}+\frac{1}{\gamma}\eta_{IK}\eta_{JL}e_{a}^{I}e_{b}^{J}F_{0c}^{KL}},
\end{align*}
where we defined the 3-dimensional Levi-Civita symbol as $\ept^{abc}\equiv\ept^{0abc}$.
If we do the same in the internal indices, that is, define $\epsilon^{ijk}\equiv\epsilon^{0ijk}$,
we get\footnote{For the first two terms, we simply take $\left(IJKL\right)=\left(0ijk\right),\left(i0jk\right),\left(ij0k\right),\left(ijk0\right)$
in the sum, which we can do due to the Levi-Civita symbol $\epsilon_{IJKL}$.
For the next two terms, we split into the following four distinct
cases:
\[
\eta_{IJ}=\begin{cases}
-1 & I=J=0,\\
+1 & I=J\ne0,\\
0 & \textrm{otherwise},
\end{cases}\soosp\eta_{IK}\eta_{JL}=\begin{cases}
\eta_{00}\eta_{00}=+1,\\
\eta_{00}\eta_{ij}=-\delta_{ij},\\
\eta_{ij}\eta_{00}=-\delta_{ij},\\
\eta_{ik}\eta_{jl}=+\delta_{ik}\delta_{jl},
\end{cases}
\]
and use the fact that $F^{00}=0$ since it's anti-symmetric.}
\[
\hf\ept^{abc}\epsilon_{IJKL}e_{0}^{I}e_{a}^{J}F_{bc}^{KL}=\hf\ept^{abc}\epsilon_{ijk}\left(\left(e_{0}^{0}e_{a}^{i}-e_{0}^{i}e_{a}^{0}\right)F_{bc}^{jk}+2e_{0}^{i}e_{a}^{j}F_{bc}^{0k}\right),
\]
\[
\hf\ept^{abc}\epsilon_{IJKL}e_{a}^{I}e_{b}^{J}F_{0c}^{KL}=\ept^{abc}\epsilon_{ijk}\left(e_{a}^{0}e_{b}^{i}F_{0c}^{jk}+e_{a}^{i}e_{b}^{j}F_{0c}^{0k}\right),
\]
\[
\frac{1}{\gamma}\ept^{abc}\eta_{IK}\eta_{JL}e_{0}^{I}e_{a}^{J}F_{bc}^{KL}=\frac{1}{\gamma}\ept^{abc}\left(\delta_{ij}\left(e_{0}^{i}e_{a}^{0}-e_{0}^{0}e_{a}^{i}\right)F_{bc}^{0j}+\delta_{ik}\delta_{jl}e_{0}^{i}e_{a}^{j}F_{bc}^{kl}\right),
\]
\[
\frac{1}{\gamma}\ept^{abc}\eta_{IK}\eta_{JL}e_{a}^{I}e_{b}^{J}F_{0c}^{KL}=\frac{1}{\gamma}\ept^{abc}\left(2\delta_{ij}e_{a}^{i}e_{b}^{0}F_{0c}^{0j}+\delta_{ik}\delta_{jl}e_{a}^{i}e_{b}^{j}F_{0c}^{kl}\right).
\]
Now, as indicated above, we impose the \emph{time gauge }(\ref{eq:time-gauge})
and define the \emph{lapse }and \emph{shift }(\ref{eq:lapse-shift}):
\[
e_{a}^{0}=0\sp e_{0}^{0}\equiv N\sp e_{0}^{i}\equiv N^{i}\equiv N^{d}e_{d}^{i},
\]
where we have converted the shift into a spatial vector $N^{d}$ instead
of an internal space vector. Plugging in, we get
\[
\hf\ept^{abc}\epsilon_{IJKL}e_{0}^{I}e_{a}^{J}F_{bc}^{KL}=\hf\ept^{abc}\epsilon_{ijk}\left(Ne_{a}^{i}F_{bc}^{jk}+2N^{d}e_{d}^{i}e_{a}^{j}F_{bc}^{0k}\right),
\]
\[
\hf\ept^{abc}\epsilon_{IJKL}e_{a}^{I}e_{b}^{J}F_{0c}^{KL}=\ept^{abc}\epsilon_{ijk}e_{a}^{i}e_{b}^{j}F_{0c}^{0k},
\]
\[
\frac{1}{\gamma}\ept^{abc}\eta_{IK}\eta_{JL}e_{0}^{I}e_{a}^{J}F_{bc}^{KL}=\frac{1}{\gamma}\ept^{abc}\left(\delta_{ik}\delta_{jl}N^{d}e_{d}^{i}e_{a}^{j}F_{bc}^{kl}-\delta_{ij}Ne_{a}^{i}F_{bc}^{0j}\right),
\]
\[
\frac{1}{\gamma}\ept^{abc}\eta_{IK}\eta_{JL}e_{a}^{I}e_{b}^{J}F_{0c}^{KL}=\frac{1}{\gamma}\ept^{abc}\delta_{ik}\delta_{jl}e_{a}^{i}e_{b}^{j}F_{0c}^{kl}.
\]
The action thus becomes, after taking out a factor of $1/\gamma$
and isolating terms proportional to $N$ and $N^{d}$:
\begin{align*}
S & =-\frac{1}{2\gamma}\int\d t\int\d^{3}x\thinspace\ept^{abc}\multisql{\left(\delta_{ik}\delta_{jl}e_{a}^{i}e_{b}^{j}F_{0c}^{kl}+\gamma\epsilon_{ijk}e_{a}^{i}e_{b}^{j}F_{0c}^{0k}\right)+}\\
 & \qquad+N^{d}\left(\delta_{ik}\delta_{jl}e_{d}^{i}e_{a}^{j}F_{bc}^{kl}+\gamma\epsilon_{ijk}e_{d}^{i}e_{a}^{j}F_{bc}^{0k}\right)+\\
 & \qquad\multisqr{-N\left(\delta_{ik}e_{a}^{i}F_{bc}^{0k}-\hf\gamma\epsilon_{ikl}e_{a}^{i}F_{bc}^{kl}\right)}.
\end{align*}

\subsection{\label{subsec:The-Ashtekar-Variables}The Ashtekar Variables}

\subsubsection{The Densitized Triad and Related Identities}

Let us define the \emph{densitized triad}, which is a rank $\left(1,0\right)$
tensor of density weight\footnote{See Footnote \ref{fn:Density} for the definition of a tensor density.
The densitized triad has weight $-1$ since $\det\left(e\right)=\sqrt{\det\left(g\right)}$
has weight $-1$.} $-1$:
\[
\Et_{i}^{a}\equiv\det\left(e\right)e_{i}^{a}.
\]
The inverse triad $e_{i}^{a}$ is related to the inverse metric $g^{ab}$
via
\[
g^{ab}=e_{i}^{a}e_{j}^{b}\delta^{ij}.
\]
Multiplying by $\det\left(g\right)=\det\left(e\right)^{2}$ we get
\[
\det\left(g\right)g^{ab}=\Et_{i}^{a}\Et_{j}^{b}\delta^{ij}.
\]
We now prove some identities. First, consider the determinant identity
for a 3-dimensional matrix,
\[
\epsilon_{ijk}e_{a}^{i}e_{b}^{j}e_{c}^{k}=\det\left(e\right)\ept_{abc}.
\]
Multiplying by $e_{l}^{a}$ and using $e_{a}^{i}e_{l}^{a}=\delta_{l}^{i}$,
we get
\[
\epsilon_{ljk}e_{b}^{j}e_{c}^{k}=\epsilon_{ijk}e_{a}^{i}e_{b}^{j}e_{c}^{k}e_{l}^{a}=\det\left(e\right)e_{l}^{a}\ept_{abc}=\Et_{l}^{a}\ept_{abc}.
\]
Next, multiplying by $\ept^{bcd}$ and using the identity
\[
\ept_{abc}\ept^{bcd}=2\delta_{a}^{d},
\]
we get
\[
\ept^{bcd}\epsilon_{ljk}e_{b}^{j}e_{c}^{k}=\Et_{l}^{a}\ept_{abc}\ept^{bcd}=2\Et_{l}^{d}.
\]
Renaming indices, we obtain the identity
\[
\Et_{i}^{a}=\hf\ept^{abc}\epsilon_{ijk}e_{b}^{j}e_{c}^{k}.
\]
Similarly, one may prove the identity\textbf{
\[
e_{a}^{i}=\frac{\epsilon^{ijk}\ept_{abc}\Et_{j}^{b}\Et_{k}^{c}}{2\det\left(e\right)}.
\]
}Since
\[
\det\left(\Et\right)=\det\left(\det\left(e\right)e_{i}^{a}\right)=\left(\det\left(e\right)\right)^{2},
\]
we obtain an expression for the triad 1-form solely in terms of the
densitized triad:
\[
e_{a}^{i}=\frac{\epsilon^{ijk}\ept_{abc}\Et_{j}^{b}\Et_{k}^{c}}{2\sqrt{\det\left(\Et\right)}}.
\]
Contracting with $\ept^{ade}$, we get 
\[
\ept^{ade}e_{a}^{i}=\frac{\epsilon^{imn}\left(\delta_{b}^{d}\delta_{c}^{e}-\delta_{c}^{d}\delta_{b}^{e}\right)\Et_{m}^{b}\Et_{n}^{c}}{2\sqrt{\det\left(\Et\right)}}=\frac{\epsilon^{imn}\Et_{m}^{d}\Et_{n}^{e}}{\sqrt{\det\left(\Et\right)}},
\]
from which we find that
\[
\ept^{abc}e_{a}^{i}=\frac{\epsilon^{ijk}\Et_{j}^{b}\Et_{k}^{c}}{\sqrt{\det\left(E\right)}}.
\]
In conclusion, we have the following definitions and identities:
\[
\Et_{i}^{a}\equiv\det\left(e\right)e_{i}^{a}=\hf\ept^{abc}\epsilon_{ijk}e_{b}^{j}e_{c}^{k},
\]
\[
\epsilon^{ijm}\Et_{m}^{c}=\ept^{abc}e_{a}^{i}e_{b}^{j}\sp\ept^{abc}e_{a}^{j}=e_{p}^{b}e_{q}^{c}\epsilon^{jpq}\det\left(e\right),
\]
\begin{equation}
e_{a}^{i}=\frac{\epsilon^{ijk}\ept_{abc}\Et_{j}^{b}\Et_{k}^{c}}{2\sqrt{\det\left(E\right)}}\sp\ept^{abc}e_{a}^{i}=\frac{\epsilon^{ijk}\Et_{j}^{b}\Et_{k}^{c}}{\sqrt{\det\left(E\right)}}.\label{eq:epsilon-e-id}
\end{equation}

\subsubsection{\label{subsec:The-Ashtekar-Barbero-Connection}The Ashtekar-Barbero
Connection}

Since we have performed a 3+1 split of the spin connection $\omega_{\mu}^{IJ}$,
we can use its individual components to define a new connection on
the spatial slice.

First, we use the fact that the spatial part of the spin connection,
$\omega_{a}^{ij}$, is anti-symmetric in the internal indices, and
thus it behaves as a 2-form on the internal space. This means that
we can take its Hodge dual\footnote{Please see Footnote \ref{fn:The-Hodge-dual} for the definition of
the Hodge dual.}, and obtain a \emph{dual spin connection}\footnote{The minus sign here is meant to make the Gauss law, which we will
derive shortly, have the same relative sign as the Gauss law from
2+1D gravity and Yang-Mills theory; note that, in some other sources,
$\Gamma_{a}^{i}$ is defined without this minus sign.} $\Gamma_{a}^{i}$:
\[
\Gamma_{a}^{i}\equiv-\hf\epsilon_{jk}^{i}\omega_{a}^{jk}\soossp\omega_{a}^{jk}=-\epsilon_{i}^{jk}\Gamma_{a}^{i}.
\]

Importantly, instead of two internal indices, $\Gamma_{a}^{i}$ only
has one\footnote{We can do this only in 3 dimensions, since the Hodge dual takes a
$k$-form into a $\left(3-k\right)$-form. We are lucky that we do,
in fact, live in a 3+1-dimensional spacetime, otherwise this simplification
would not have been possible!}. 

Next, we define the \emph{extrinsic curvature}\footnote{Again, this definition differs by a minus sign from some other sources.}
$K_{a}^{i}$:
\[
K_{a}^{i}\equiv\omega_{a}^{i0}=-\omega_{a}^{0i}.
\]
Note that we will extend both definitions to $a=0$, for brevity only;
$\Gamma_{0}$ and $K_{0}$ will not be dynamical variables, as we
shall see.

Using the dual spin connection and the extrinsic curvature, we may
now define the \emph{Ashtekar-Barbero connection} $A_{a}^{i}$:
\[
A_{a}^{i}\equiv\Gamma_{a}^{i}+\gamma K_{a}^{i}.
\]
The original spin connection $\omega_{\mu}^{IJ}$ was 1-form on spacetime
which had two internal indices, and was valued in the Lie algebra
of the Lorentz group, also known as $\mathfrak{so}\left(3,1\right)$.
In short, it was an $\mathfrak{so}\left(3,1\right)$-valued 1-form
on spacetime\footnote{The generators of the Lorentz algebra are $L^{IJ}$ with $I,J\in\left\{ 0,1,2,3\right\} $,
and they are anti-symmetric in $I$ and $J$. They are related to
rotations $J^{I}$ and boosts $K^{I}$ by $J^{I}=\hf\udi{\epsilon}I{JK}L^{JK}$
and $K^{I}=L^{0I}$.}. The three quantities we have defined, $\Gamma_{a}^{i}$, $K_{a}^{i}$,
and $A_{a}^{i}$, resulted from reducing both spacetime and the internal
space from 3+1 dimensions to 3 dimensions. Thus, they are 1-forms
on 3-dimensional space, not spacetime, and the internal space is now
invariant under $\mathfrak{so}\left(3\right)$ only.

Since the Lie algebras $\mathfrak{so}\left(3\right)$ and $\sut$
are isomorphic, and since in Yang-Mills theory we use $\sut$, we
might as well use $\sut$ as the symmetry of our internal space instead
of $\mathfrak{so}\left(3\right)$. Thus, the quantities $\Gamma_{a}^{i}$,
$K_{a}^{i}$ and $A_{a}^{i}$ are all $\sut$-valued 1-forms on 3-dimensional
space. We can also, however, work more generally with some unspecified
(compact) Lie algebra $\mfg$. We will use index-free notation, as
defined in (\ref{eq:index-free-notation}). In particular, we will
write for the connection, frame field, dual connection and extrinsic
curvature:
\[
\A\equiv A_{a}^{i}\ta_{i}\thinspace\d x^{a}\sp\ee\equiv e_{a}^{i}\ta_{i}\thinspace\d x^{a}\sp\Ga\equiv\Gamma_{a}^{i}\ta_{i}\thinspace\d x^{a}\sp\K\equiv K_{a}^{i}\ta_{i}\thinspace\d x^{a},
\]
where $\ta_{i}$ are the generators of $\mfg$.

\subsubsection{The Dual Spin Connection in Terms of the Frame Field}

Recall that in the Lagrangian formulation we had the torsion equation
of motion
\[
T^{I}\equiv\d_{\omega}e^{I}=\d e^{I}+\omega_{\ J}^{I}\wedge e^{J}=0.
\]
Explicitly, the components of the 2-form $T^{I}$ are:
\[
\hf T_{\mu\nu}^{I}=\partial_{[\mu}e_{\nu]}^{I}+\eta_{JK}\omega_{[\mu}^{IJ}e_{\nu]}^{K}.
\]
Taking the spatial components after a 3+1 split in both spacetime
and the internal space, we get
\[
\hf T_{ab}^{i}=\partial_{[a}e_{b]}^{i}-\omega_{[a}^{i0}e_{b]}^{0}+\delta_{jk}\omega_{[a}^{ij}e_{b]}^{k}.
\]
However, after imposing the time gauge $e_{b}^{0}=0$ the middle term
vanishes:
\[
\hf T_{ab}^{i}=\partial_{[a}e_{b]}^{i}+\delta_{jk}\omega_{[a}^{ij}e_{b]}^{k}.
\]
Let us now plug in
\[
\omega_{a}^{ij}=-\epsilon_{l}^{ij}\Gamma_{a}^{l},
\]
to get
\[
\hf T_{ab}^{i}=\partial_{[a}e_{b]}^{i}+\epsilon_{kl}^{i}\Gamma_{[a}^{k}e_{b]}^{l}\equiv D_{[a}e_{b]}^{i},
\]
where we have defined the \emph{covariant derivative} $D_{a}$, which
acts on $\mfg$-valued 1-forms $e_{b}^{i}$ as
\[
D_{a}e_{b}^{i}\equiv\partial_{a}e_{b}^{i}+\epsilon_{kl}^{i}\Gamma_{a}^{k}e_{b}^{l}.
\]
The equation $D_{a}e_{b}^{i}=0$ can be seen as the definition of
$\Gamma_{a}^{i}$ in terms of $e_{a}^{i}$, just as $\d_{\omega}e^{I}=0$
defines $\omega^{IJ}$ in terms of $e^{I}$.

In index-free notation, the spatial torsion equation of motion is
simply
\[
\T=\d_{\Ga}\ee=\d\ee+\left[\Ga,\ee\right]=0,
\]
where 
\[
\T\equiv\hf T_{ab}^{i}\ta_{i}\d x^{a}\wedge\d x^{b}
\]
is a $\mfg$-valued 2-form.

\subsubsection{The ``Electric Field''}

We now define the \emph{electric field} 2-form $\E$ as (half) the
commutator of two frame fields:
\[
\E\equiv\hf\left[\ee,\ee\right].
\]
This is analogous to the electric field in electromagnetism and Yang-Mills
theory. In terms of components, we have
\[
\E\equiv\hf E_{ab}^{i}\ta_{i}\d x^{a}\wedge\d x^{b}\sp E_{ab}^{i}=\hf\left[\ee,\ee\right]_{ab}^{i}=\epsilon_{jk}^{i}e_{a}^{j}e_{b}^{k}.
\]
Alternatively, starting from the definition $\Et_{i}^{c}\equiv\hf\ept^{abc}\epsilon_{ijk}e_{a}^{j}e_{b}^{k}$
of the densitized triad, we multiply both sides by $\ept_{cde}$ and
get:
\[
\ept_{cde}\Et_{i}^{c}=\hf\left(\ept_{cde}\ept^{abc}\right)\epsilon_{ijk}e_{a}^{j}e_{b}^{k}=\hf\left(\delta_{d}^{a}\delta_{e}^{b}-\delta_{e}^{a}\delta_{d}^{b}\right)\epsilon_{ijk}e_{a}^{j}e_{b}^{k}=\epsilon_{ijk}e_{d}^{j}e_{e}^{k},
\]
which gives us the electric field in terms of the densitized triad:
\[
E_{ab}^{i}=\ept_{abc}\delta^{ij}\Et_{j}^{c}.
\]
Note that in the definition we ``undensitize'' the densitized triad,
which is a tensor density of weight $-1$, by contracting it with
the Levi-Civita tensor density, which has weight $1$. The 2-form
$\E$ is thus a proper tensor.

Now, since $\E=\left[\ee,\ee\right]/2$, we have
\begin{equation}
\d_{\Ga}\E=\hf\d_{\Ga}\left[\ee,\ee\right]=\left[\d_{\Ga}\ee,\ee\right]=\left[\T,\ee\right]=0.\label{eq:dGammaE}
\end{equation}
Therefore, just like the frame field $\ee$, the electric field $\E$
is also torsionless with respect to the connection $\Ga$.

\subsection{\label{subsec:The-Action-in}The Action in Terms of the Ashtekar
Variables}

\subsubsection{The Curvature}

The spacetime components $F_{\mu\nu}^{IJ}$ of the curvature 2-form,
related to the partially-index-free quantity $F^{IJ}$ by
\[
F^{IJ}\equiv\hf F_{\mu\nu}^{IJ}\d x^{\mu}\wedge\d x^{\nu},
\]
are
\[
\hf F_{\mu\nu}^{IJ}=\partial_{[\mu}\omega_{\nu]}^{IJ}+\eta_{KL}\omega_{[\mu}^{IK}\omega_{\nu]}^{LJ}.
\]
Let us write the 3+1 decomposition in spacetime:
\[
\hf F_{0c}^{IJ}=\partial_{[0}\omega_{c]}^{IJ}+\eta_{KL}\omega_{[0}^{IK}\omega_{c]}^{LJ},
\]
\[
\hf F_{bc}^{IJ}=\partial_{[b}\omega_{c]}^{IJ}+\eta_{KL}\omega_{[b}^{IK}\omega_{c]}^{LJ}.
\]
We can further decompose it in the internal space, remembering that
$\eta_{00}=-1$, $\eta_{ij}=\delta_{ij}$ and $\omega^{00}=0$:
\[
\hf F_{0c}^{0k}=\partial_{[0}\omega_{c]}^{0k}+\delta_{mn}\omega_{[0}^{0m}\omega_{c]}^{nk},
\]
\[
\hf F_{0c}^{kl}=\partial_{[0}\omega_{c]}^{kl}-\omega_{[0}^{k0}\omega_{c]}^{0l}+\delta_{mn}\omega_{[0}^{km}\omega_{c]}^{nl},
\]
\[
\hf F_{bc}^{0k}=\partial_{[b}\omega_{c]}^{0k}+\delta_{mn}\omega_{[b}^{0m}\omega_{c]}^{nk},
\]
\[
\hf F_{bc}^{kl}=\partial_{[b}\omega_{c]}^{kl}-\omega_{[b}^{k0}\omega_{c]}^{0l}+\delta_{mn}\omega_{[b}^{km}\omega_{c]}^{nl}.
\]
Plugging the definitions of $\Gamma_{a}^{i}$ and $K_{a}^{i}$ into
these expressions, we obtain:
\[
-\hf F_{0c}^{0k}=\partial_{[0}K_{c]}^{k}+\epsilon_{pq}^{k}K_{[0}^{p}\Gamma_{c]}^{q},
\]
\[
-\hf F_{0c}^{kl}=\epsilon_{p}^{kl}\partial_{[0}\Gamma_{c]}^{p}-K_{[0}^{k}K_{c]}^{l}+\Gamma_{[0}^{k}\Gamma_{c]}^{l},
\]
\[
-\hf F_{bc}^{0k}=\partial_{[b}K_{c]}^{k}+\epsilon_{pq}^{k}K_{[b}^{p}\Gamma_{c]}^{q},
\]
\[
-\hf F_{bc}^{kl}=\epsilon_{p}^{kl}\partial_{[b}\Gamma_{c]}^{p}-K_{[b}^{k}K_{c]}^{l}+\Gamma_{[b}^{k}\Gamma_{c]}^{l}.
\]
Note that, in arriving at these expressions, we obtained terms proportional
to $\delta^{kl}$, but they must vanish, since $F^{kl}$ must be anti-symmetric
in $k,l$.

Now we are finally ready to plug the curvature into the action. For
clarity, we define
\[
S=\frac{1}{\gamma}\int\d t\int\d^{3}x\left(L_{1}+L_{2}+L_{3}\right),
\]
where
\[
L_{1}\equiv-\hf\ept^{abc}e_{a}^{i}e_{b}^{j}\left(\delta_{ik}\delta_{jl}F_{0c}^{kl}+\gamma\epsilon_{ijk}F_{0c}^{0k}\right),
\]
\[
L_{2}\equiv-\hf N^{d}\ept^{abc}e_{d}^{i}e_{a}^{j}\left(\delta_{ik}\delta_{jl}F_{bc}^{kl}+\gamma\epsilon_{ijk}F_{bc}^{0k}\right),
\]
\[
L_{3}\equiv\hf N\ept^{abc}e_{a}^{i}\left(\delta_{ik}F_{bc}^{0k}-\hf\gamma\epsilon_{ikl}F_{bc}^{kl}\right).
\]
Let us calculate these terms one by one. In the interest of conciseness,
we will skip many steps; a more detailed derivation, showing all intermediate
steps, may be found in \cite{PhDThesis}.

\subsubsection{$L_{1}$: The Kinetic Term and the Gauss Constraint}

Plugging the curvature into $L_{1}$, we find:
\begin{align*}
L_{1} & =\hf\ept^{abc}e_{a}^{i}e_{b}^{j}\delta_{ik}\delta_{jl}\left(\epsilon_{p}^{kl}\left(\partial_{0}\Gamma_{c}^{p}-\partial_{c}\Gamma_{0}^{p}\right)-K_{0}^{k}K_{c}^{l}+K_{c}^{k}K_{0}^{l}+\Gamma_{0}^{k}\Gamma_{c}^{l}-\Gamma_{c}^{k}\Gamma_{0}^{l}\right)+\\
 & \qquad+\hf\gamma\ept^{abc}\epsilon_{ijk}e_{a}^{i}e_{b}^{j}\left(\partial_{0}K_{c}^{k}-\partial_{c}K_{0}^{k}+\epsilon_{pq}^{k}\left(K_{0}^{p}\Gamma_{c}^{q}-K_{c}^{p}\Gamma_{0}^{q}\right)\right).
\end{align*}
The densitized triad appears in both lines of $L_{1}$:
\begin{align*}
L_{1} & =\Et_{k}^{c}\partial_{0}\left(\Gamma_{c}^{k}+\gamma K_{c}^{k}\right)\\
 & \qquad+\left(\Gamma_{0}^{i}-\frac{1}{\gamma}K_{0}^{i}\right)\left(\partial_{c}\Et_{i}^{c}+\epsilon_{ij}^{k}\left(\Gamma_{c}^{j}+\gamma K_{c}^{j}\right)\Et_{k}^{c}\right)+\left(\frac{1}{\gamma}+\gamma\right)K_{0}^{i}\left(\partial_{c}\Et_{i}^{c}+\epsilon_{ij}^{k}\Gamma_{c}^{j}\Et_{k}^{c}\right),
\end{align*}
where we used the identity $\epsilon_{kl}^{m}\epsilon_{p}^{kl}=2\delta_{p}^{m}$,
integrated by parts the expressions $\Et_{p}^{c}\partial_{c}\Gamma_{0}^{p}$
and $\gamma\Et_{k}^{c}\partial_{c}K_{0}^{k}$, and then relabeled
indices and rearranged terms. Finally, we plug in the Ashtekar-Barbero
connection:
\[
A_{c}^{k}\equiv\Gamma_{c}^{k}+\gamma K_{c}^{k},
\]
define two Lagrange multipliers:
\[
\lambda^{i}\equiv\Gamma_{0}^{i}-\frac{1}{\gamma}K_{0}^{i}\sp\alpha^{i}\equiv\left(\frac{1}{\gamma}+\gamma\right)K_{0}^{i},
\]
and the \emph{Gauss constraint}:
\begin{equation}
G_{i}\equiv\partial_{c}\Et_{i}^{c}+\epsilon_{ij}^{k}A_{c}^{j}\Et_{k}^{c}.\label{eq:Gauss-index}
\end{equation}
The complete expression can now be written simply as:
\begin{equation}
L_{1}=\Et_{k}^{c}\partial_{0}A_{c}^{k}+\lambda^{i}G_{i}+\left(\frac{1}{\gamma}+\gamma\right)K_{0}^{i}\left(\partial_{c}\Et_{i}^{c}+\epsilon_{ij}^{k}\Gamma_{c}^{j}\Et_{k}^{c}\right).\label{eq:L_1}
\end{equation}
The first term is clearly a \emph{kinetic term}, indicating that $A_{c}^{k}$
and $\Et_{k}^{c}$ are \emph{conjugate variables}. The second term
imposes the Gauss constraint, which, as we will see in Section \ref{subsec:The-Constraints-as},
generates $\SUT$ gauge transformations. As for the third term, we
will show in the next subsection that it vanishes by the definition
of $\Gamma_{c}^{j}$.

\subsubsection{The Gauss Constraint in Index-Free Notation}

We can write the Gauss constraint in index-free notation. The covariant
differential of $\E$ in terms of the connection $\A$ is given by
\[
\d_{\A}\E\equiv\d\E+\left[\A,\E\right].
\]
The components of this 3-form are given by
\[
\d_{\A}\E=\hf\ept_{d[bc}\left(\delta^{il}\partial_{a]}\Et_{l}^{d}+\epsilon_{jk}^{i}A_{a]}^{j}\delta^{kl}\Et_{l}^{d}\right)\ta_{i}\d x^{a}\wedge\d x^{b}\wedge\d x^{c}.
\]
Next, we use the relation
\[
\d x^{a}\wedge\d x^{b}\wedge\d x^{c}=\ept^{abc}\d x^{1}\wedge\d x^{2}\wedge\d x^{3}\equiv\ept^{abc}\d^{3}x,
\]
along with the identity $\ept_{dbc}\ept^{abc}=2\delta_{d}^{a}$, to
find that
\[
\d_{\A}\E=\left(\partial_{a}\Et_{i}^{a}+\epsilon_{ij}^{k}A_{a}^{j}\Et_{k}^{a}\right)\ta^{i}\d^{3}x.
\]
Finally, we \emph{smear} this 3-form inside a 3-dimensional integral,
with a Lagrange multiplier $\la\equiv\lambda^{i}\ta_{i}$:
\[
\int\la\cdot\d_{\A}\E=\int\lambda^{i}\left(\partial_{a}\Et_{i}^{a}+\epsilon_{ij}^{k}A_{a}^{j}\Et_{k}^{a}\right)\d^{3}x.
\]
We thus see that demanding $\d_{\A}\E=0$ is equivalent to demanding
that (\ref{eq:Gauss-index}) vanishes:
\[
\G=\d_{\A}\E=0\soossp G_{i}\equiv\partial_{a}\Et_{i}^{a}+\epsilon_{ij}^{k}A_{a}^{j}\Et_{k}^{a}=0.
\]
Let us also write (\ref{eq:dGammaE}) with indices in the same way,
replacing $A_{a}^{j}$ with $\Gamma_{a}^{j}$:
\begin{equation}
\d_{\Ga}\E=0\soossp\partial_{a}\Et_{i}^{a}+\epsilon_{ij}^{k}\Gamma_{a}^{j}\Et_{k}^{a}=0.\label{eq:dGammaE2}
\end{equation}
Taking the difference of the two constraints, we get
\[
\d_{\A}\E-\d_{\Ga}\E=\gamma\left[\K,\E\right]=0\soosp\epsilon_{ki}^{j}K_{a}^{i}\Et_{j}^{a}=0.
\]
Now, the extrinsic curvature with two spatial indices is symmetric,
and it is related to $K_{a}^{i}$ by $K_{ab}=K_{a}^{i}e_{b}^{j}\delta_{ij}$.
Thus, the condition that its anti-symmetric part vanishes is
\[
K_{[ab]}=K_{[a}^{i}e_{b]}^{j}\delta_{ij}=0.
\]
Contracting with $\det\left(e\right)\epsilon^{klm}e_{k}^{a}e_{l}^{b}$,
we get
\[
\epsilon_{i}^{mk}K_{a}^{i}\Et_{k}^{a}=0.
\]
Therefore, $G_{k}=0$ is also equivalent to $K_{\left[ab\right]}=0$.
Yet another way to write this constraint, in index-free notation,
is to define a new quantity \cite{Freidel2019}
\begin{equation}
\P\equiv\d_{\A}\ee,\label{eq:new-quantity}
\end{equation}
such that
\[
\d_{\A}\E=\hf\d_{\A}\left[\ee,\ee\right]=\left[\d_{\A}\ee,\ee\right]=\left[\P,\ee\right].
\]

Finally, given (\ref{eq:dGammaE2}) we can simplify (\ref{eq:L_1})
to
\[
L_{1}=\Et_{k}^{c}\partial_{0}A_{c}^{k}+\lambda^{i}G_{i}.
\]

\subsubsection{$L_{2}$: The Vector (Spatial Diffeomorphism) Constraint}

Plugging the curvature into $L_{2}$, we find:
\[
L_{2}=N^{d}\ept^{abc}e_{d}^{i}e_{a}^{j}\left(\epsilon_{ijk}\partial_{b}A_{c}^{k}+\delta_{il}\delta_{jm}\left(\left(\Gamma_{b}^{l}\Gamma_{c}^{m}-K_{b}^{l}K_{c}^{m}\right)+\gamma\left(K_{b}^{l}\Gamma_{c}^{m}+\Gamma_{b}^{l}K_{c}^{m}\right)\right)\right).
\]
The curvature 2-form of the Ashtekar-Barbero connection, for which
we will also use the letter $F$ but with only one internal index,
is defined as:
\[
\hf F_{bc}^{k}\equiv\partial_{[b}A_{c]}^{k}+\hf\epsilon_{lm}^{k}A_{b}^{l}A_{c}^{m}.
\]
Expanding $A_{c}^{k}\equiv\Gamma_{c}^{k}+\gamma K_{c}^{k}$ and contracting
with $\ept^{abc}\epsilon_{ijk}$, we get
\[
\hf\ept^{abc}\epsilon_{ijk}F_{bc}^{k}=\ept^{abc}\left(\epsilon_{ijk}\partial_{b}A_{c}^{k}+\delta_{il}\delta_{jm}\left(\Gamma_{b}^{l}\Gamma_{c}^{m}+\gamma\left(K_{b}^{l}\Gamma_{c}^{m}+\Gamma_{b}^{l}K_{c}^{m}\right)+\gamma^{2}K_{b}^{l}K_{c}^{m}\right)\right).
\]
Therefore
\begin{align*}
 & \ept^{abc}\left(\hf\epsilon_{ijk}F_{bc}^{k}-\delta_{il}\delta_{jm}\left(1+\gamma^{2}\right)K_{b}^{l}K_{c}^{m}\right)=\\
 & =\ept^{abc}\left(\epsilon_{ijk}\partial_{b}A_{c}^{k}+\delta_{il}\delta_{jm}\left(\left(\Gamma_{b}^{l}\Gamma_{c}^{m}-K_{b}^{l}K_{c}^{m}\right)+\gamma\left(K_{b}^{l}\Gamma_{c}^{m}+\Gamma_{b}^{l}K_{c}^{m}\right)\right)\right).
\end{align*}
Plugging into $L_{2}$, we get
\[
L_{2}=N^{d}\ept^{abc}e_{d}^{i}e_{a}^{j}\left(\hf\epsilon_{ijk}F_{bc}^{k}-\delta_{il}\delta_{jm}\left(1+\gamma^{2}\right)K_{b}^{l}K_{c}^{m}\right).
\]
For the next step, we use the identity
\[
\ept^{abc}e_{a}^{j}=e_{p}^{b}e_{q}^{c}\epsilon^{jpq}\det\left(e\right).
\]
Plugging in, we obtain
\[
L_{2}=-N^{a}\det\left(e\right)\left(e_{p}^{b}F_{ab}^{p}+e_{a}^{i}e_{p}^{b}e_{q}^{c}\epsilon_{m}^{pq}\delta_{il}\left(1+\gamma^{2}\right)K_{b}^{l}K_{c}^{m}\right).
\]
Next, we use the definition of the densitized triad $\Et_{i}^{a}\equiv\det\left(e\right)e_{i}^{a}$:
\[
L_{2}=-N^{a}\left(\Et_{p}^{b}F_{ab}^{p}+\left(1+\gamma^{2}\right)e_{a}^{i}e_{p}^{b}\delta_{il}K_{b}^{l}\epsilon_{m}^{pq}K_{c}^{m}\Et_{q}^{c}\right).
\]
Recall that the Gauss constraint is equivalent to $G_{i}=\gamma\epsilon_{ij}^{k}K_{c}^{j}\Et_{k}^{c}$,
or, relabeling indices and rearranging,
\[
\epsilon_{m}^{pq}K_{c}^{m}\Et_{q}^{c}=-\frac{1}{\gamma}G^{p}.
\]
Plugging into $L_{2}$, we get
\[
L_{2}=-N^{a}\Et_{p}^{b}F_{ab}^{p}+\left(\frac{1}{\gamma}+\gamma\right)N_{a}K_{p}^{a}G^{p}.
\]
The part with $G^{p}$ is redundant -- the Gauss constraint is already
enforced by $L_{1}$, and we can combine the second term of $L_{2}$
with $L_{1}$ by redefining some fields. Thus we get 
\[
L_{2}=-N^{a}\Et_{p}^{b}F_{ab}^{p}.
\]
We can now define the \emph{vector (or momentum) constraint}:
\[
V_{a}\equiv\Et_{p}^{b}F_{ab}^{p}.
\]
Then $L_{2}$ simply enforces this constraint with the Lagrange multiplier
$N^{a}$:
\[
L_{2}=-N^{a}V_{a}.
\]
In Section \ref{subsec:The-Constraints-as} we will discuss how this
constraint is related to spatial diffeomorphisms.

To write the vector constraint in index-free notation, we note that
\begin{align*}
N^{i}\left[\ee,\F\right]_{i} & =-N^{b}\Et_{k}^{c}F_{bc}^{k}\d^{3}x.
\end{align*}
Thus, in terms of differential forms, we can write the vector constraint
as
\[
\N\cdot\left[\ee,\F\right]=0.
\]

\subsubsection{$L_{3}$: The Scalar (Hamiltonian) Constraint}

Finally, we plug the curvature into the last term in the action:
\[
L_{3}=-N\ept^{abc}e_{a}^{i}\left(\delta_{ik}\left(\partial_{b}K_{c}^{k}+\epsilon_{pq}^{k}K_{b}^{p}\Gamma_{c}^{q}\right)-\hf\gamma\epsilon_{ikl}\left(\epsilon_{p}^{kl}\partial_{b}\Gamma_{c}^{p}-K_{b}^{k}K_{c}^{l}+\Gamma_{b}^{k}\Gamma_{c}^{l}\right)\right).
\]
Using the identity for $\ept^{abc}e_{a}^{i}$ in (\ref{eq:epsilon-e-id}),
we get
\begin{align*}
L_{3} & =N\frac{\epsilon^{imn}\Et_{m}^{b}\Et_{n}^{c}}{\sqrt{\det\left(E\right)}}\left(\gamma\delta_{ij}\partial_{b}\left(\Gamma_{c}^{j}-\frac{1}{\gamma}K_{c}^{j}\right)-\epsilon_{ijk}\left(K_{b}^{j}\Gamma_{c}^{k}+\hf\gamma\left(K_{b}^{j}K_{c}^{k}-\Gamma_{b}^{j}\Gamma_{c}^{k}\right)\right)\right).
\end{align*}
Substituting $K_{c}^{j}=\frac{1}{\gamma}\left(A_{c}^{j}-\Gamma_{c}^{j}\right)$,
we obtain
\begin{align*}
L_{3} & =-N\frac{\epsilon_{i}^{mn}\Et_{m}^{b}\Et_{n}^{c}}{2\gamma\sqrt{\det\left(E\right)}}\left(F_{bc}^{i}-\left(1+\gamma^{2}\right)R_{bc}^{i}\right),
\end{align*}
where we identified the curvature of the Ashtekar-Barbero connection:
\[
\hf F_{bc}^{i}\equiv\partial_{[b}A_{c]}^{i}+\hf\epsilon_{jk}^{i}A_{b}^{j}A_{c}^{k},
\]
as well as the curvature of the spin connection:
\[
\hf R_{bc}^{i}\equiv\partial_{[b}\Gamma_{c]}^{i}+\hf\epsilon_{jk}^{i}\Gamma_{b}^{j}\Gamma_{c}^{k}.
\]
If we define the \emph{scalar} (or \emph{Hamiltonian}) \emph{constraint}:
\begin{equation}
C\equiv-\frac{\epsilon_{i}^{mn}\Et_{m}^{b}\Et_{n}^{c}}{2\gamma\sqrt{\det\left(E\right)}}\left(F_{bc}^{i}-\left(1+\gamma^{2}\right)R_{bc}^{i}\right),\label{eq:scalar-R}
\end{equation}
then $L_{3}$ is simply
\[
L_{3}=NC,
\]
and it imposes the scalar constraint via the Lagrange multiplier $N$.
The scalar constraint generates the time evolution of the theory,
that is, from one spatial slice to another.

\subsubsection{The Scalar Constraint in Terms of the Extrinsic Curvature}

We can rewrite the scalar constraint in another way which is more
commonly encountered. First we plug $A_{b}^{j}=\Gamma_{b}^{j}+\gamma K_{b}^{j}$
into the curvature of the Ashtekar-Barbero connection and get
\[
F_{bc}^{i}=R_{bc}^{i}+\gamma^{2}\epsilon_{jk}^{i}K_{b}^{j}K_{c}^{k}+2\gamma D_{[b}\left(\Gamma\right)K_{c]}^{i},
\]
where we defined the covariant derivative of $K_{c}^{i}$ with respect
to the spin connection:
\[
D_{[b}\left(\Gamma\right)K_{c]}^{i}\equiv\partial_{[b}K_{c]}^{i}+\epsilon_{jk}^{i}\Gamma_{b}^{j}K_{c}^{k}.
\]
Now, since the spin connection $\Gamma_{a}^{k}$ is compatible with
the triad, we have:
\[
\hf T_{ab}^{i}\equiv D_{a}\left(\Gamma\right)e_{b}^{i}\equiv\partial_{a}e_{b}^{i}+\epsilon_{kl}^{i}\Gamma_{a}^{k}e_{b}^{l}=0.
\]
However,
\[
D_{a}\left(\Gamma\right)\left(e_{b}^{i}e_{i}^{b}\right)=D_{a}\left(\Gamma\right)\left(3\right)=0\soosp e_{i}^{b}D_{a}\left(\Gamma\right)e_{b}^{i}=-e_{b}^{i}D_{a}\left(\Gamma\right)e_{i}^{b}.
\]
Therefore
\[
0=e_{b}^{i}D_{a}\left(\Gamma\right)e_{i}^{b}=-e_{i}^{b}D_{a}\left(\Gamma\right)e_{b}^{i}=e_{b}^{i}\left(\partial_{a}e_{i}^{b}+\epsilon_{ik}^{l}\Gamma_{a}^{k}e_{l}^{b}\right),
\]
and we obtain that
\[
D_{a}\left(\Gamma\right)e_{i}^{b}\equiv\partial_{a}e_{i}^{b}+\epsilon_{ik}^{l}\Gamma_{a}^{k}e_{l}^{b}=0.
\]
Multiplying by $\det\left(e\right)$ and using the definition of $\Et_{i}^{b}$,
we get
\[
D_{a}\left(\Gamma\right)\Et_{i}^{b}\equiv\partial_{a}\Et_{i}^{b}+\epsilon_{ik}^{l}\Gamma_{a}^{k}\Et_{l}^{b}=0.
\]
Thus, when we contract $F_{bc}^{i}$ with $\epsilon_{i}^{mn}\Et_{m}^{b}\Et_{n}^{c}$,
we can insert $\gamma\epsilon_{i}^{mn}\Et_{n}^{c}$ into the covariant
derivative: 
\[
\epsilon_{i}^{mn}\Et_{m}^{b}\Et_{n}^{c}F_{bc}^{i}=\epsilon_{i}^{mn}\Et_{m}^{b}\Et_{n}^{c}\left(R_{bc}^{i}+\gamma^{2}\epsilon_{jk}^{i}K_{b}^{j}K_{c}^{k}\right)-2\Et_{m}^{b}D_{b}\left(\Gamma\right)G^{m},
\]
where we used the identity $G^{m}=-\gamma\epsilon_{i}^{mn}K_{c}^{i}\Et_{n}^{c}$
again. Rearranging terms, we get
\[
\epsilon_{i}^{mn}\Et_{m}^{b}\Et_{n}^{c}R_{bc}^{i}=\epsilon_{i}^{mn}\Et_{m}^{b}\Et_{n}^{c}\left(F_{bc}^{i}-\gamma^{2}\epsilon_{jk}^{i}K_{b}^{j}K_{c}^{k}\right)+2\Et_{m}^{b}D_{b}\left(\Gamma\right)G^{m}.
\]
Plugging into $C$, we obtain
\[
C=\frac{\gamma\epsilon_{i}^{mn}\Et_{m}^{b}\Et_{n}^{c}}{2\sqrt{\det\left(E\right)}}\left(F_{bc}^{i}-\left(1+\gamma^{2}\right)\epsilon_{jk}^{i}K_{b}^{j}K_{c}^{k}\right)+\frac{\left(1+\gamma^{2}\right)\Et_{m}^{b}}{\gamma\sqrt{\det\left(E\right)}}D_{b}\left(\Gamma\right)G^{m}.
\]
Now, if the Gauss constraint is satisfied, then the second term is
redundant, and we can get rid of it. We obtain the familiar expression
for the scalar or Hamiltonian constraint:
\[
C=\frac{\gamma\epsilon_{i}^{mn}\Et_{m}^{b}\Et_{n}^{c}}{2\sqrt{\det\left(E\right)}}\left(F_{bc}^{i}-\left(1+\gamma^{2}\right)\epsilon_{jk}^{i}K_{b}^{j}K_{c}^{k}\right).
\]

\subsubsection{The Scalar Constraint in Index-Free Notation}

Finally, let us write the scalar constraint in index-free notation.
We first use the identity (\ref{eq:epsilon-e-id}):
\[
\frac{\epsilon^{imn}\Et_{m}^{b}\Et_{n}^{c}}{\sqrt{\det\left(E\right)}}=\ept^{abc}e_{a}^{i}.
\]
Plugging in, and ignoring the overall factor of $\gamma$, we get
\[
C=\hf\ept^{abc}\delta_{il}e_{a}^{l}\left(F_{bc}^{i}-\left(1+\gamma^{2}\right)\epsilon_{jk}^{i}K_{b}^{j}K_{c}^{k}\right).
\]
Now, from our definition of the graded dot product (see Section \ref{subsec:The-Graded-Dot})
we have:
\[
\ee\cdot\F=\hf\delta_{il}e_{a}^{l}F_{bc}^{i}\ept^{abc}\d^{3}x.
\]
Furthermore, from our definition (\ref{eq:triple-product}) of the
triple product we have
\[
\ee\cdot\left[\K,\K\right]=\delta_{il}\epsilon_{jk}^{i}e_{a}^{l}K_{b}^{j}K_{c}^{k}\ept^{abc}\d^{3}x.
\]
Thus we can write
\[
C\equiv\ee\cdot\left(\F-\frac{1+\gamma^{2}}{2}\left[\K,\K\right]\right).
\]
If we smear this 3-form inside a 3-dimensional integral, with a Lagrange
multiplier $N$, we get the appropriate expression for the scalar
constraint.

Furthermore, let us consider again the new quantity defined in (\ref{eq:new-quantity}).
We find that
\[
\P\equiv\d_{\A}\ee=\d\ee+\left[\A,\ee\right]=\gamma\left[\K,\ee\right],
\]
since $\d_{\Ga}\ee=0$. Thus we can write
\[
\ee\cdot\left[\K,\K\right]=\K\cdot\left[\K,\ee\right]=\frac{1}{\gamma}\K\cdot\P,
\]
and the scalar constraint becomes
\[
C\equiv\ee\cdot\F-\hf\left(\frac{1}{\gamma}+\gamma\right)\K\cdot\P.
\]
In this form of the constraint, it is clear that it is automatically
satisfied if $\F=\P=0$.

Similarly, for (\ref{eq:scalar-R}),
\[
C=-\frac{\epsilon_{i}^{mn}\Et_{m}^{b}\Et_{n}^{c}}{2\gamma\sqrt{\det\left(E\right)}}\left(F_{bc}^{i}-\left(1+\gamma^{2}\right)R_{bc}^{i}\right),
\]
we again use (\ref{eq:epsilon-e-id}) to get, ignoring the overall
factor of $-1/\gamma$,
\[
C=\hf\ept^{abc}\delta_{il}e_{a}^{l}\left(F_{bc}^{i}-\left(1+\gamma^{2}\right)R_{bc}^{i}\right).
\]
Then, we have as before
\[
\ee\cdot\F=\hf\delta_{il}e_{a}^{l}F_{bc}^{i}\ept^{abc}\d^{3}x,
\]
and similarly
\[
\ee\cdot\RR=\hf\delta_{il}e_{a}^{l}R_{bc}^{i}\ept^{abc}\d^{3}x,
\]
where
\[
\RR\equiv\d_{\Ga}\Ga=\d\Ga+\hf\left[\Ga,\Ga\right].
\]
The scalar constraint can thus be written simply as
\[
C=\ee\cdot\left(\F-\left(1+\gamma^{2}\right)\RR\right).
\]

\subsection{\label{subsec:The-Symplectic-Potential-Ashtekar}The Symplectic Potential}

Above, we found the symplectic potential \ref{eq:symplectic-Holst}
of the Holst action:
\[
\Theta=\fr\int_{\Sigma}\left(\star+\frac{1}{\gamma}\right)e_{I}\wedge e_{J}\wedge\delta\omega^{IJ}.
\]
Let us rewrite it in terms of the 3-dimensional internal indices,
using the 3-dimesional internal-space Levi-Civita symbol $\epsilon_{ijk}\equiv\epsilon_{0ijk}$:
\[
\Theta=\fr\int_{\Sigma}\left(\epsilon_{ijk}e^{0}\wedge e^{i}\wedge\delta\omega^{jk}+\epsilon_{ijk}e^{i}\wedge e^{j}\wedge\delta\omega^{0k}+\frac{2}{\gamma}e_{0}\wedge e_{i}\wedge\delta\omega^{0i}+\frac{1}{\gamma}e_{i}\wedge e_{j}\wedge\delta\omega^{ij}\right).
\]
Since $e^{0}=0$ on $\Sigma$ due to the time gauge, the two terms
with $e^{0}$ vanish and we are left with:
\[
\Theta=\fr\int_{\Sigma}\left(\epsilon_{ijk}e^{i}\wedge e^{j}\wedge\delta\omega^{0k}+\frac{1}{\gamma}e_{i}\wedge e_{j}\wedge\delta\omega^{ij}\right).
\]
Recall that we defined the electric field as
\[
\E\equiv\hf\left[\ee,\ee\right],
\]
or with indices
\[
E_{i}=\epsilon_{ijk}e^{j}\wedge e^{k}\soossp e_{i}\wedge e_{j}=\hf\epsilon_{ijk}E^{k}.
\]
Thus our symplectic potential becomes
\[
\Theta=\frac{1}{4\gamma}\int_{\Sigma}E_{i}\wedge\delta\left(\hf\epsilon_{jk}^{i}\omega^{jk}+\gamma\omega^{0i}\right).
\]
We identify here the dual spin connection and extrinsic curvature
defined in Section \ref{subsec:The-Ashtekar-Barbero-Connection}:
\[
\Gamma_{a}^{i}\equiv-\hf\epsilon_{jk}^{i}\omega_{a}^{jk}\sp K_{a}^{i}\equiv-\omega_{a}^{0i},
\]
so the expression in parentheses is none other than the (minus) Ashtekar-Barbero
connection:
\[
\A\equiv\Ga+\gamma\K\soosp A_{a}^{i}\equiv\Gamma_{a}^{i}+\gamma K_{a}^{i}=-\hf\epsilon_{jk}^{i}\omega_{a}^{jk}-\gamma\omega_{a}^{0i}.
\]
Ignoring the irrelevant overall factor, the symplectic potential now
reaches its final form
\[
\Theta=\int_{\Sigma}\E\cdot\delta\A.
\]

\subsection{Summary}

To summarize our calculation, we find that the action for 3+1D gravity
with the Ashtekar variables is 
\[
S=\frac{1}{\gamma}\int\d t\int_{\Sigma}\d^{3}x\left(\Et_{i}^{a}\partial_{t}A_{a}^{i}+\lambda^{i}G_{i}+N^{a}V_{a}+NC\right),
\]
or in index-free notation
\[
S=\frac{1}{\gamma}\int\d t\int_{\Sigma}\left(\E\cdot\partial_{t}\A+\la\cdot\left[\ee,\P\right]+\N\cdot\left[\ee,\F\right]+N\left(\ee\cdot\F-\hf\left(\frac{1}{\gamma}+\gamma\right)\K\cdot\P\right)\right),
\]
and the symplectic potential is
\[
\Theta=\int_{\Sigma}\E\cdot\delta\A.
\]
Chapter \ref{sec:Gravity-as-a} in the main text provides a legend
and interpretation for the various quantities in these expressions.

\section{\label{sec:Cosmic-Strings-in}Cosmic Strings}

In this appendix, we consider a very simple 3+1-dimensional geometry:
a flat and torsionless spacetime with a single infinite string, which
appears as a 1-dimensional topological defect and carries curvature
and torsion degrees of freedom. This can also be interpreted as a
flat universe which is completely empty except for a single infinite
string, serving as the only source of matter.

\subsection{\label{sec:Delta-function-proof}Proof that $\protect\d^{2}\phi=2\pi\delta^{\left(2\right)}\left(r\right)$}

We begin by proving a relation between $\d^{2}\phi$ and the Dirac
delta 2-form in cylindrical coordinates. This relation will be used
later in this appendix to show that a cosmic string has distributional
curvature and torsion.

Let us define a cylinder $\Sigma$ with coordinates $\left(r,\phi,z\right)$
such that 
\[
r\in\left[0,R\right]\sp\phi\in\left[0,2\pi\right)\sp z\in\left[-\frac{L}{2},+\frac{L}{2}\right].
\]
Furthermore, let 
\[
f\equiv f_{r}\thinspace\d r+f_{\phi}\thinspace\d\phi+f_{z}\thinspace\d z
\]
be a test 1-form such that
\begin{equation}
\partial_{\phi}f_{z}\left(r=0\right)=0.\label{eq:delta-con}
\end{equation}
The condition (\ref{eq:delta-con}) means that that value of the 1-form
on the string itself, $f\left(r=0\right)$, is the same for each value
of $\phi$. This certainly makes sense, as different values of $\phi$
at $r=0$ (for a particular choice of $z$) correspond to the same
point.

We define a 2-form distribution $\delta^{\left(2\right)}\left(r\right)$
such that 
\begin{equation}
\int_{\Sigma}f\wedge\delta^{\left(2\right)}\left(r\right)=\int_{\left\{ r=0\right\} }f,\label{eq:delta-def}
\end{equation}
where $\left\{ r=0\right\} $ is the line along the $z$ axis. Let
us now show that the 2-form $\d^{2}\phi$ satisfies this definition.

Using the graded Leibniz rule we have, since $f$ is a 1-form, 
\[
f\wedge\d^{2}\phi=\d f\wedge\d\phi-\d\left(f\wedge\d\phi\right).
\]

Integrating this on $\Sigma$, we get 
\begin{equation}
\int_{\Sigma}f\wedge\d^{2}\phi=\int_{\Sigma}\d f\wedge\d\phi-\int_{\Sigma}\d\left(f\wedge\d\phi\right).\label{eq:int}
\end{equation}
The second integral in (\ref{eq:int}) can easily be integrated using
Stokes' theorem: 
\[
\int_{\Sigma}\d\left(f\wedge\d\phi\right)=\int_{\partial\Sigma}f\wedge\d\phi=\int_{\partial\Sigma}\left(f_{r}\thinspace\d r+f_{z}\thinspace\d z\right)\wedge\d\phi.
\]
The boundary of the cylinder consists of three parts: 
\[
\partial\Sigma=\left\{ r=R\right\} \cup\left\{ z=-\frac{L}{2}\right\} \cup\left\{ z=+\frac{L}{2}\right\} .
\]
Note that $\d r=0$ for the first part and $\d z=0$ for the second
and third; thus 
\begin{equation}
\int_{\Sigma}\d\left(f\wedge\d\phi\right)=\int_{\left\{ r=R\right\} }f_{z}\thinspace\d z\wedge\d\phi+\int_{\left\{ z=\pm L/2\right\} }f_{r}\thinspace\d r\wedge\d\phi.\label{eq:dfdphi}
\end{equation}
As for the first integral in (\ref{eq:int}), we have 
\begin{align*}
\int_{\Sigma}\d f\wedge\d\phi & =\int_{\Sigma}\d\left(f_{r}\thinspace\d r+f_{\phi}\thinspace\d\phi+f_{z}\thinspace\d z\right)\wedge\d\phi\\
 & =\int_{\Sigma}\left(\partial_{z}f_{r}\thinspace\d z\wedge\d r+\partial_{r}f_{z}\d r\wedge\d z\right)\wedge\d\phi.
\end{align*}
For the first term we find 
\begin{align*}
\int_{\Sigma}\partial_{z}f_{r}\thinspace\d z\wedge\d r\wedge\d\phi & =\int_{\phi=0}^{2\pi}\int_{r=0}^{R}\left(\int_{z=-L/2}^{+L/2}\partial_{z}f_{r}\thinspace\d z\right)\d r\wedge\d\phi\\
 & =\int_{\phi=0}^{2\pi}\int_{r=0}^{R}\left(f_{r}\left(z=+\frac{L}{2}\right)-f_{r}\left(z=-\frac{L}{2}\right)\right)\d r\wedge\d\phi\\
 & =\int_{\left\{ z=\pm L/2\right\} }f_{r}\thinspace\d r\wedge\d\phi,
\end{align*}
where the orientation of the boundary at $z=-L/2$ is chosen to be
opposite to that at $z=+L/2$, and for the second term we find 
\begin{align*}
\int_{\Sigma}\partial_{r}f_{z}\d r\wedge\d z\wedge\d\phi & =\int_{\phi=0}^{2\pi}\int_{z=-L/2}^{+L/2}\left(\int_{r=0}^{R}\partial_{r}f_{z}\d r\right)\d z\wedge\d\phi\\
 & =\int_{\phi=0}^{2\pi}\int_{z=-L/2}^{+L/2}\left(f_{z}\left(r=R\right)-f_{z}\left(r=0\right)\right)\d z\wedge\d\phi\\
 & =\int_{\left\{ r=R\right\} }f_{z}\thinspace\d z\wedge\d\phi-\int_{\phi=0}^{2\pi}\int_{z=-L/2}^{+L/2}f_{z}\left(r=0\right)\d z\wedge\d\phi.
\end{align*}
In conclusion, given (\ref{eq:dfdphi}) we see that 
\[
\int_{\Sigma}\d f\wedge\d\phi=\int_{\Sigma}\d\left(f\wedge\d\phi\right)-\int_{\phi=0}^{2\pi}\int_{z=-L/2}^{+L/2}f_{z}\left(r=0\right)\d z\wedge\d\phi,
\]
and therefore (\ref{eq:int}) becomes
\[
\int_{\Sigma}f\wedge\d^{2}\phi=-\int_{\phi=0}^{2\pi}\int_{z=-L/2}^{+L/2}f_{z}\left(r=0\right)\d z\wedge\d\phi.
\]
Finally, due to the condition (\ref{eq:delta-con}), we can rewrite
this as: 
\begin{align*}
\int_{\Sigma}f\wedge\d^{2}\phi & =\int_{z=-L/2}^{+L/2}\int_{\phi=0}^{2\pi}\left(f_{z}\left(r=0\right)\d\phi\right)\d z\\
 & =2\pi\int_{z=-L/2}^{+L/2}f_{z}\left(r=0\right)\d z.
\end{align*}
Noting that $\d\phi=\d r=0$ along the line $\left\{ r=0\right\} $,
we see that 
\[
\int_{\left\{ r=0\right\} }f=\int_{\left\{ r=0\right\} }f_{z}\thinspace\d z,
\]
and thus we find that 
\begin{equation}
\int_{\Sigma}f\wedge\d^{2}\phi=2\pi\int_{\left\{ r=0\right\} }f.\label{eq:d2phi-delta}
\end{equation}
Given (\ref{eq:delta-def}), we see that indeed $\d^{2}\phi=2\pi\delta^{\left(2\right)}\left(r\right)$,
as we wanted to prove.

Note that the delta 2-form distribution may be written as
\[
\delta^{\left(2\right)}\left(r\right)=\delta\left(r\right)\d x\wedge\d y=\delta\left(r\right)r\thinspace\d r\wedge\d\phi,
\]
where $\delta\left(r\right)$ is the usual 1-dimensional delta function.
Therefore we have
\[
\d^{2}\phi=2\pi\delta\left(r\right)r\thinspace\d r\wedge\d\phi.
\]

\subsection{The Frame Field and Spin Connection}

To describe a cosmic string, we use cylindrical coordinates $\left(t,r,\phi,z\right)$
with the infinite string lying along the $z$ axis. This metric can
then be embedded in a larger space to represent a finite string. The
metric will be
\[
\d s^{2}=-\d t^{2}+\frac{\d r^{2}}{\left(1-M\right)^{2}}+r^{2}\thinspace\d\phi^{2}+\left(\d z+S\thinspace\d\phi\right)^{2}.
\]
We can define new coordinates
\begin{equation}
T\equiv t\sp R\equiv\frac{r}{1-M}\sp\Phi\equiv\left(1-M\right)\phi\sp Z\equiv z+S\phi,\label{eq:flat-coords-3p1}
\end{equation}
where $M$ will be referred to as the ``mass'' and $S$ the ``spin''.
Then the metric becomes flat:
\[
\d s^{2}=-\d T^{2}+\d R^{2}+R^{2}\thinspace\d\Phi^{2}+\d Z^{2}.
\]
However, the periodicity condition $\phi\sim\phi+2\pi$ becomes
\[
\Phi\sim\Phi+2\pi\left(1-M\right)\sp Z\sim Z+2\pi S.
\]
The identification $\Phi\sim\Phi+2\pi\left(1-M\right)$ means that
in a slice of constant $T$, as we go around the origin at $r=0$,
we find that it only takes us $2\pi\left(1-M\right)$ radians to complete
a full circle, rather than $2\pi$ radians. Therefore we have obtained
a ``Pac-Man''-like surface, where the angle of the ``mouth'' is
$2\pi M$, and both ends of the ``mouth'' are glued to each other.
This produces a cone with \emph{deficit angle }$2\pi M$. Note that
if $M=0$ and $S=0$, the particle is indistinguishable from flat
spacetime\footnote{It is interesting to compare this to the 2+1D case, which is discussed
in \cite{Deser:1983tn,SousaGerbert1989,deSousaGerbert:1990yp,Meusburger:2005in,Hooft1988}.
There we have point particles instead of strings. The metric is:
\[
\d s^{2}=-\left(\d t+S\thinspace\d\phi\right)^{2}+\frac{\d r^{2}}{\left(1-M\right)^{2}}+r^{2}\thinspace\d\phi^{2},
\]
and it becomes flat upon defining
\[
T\equiv t+S\phi\sp R\equiv\frac{r}{1-M}\sp\Phi\equiv\left(1-M\right)\phi,
\]
which yields the periodicity conditions
\[
T\sim T+2\pi S\sp\Phi\sim\Phi+2\pi\left(1-M\right).
\]

If the spin $S$ is non-zero, one end of the mouth is identified with
the other end, but at a different point in time -- there is a time
shift of $2\pi S$. This seems like it might create \emph{closed timelike
curves}, which would lead to causality violations \cite{FTL}. However,
when the spin is due to internal orbital angular momentum, the source
itself would need to be larger than the radius of any closed timelike
curves \cite{deSousaGerbert:1990yp,hooft1992causality}; thus, no
causality violations take place.

In the 3+1D case discussed here, since we do not have a time shift,
we will not create closed timelike curves. Instead, the periodicity
is in the $Z$ direction. When we foliate spacetime into 3-dimensional
spatial slices in order to go to the Hamiltonian formulation, $Z$
will play a role analogous to the one $T$ plays in the 2+1D case
(which does \textbf{not} involve a foliation).}.

Next, we define the frame fields:
\begin{equation}
e^{0}=\d T\sp e^{1}=\d R\sp e^{2}=R\thinspace\d\Phi\sp e^{3}=\d Z.\label{eq:frame-field}
\end{equation}
The torsion 2-form is
\[
T^{I}\equiv\d_{\omega}e^{I}=\d e^{I}+\udi{\omega}IJ\wedge e^{J},
\]
and its four components are:
\[
T^{0}=\udi{\omega}01\wedge\d R+\udi{\omega}02\wedge R\thinspace\d\Phi+\udi{\omega}03\wedge\d Z,
\]
\[
T^{1}=\udi{\omega}10\wedge\d T+\udi{\omega}12\wedge R\thinspace\d\Phi+\udi{\omega}13\wedge\d Z,
\]
\[
T^{2}=\d R\wedge\d\Phi+\udi{\omega}20\wedge\d T+\udi{\omega}21\wedge\d R+\udi{\omega}23\wedge\d Z,
\]
\[
T^{3}=\udi{\omega}30\wedge\d T+\udi{\omega}31\wedge\d R+\udi{\omega}32\wedge R\thinspace\d\Phi.
\]
The spin connection is the one for which $T^{I}=0$. In order for
the torsion to vanish, all of the components of $\udi{\omega}IJ$
must be set to zero\footnote{Indeed, one can check that there are no solutions to this system of
equations where any $\udi{\omega}IJ$ other than $\udi{\omega}21$
can be non-zero.} except
\[
\udi{\omega}21=\d\Phi,
\]
which is needed in order to cancel the $\d R\wedge\d\Phi$ term in
$T^{2}$. Note that, since the metric on the internal space is flat,
we also have that $\udi{\omega}21=\omega^{21}=-\omega^{12}=\d\Phi$.
Finally, we go back to the original coordinates using (\ref{eq:flat-coords-3p1}):
\begin{equation}
\udi{\omega}21=\left(1-M\right)\d\phi=-\udi{\omega}12\sp e^{0}=\d t\sp e^{1}=\frac{\d r}{1-M}\sp e^{2}=r\thinspace\d\phi\sp e^{3}=\d z+S\thinspace\d\phi.\label{eq:spin-frame}
\end{equation}

Then the torsion becomes
\[
T^{0}=T^{1}=T^{2}=0\sp T^{3}=S\thinspace\d^{2}\phi=2\pi S\delta\left(r\right)\d r\wedge\d\phi,
\]
where we used the fact that $\d^{2}\phi=2\pi\delta\left(r\right)\d r\wedge\d\phi$,
as proven above, and thus
\[
\d e^{3}=\d\left(\d z+S\thinspace\d\phi\right)=S\thinspace\d^{2}\phi=2\pi S\delta\left(r\right)\d r\wedge\d\phi.
\]
We may also calculate the curvature of the spin connection, which
is defined as
\[
\udi RIJ\equiv\d_{\omega}\udi{\omega}IJ=\d\udi{\omega}IJ+\udi{\omega}IK\wedge\udi{\omega}KJ.
\]
Its components will all be zero, except for
\[
\udi R12=-\udi R21=-\left(1-M\right)\d^{2}\phi=-2\pi\left(1-M\right)\delta\left(r\right)\d r\wedge\d\phi.
\]

\subsection{The Foliation of Spacetime and the Ashtekar Variables}

To go to the Hamiltonian formulation, we perform the 3+1 split and
impose the time gauge, as detailed in Section \ref{subsec:The-3+1-Split}:
\[
e^{0}=N\thinspace\d t\sp e^{i}=N^{i}\thinspace\d t+e_{a}^{i}\thinspace\d x^{a}.
\]
From (\ref{eq:spin-frame}) we see that we are already in the time
gauge, and the lapse and shift are trivial, $N=1$ and $N^{i}=0$,
as one would indeed expect from a flat spacetime.

Next, we turn the two internal-space indices on the spin connection
into one index by defining the \emph{dual spin connection} as in Section
\ref{subsec:The-Ashtekar-Barbero-Connection}:
\[
\Gamma^{i}\equiv-\hf\udi{\epsilon}i{jk}\omega^{jk}.
\]
Since the only non-zero components of $\omega^{ij}$ are $\omega^{21}=-\omega^{12}=\left(1-M\right)\d\phi$,
we get:
\[
\Gamma^{1}=\Gamma^{2}=0\sp\Gamma^{3}=\left(1-M\right)\d\phi.
\]
The frame field on each spatial slice is simply
\[
e^{1}=\frac{\d r}{1-M}\sp e^{2}=r\thinspace\d\phi\sp e^{3}=\d z+S\thinspace\d\phi.
\]
We can now use index-free notation again:
\[
\Ga=\left(1-M\right)\J_{3}\thinspace\d\phi\sp\ee=\frac{\P_{1}\thinspace\d r}{1-M}+\P_{3}\thinspace\d z+\left(S\P_{3}+r\P_{2}\right)\d\phi.
\]
The torsion will be
\[
\T\equiv\d_{\Ga}\ee=\d\ee+\left[\Ga,\ee\right]=S\P_{3}\d^{2}\phi=2\pi S\thinspace\delta\left(r\right)\ta_{3}\thinspace\d r\wedge\d\phi.
\]
The first Ashtekar variable is the electric field $\E$, defined as
\[
\E\equiv\hf\left[\ee,\ee\right]\soosp E^{i}=\hf\udi{\epsilon}i{jk}e^{j}\wedge e^{k}.
\]
Calculating it, we get
\[
\E=\frac{\left(r\P_{3}-S\P_{2}\right)\d r\wedge\d\phi+\left(\P_{1}+\P_{2}\right)\d z\wedge\d r}{1-M}+r\P_{1}\d\phi\wedge\d z.
\]
The second Ashtekar variables is the Ashtekar-Barbero connection $\A$,
defined as
\[
\A\equiv\Ga+\gamma\K,
\]
where $\gamma$ is the \emph{Barbero-Immirzi parameter} and $\K$
is the extrinsic curvature, defined as
\[
K_{a}^{i}\equiv\omega_{a}^{i0}.
\]
In our case, it is clear that the extrinsic curvature vanishes; this
makes sense, as we are on equal-time slices in an essentially flat
spacetime. Therefore the Ashtekar connection is in fact identical
to the dual spin connection:
\[
\A=\Ga=\left(1-M\right)\J_{3}\thinspace\d\phi.
\]
The curvatures of these connections are:
\[
\RR\equiv\d_{\Ga}\Ga=\d\Ga+\hf\left[\Ga,\Ga\right],
\]
\[
\F\equiv\d_{\A}\A=\d\A+\hf\left[\A,\A\right],
\]
and they are both equal to:
\[
\RR=\F=\left(1-M\right)\J_{3}\thinspace\d^{2}\phi=2\pi\left(1-M\right)\delta\left(r\right)\J_{3}\thinspace\d r\wedge\d\phi.
\]
If we define $\m\equiv\left(1-M\right)\J_{3}\in\mfg$ and $\s\equiv S\P_{3}\in\mfg^{*}$,
then we may write
\[
\Ga=\A=\m\thinspace\d\phi\sp\ee=\frac{\P_{1}\thinspace\d r}{1-M}+\P_{3}\thinspace\d z+\left(\s+r\P_{2}\right)\d\phi,
\]
\[
\T=\P=2\pi\s\thinspace\delta\left(r\right)\thinspace\d r\wedge\d\phi\sp\RR=\F=2\pi\m\thinspace\delta\left(r\right)\d r\wedge\d\phi.
\]

\subsection{The Dressed Quantities}

The expressions for $\A$ and $\ee$ are not invariant under the $G\ltimes\mfg^{*}$
gauge transformation
\[
\A\mt h^{-1}\A h+h^{-1}\d h\sp\ee\mt h^{-1}\left(\ee+\d_{\A}\x\right)h,
\]
\[
\F\mt h^{-1}\F h\sp\T\mt h^{-1}\left(\T+\left[\F,\x\right]\right)h,
\]
where the gauge parameters are a $G$-valued 0-form $h$ and a $\mfg^{*}$-valued
0-form $\x$. When we apply these transformations, we get:
\[
\Ga=\A=h^{-1}\m h\thinspace\d\phi+h^{-1}\d h,
\]
\[
\ee=h^{-1}\left(\d\x+\left(\s+\left[\m,\x\right]\right)\d\phi\right)h,
\]
\[
\E=h^{-1}\left(\hf\left[\d\x,\d\x\right]+\left[\d\x,\left(\s+\left[\m,\x\right]\right)\d\phi\right]\right)h,
\]
\[
\RR=\F=2\pi h^{-1}\m h\thinspace\delta\left(r\right)\d r\wedge\d\phi,
\]
\[
\T=\P=2\pi h^{-1}\left(\s+\left[\m,\x\right]\right)h\thinspace\delta\left(r\right)\d r\wedge\d\phi.
\]
These expressions are gauge-invariant, since any additional gauge
transformation will produce the same expression with the new gauge
parameters composed with the old ones in a well-defined way. The process
of adding new degrees of freedom in order to make variables invariant
under gauge transformations is called \emph{dressing}.

Finally, let us define a \emph{momentum }$\p$ and \emph{angular momentum
}$\j$:
\[
\p\equiv h^{-1}\m h,\qquad\j\equiv h^{-1}\left(\s+\left[\m,\x\right]\right)h,
\]
which satisfy, as one would expect, the relations
\[
\p^{2}\equiv\m^{2}\sp\p\cdot\j=\m\cdot\s.
\]
Then we have
\[
\F=2\pi\p\thinspace\delta\left(r\right)\d r\wedge\d\phi\sp\T=2\pi\j\thinspace\delta\left(r\right)\d r\wedge\d\phi.
\]
We see that the source of curvature is momentum, while the source
of torsion is angular momentum, as discussed in \cite{Meusburger:2003ta,Freidel2004,Meusburger:2005in,Meusburger:2005mg,Schroers2007,Meusburger:2008dc}
for the 2+1D case.

In conclusion, in the Ashtekar formulation, a cosmic string is a distributional
source of both curvature and torsion in an otherwise flat and torsionless
spacetime.

\bibliographystyle{Utphys}
\phantomsection\addcontentsline{toc}{section}{\refname}\bibliography{Spin_Networks_and_Cosmic_Strings}

\end{document}